**Figure 1. An m=3 strategy.**

| m-string | Prediction |
|---|---|
| 000 | 0 |
| 001 | 1 |
| 010 | 1 |
| 011 | 0 |
| 100 | 0 |
| 101 | 1 |
| 110 | 0 |
| 111 | 1 |

**Figure 2: N=101 s=1**

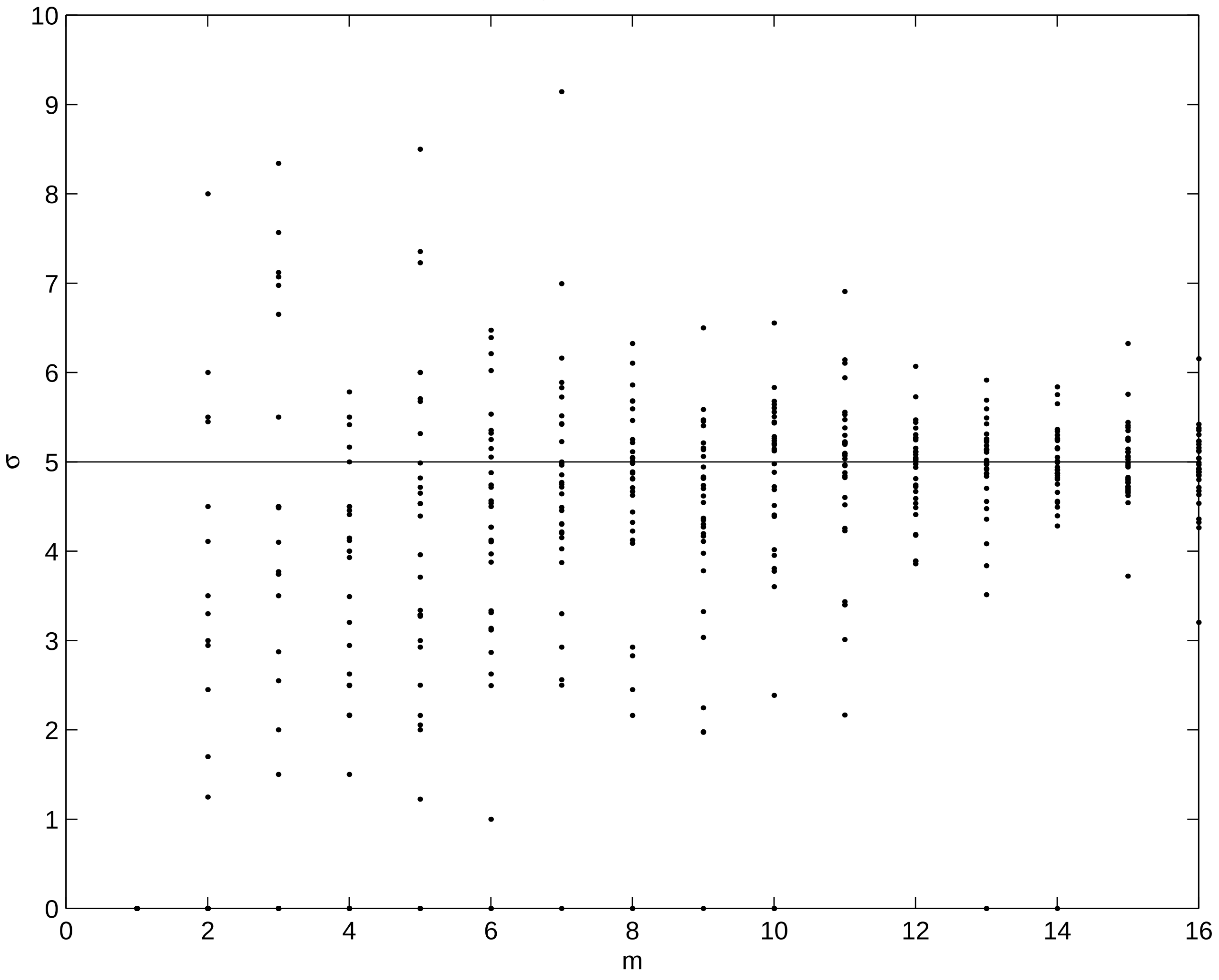

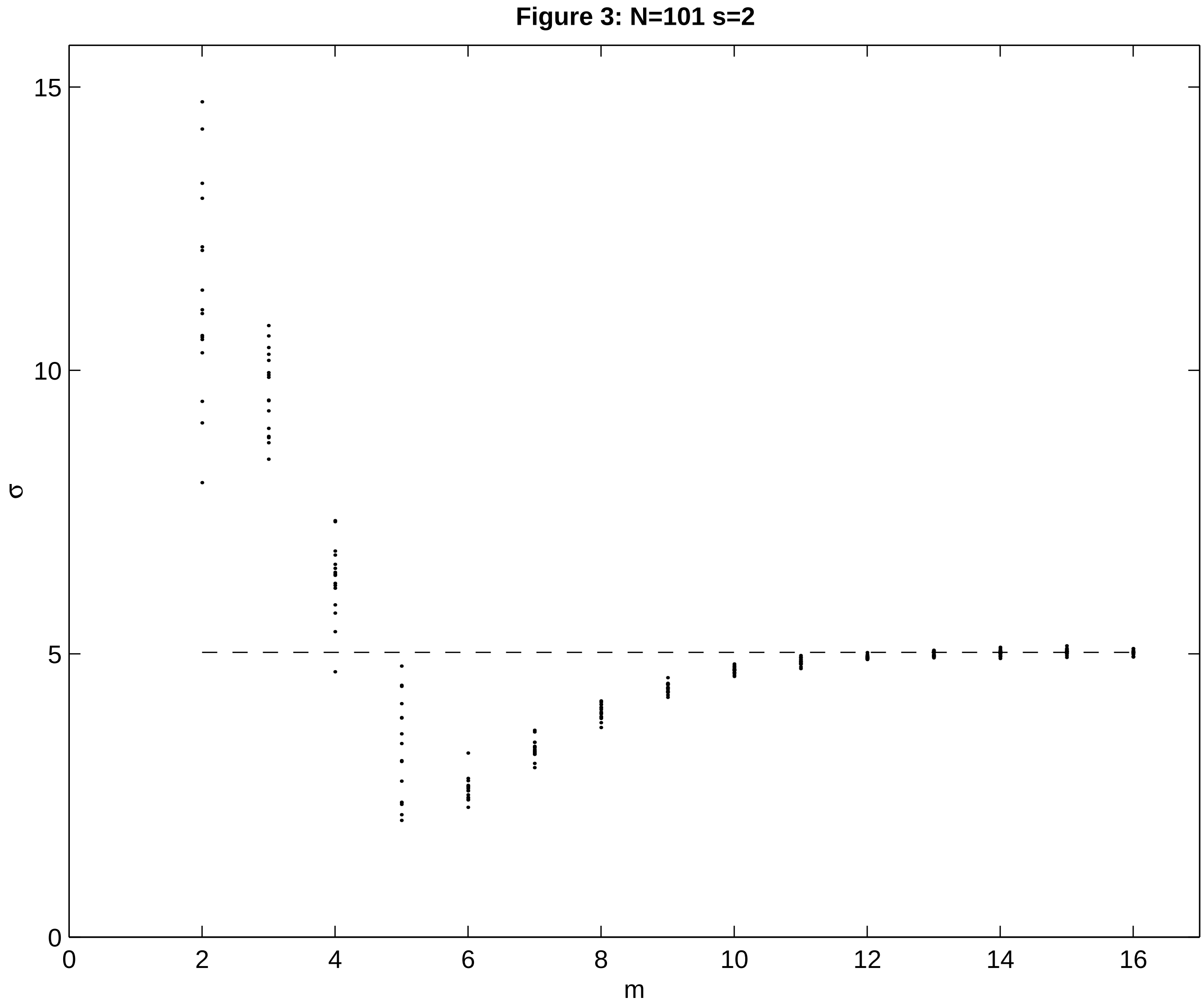

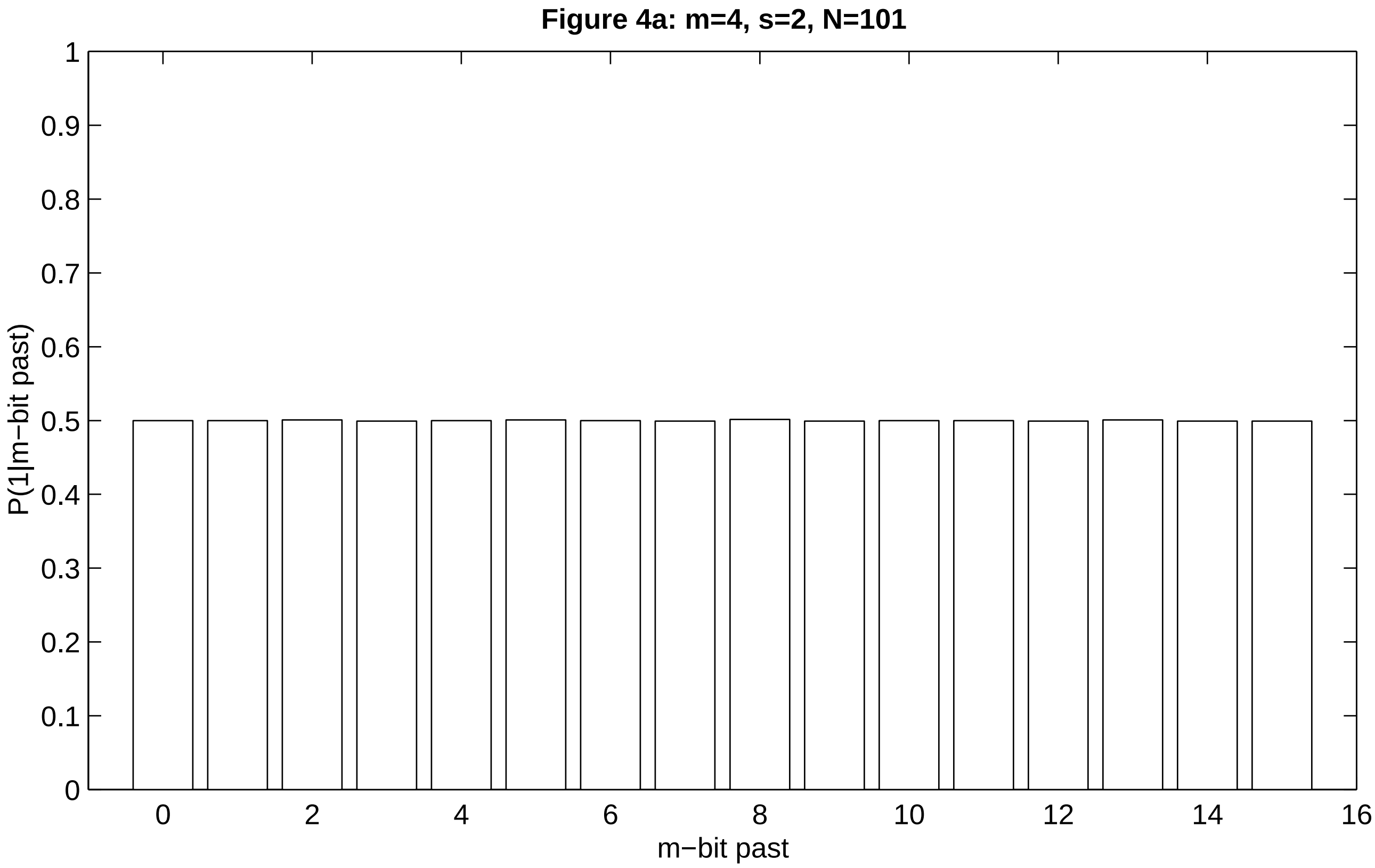

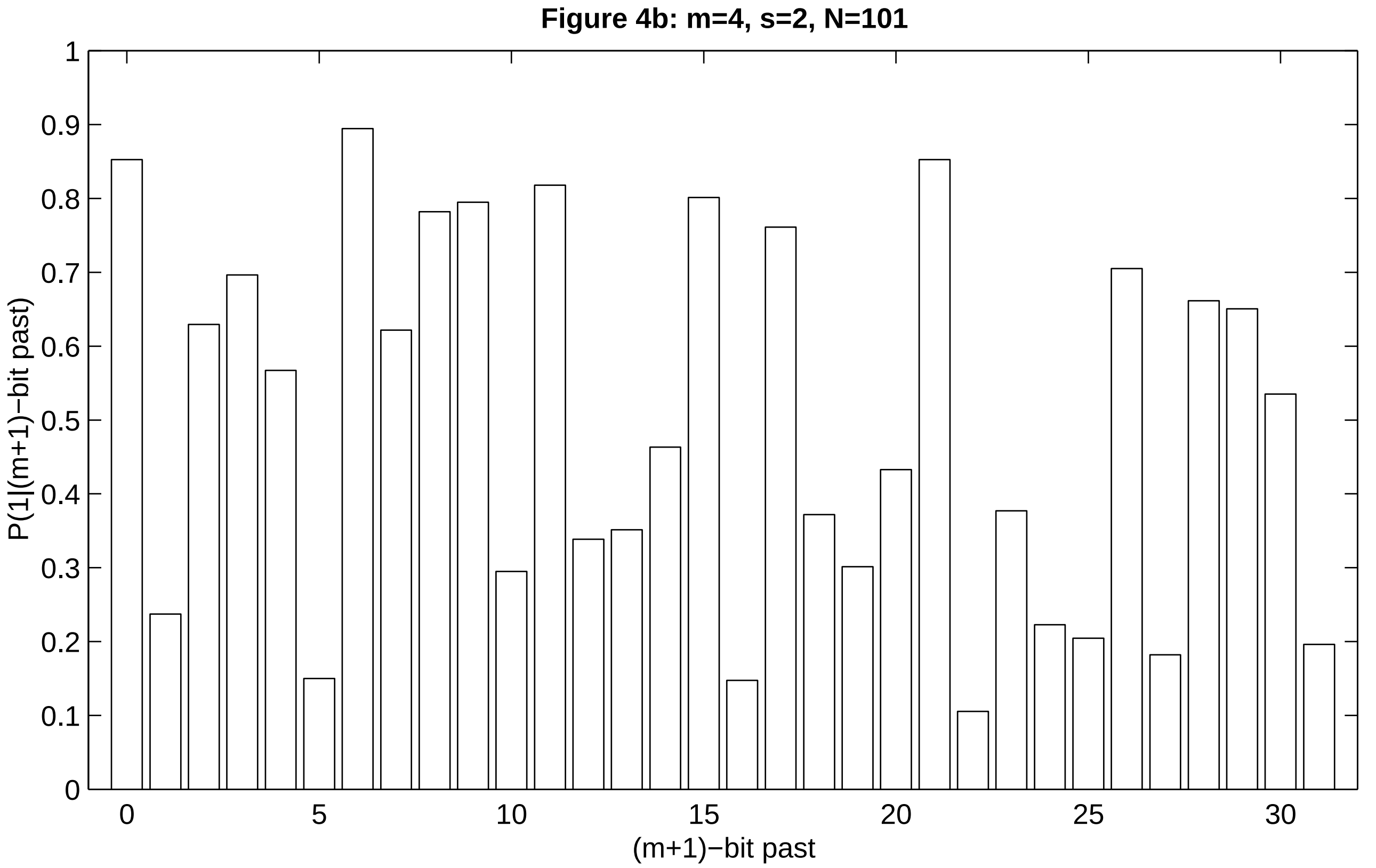

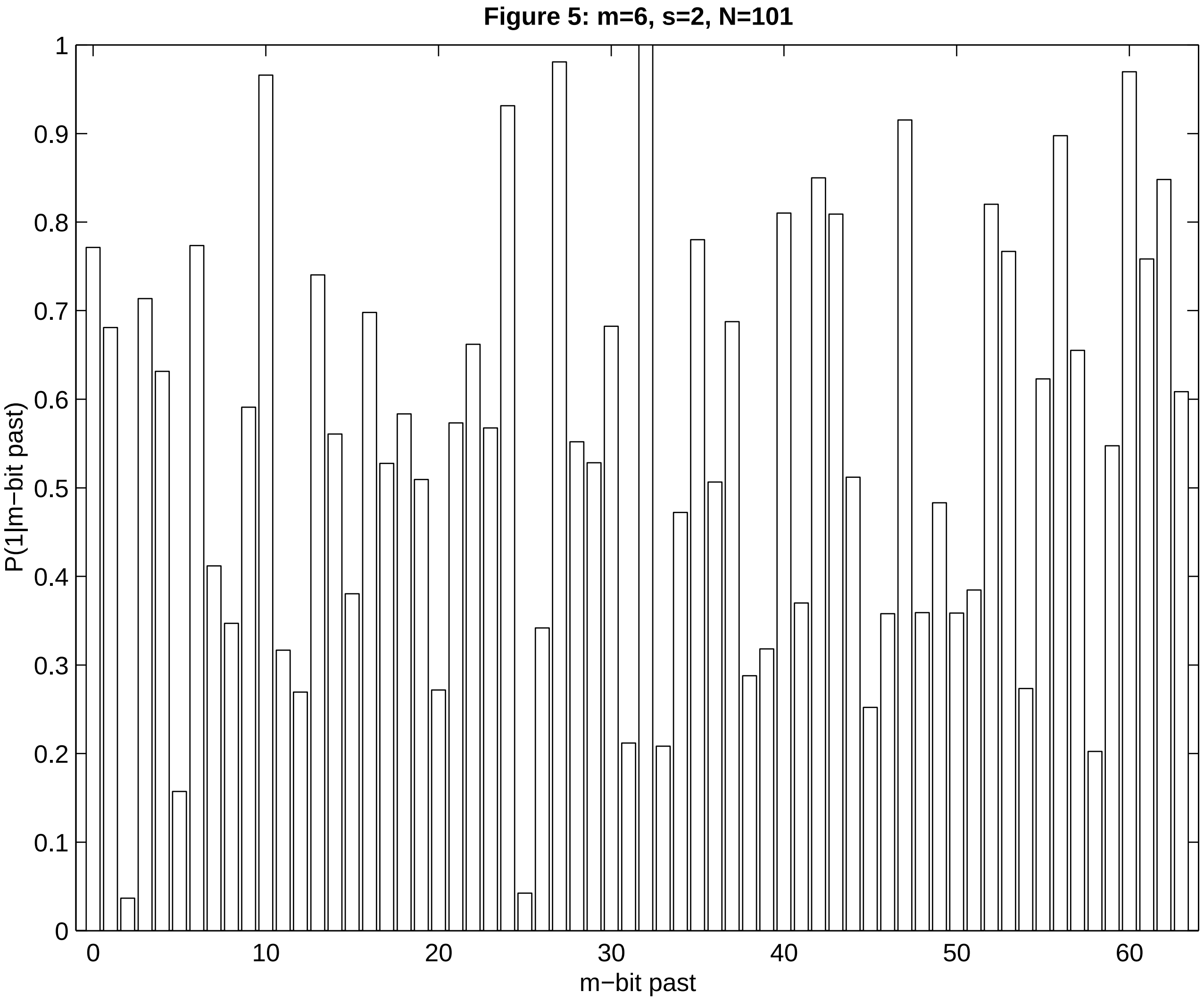

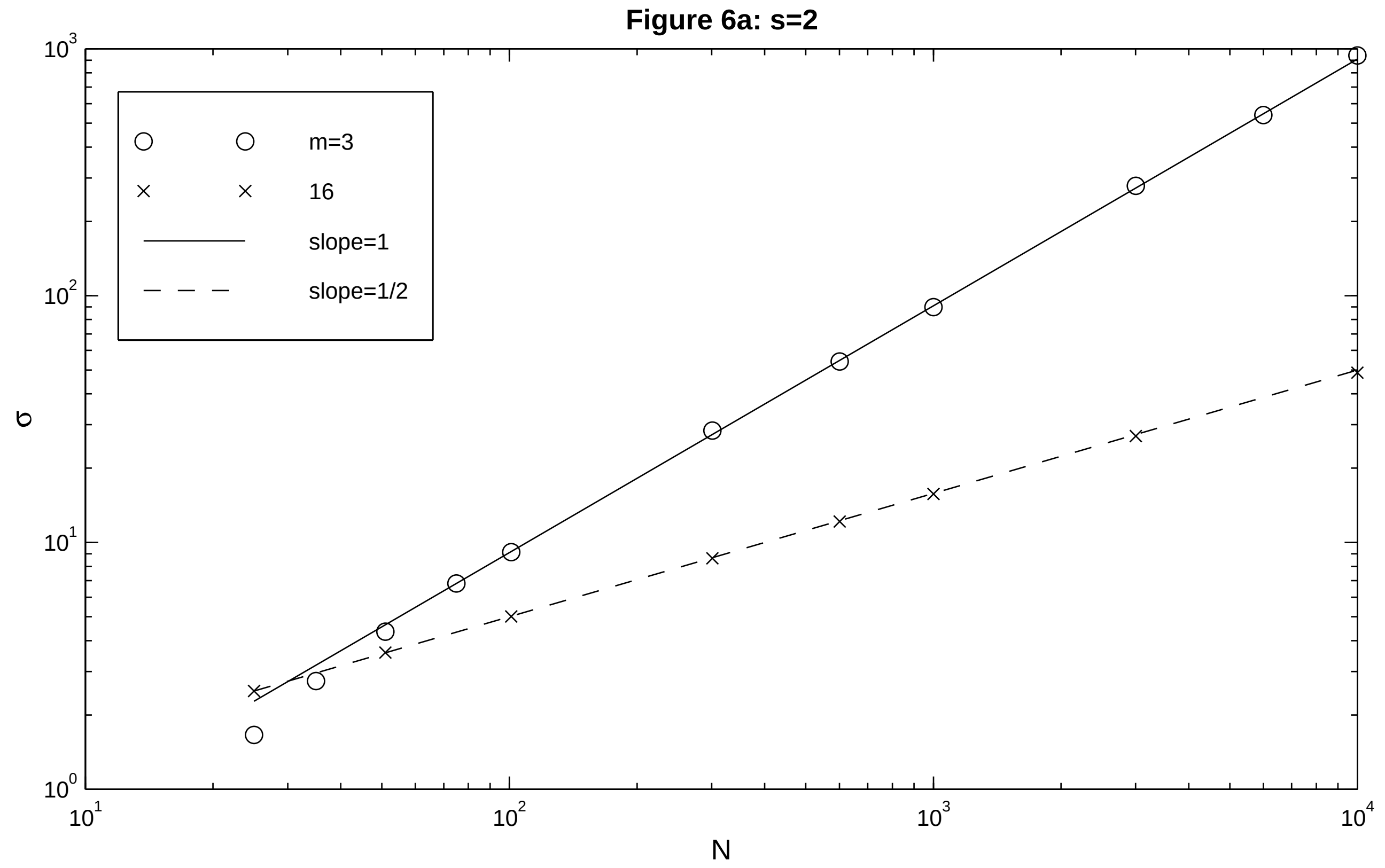

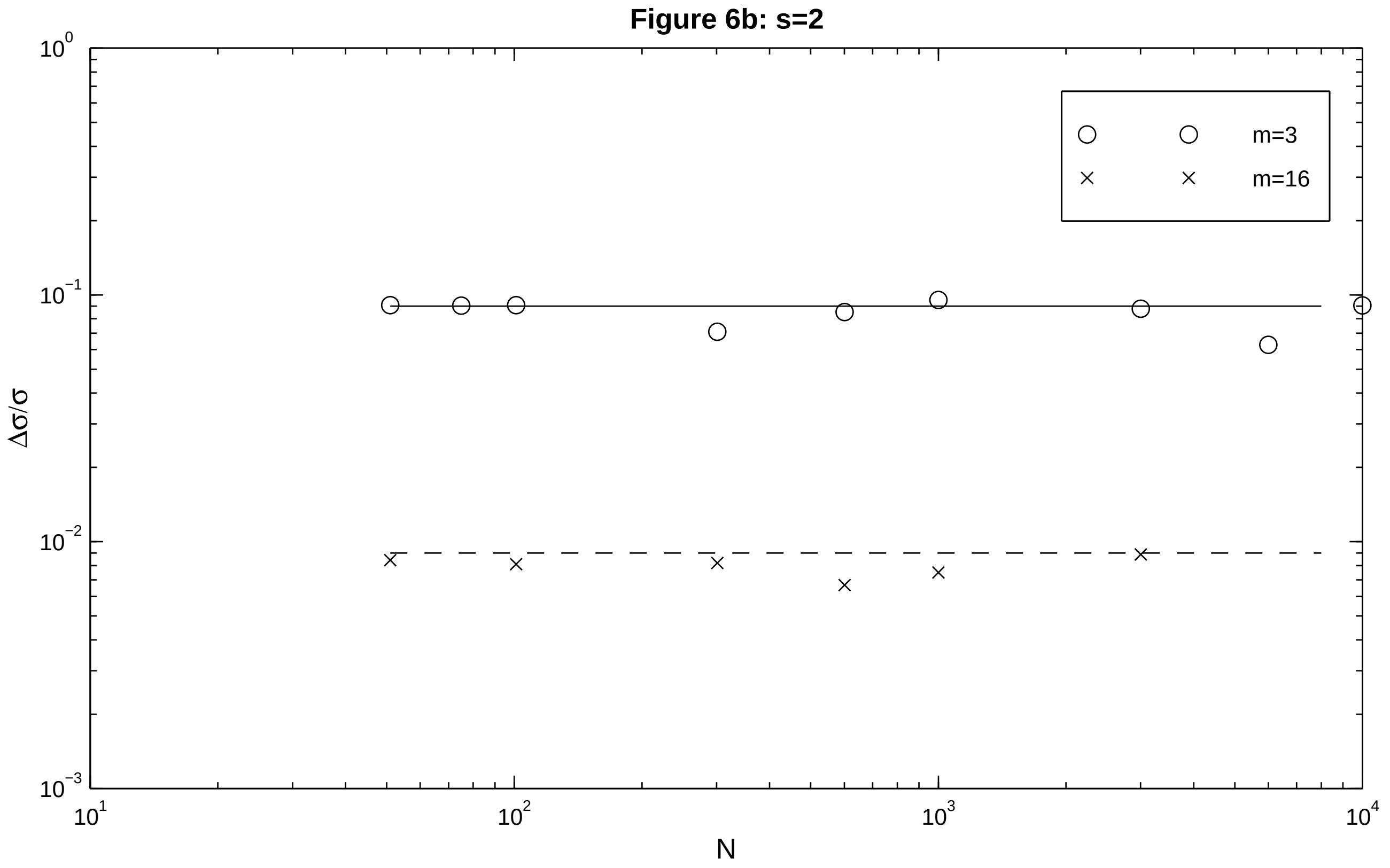

**Figure 7: s=2**

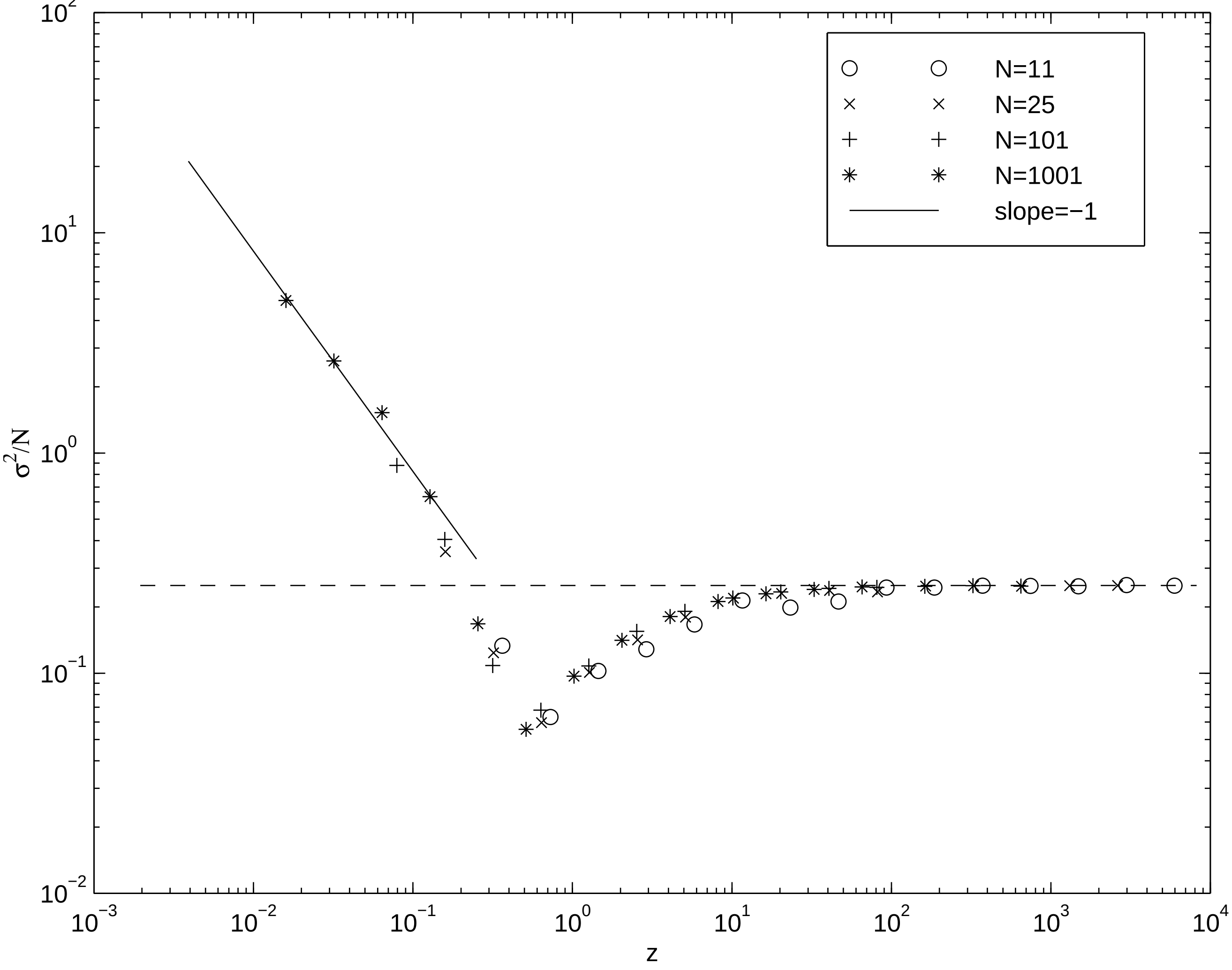

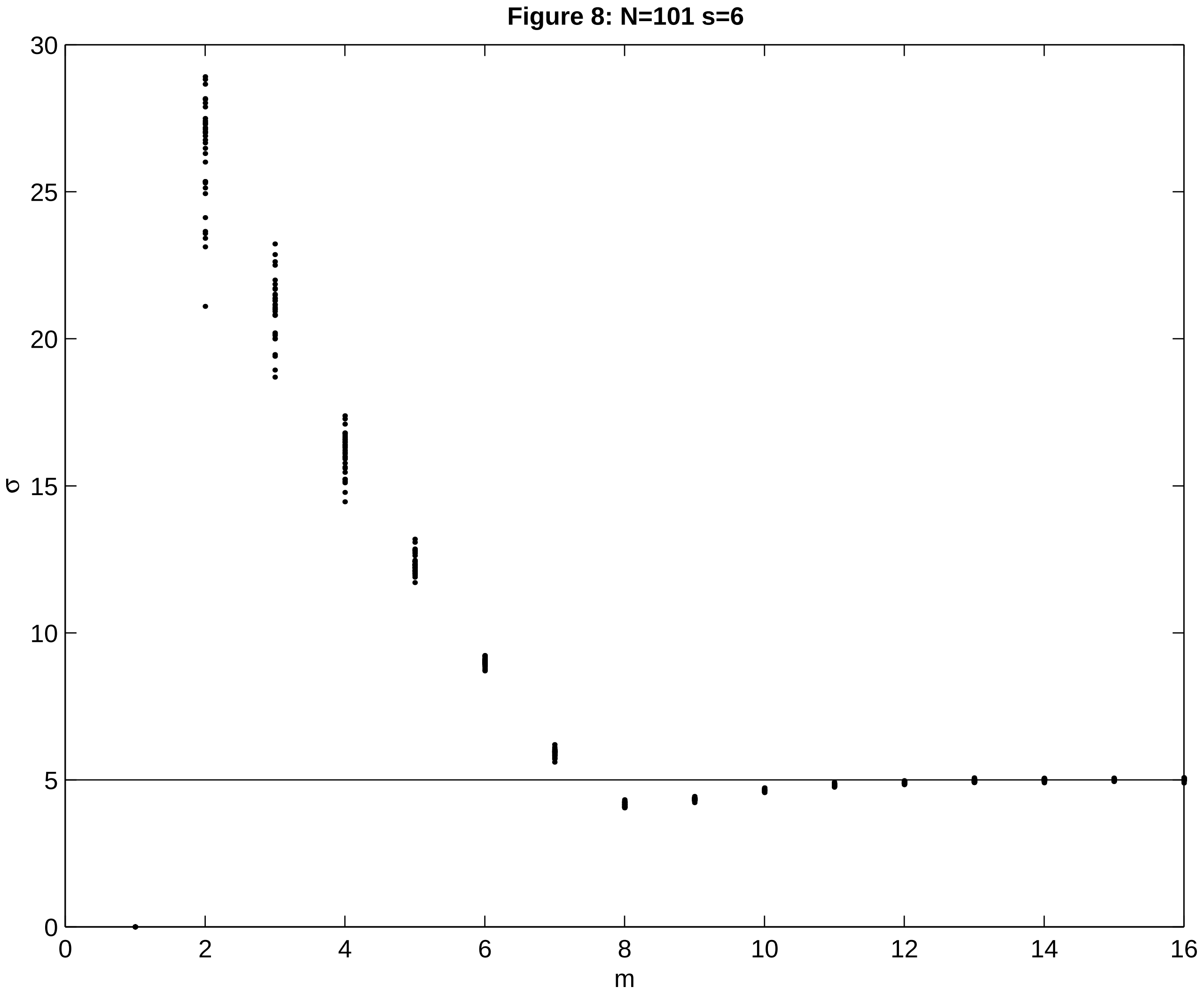

**Figure 9a: s=6**

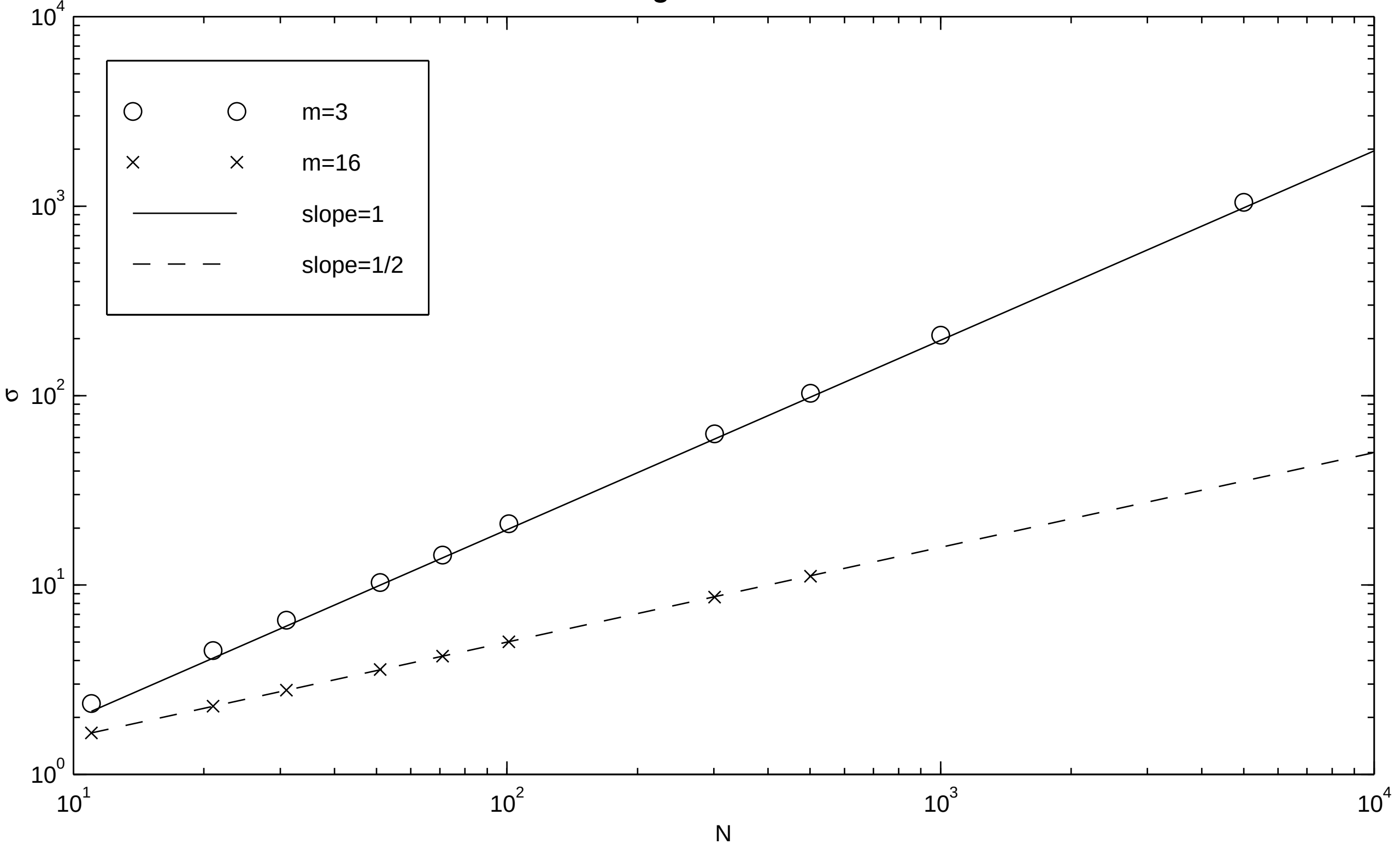

**Figure 9b: s=6**

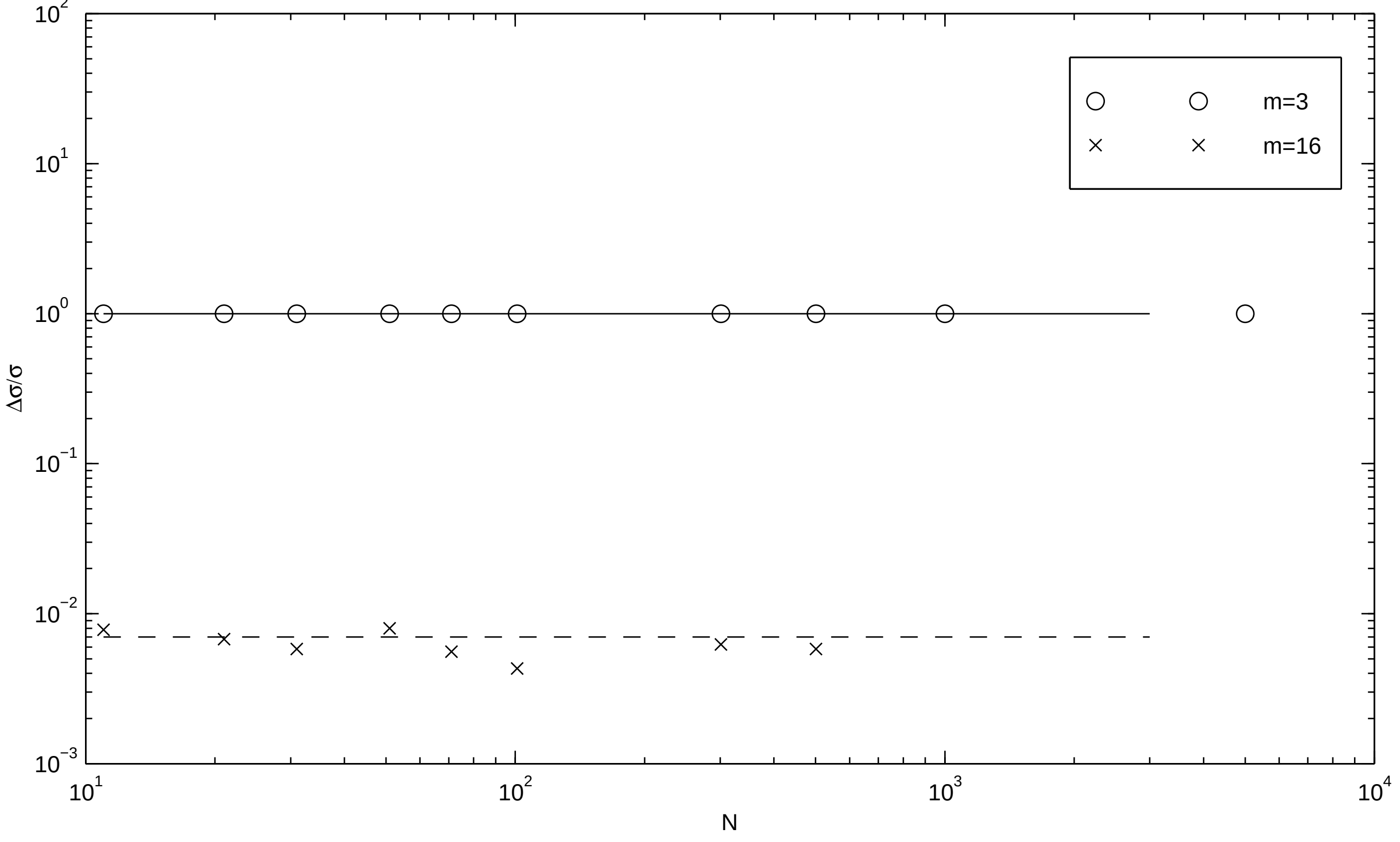

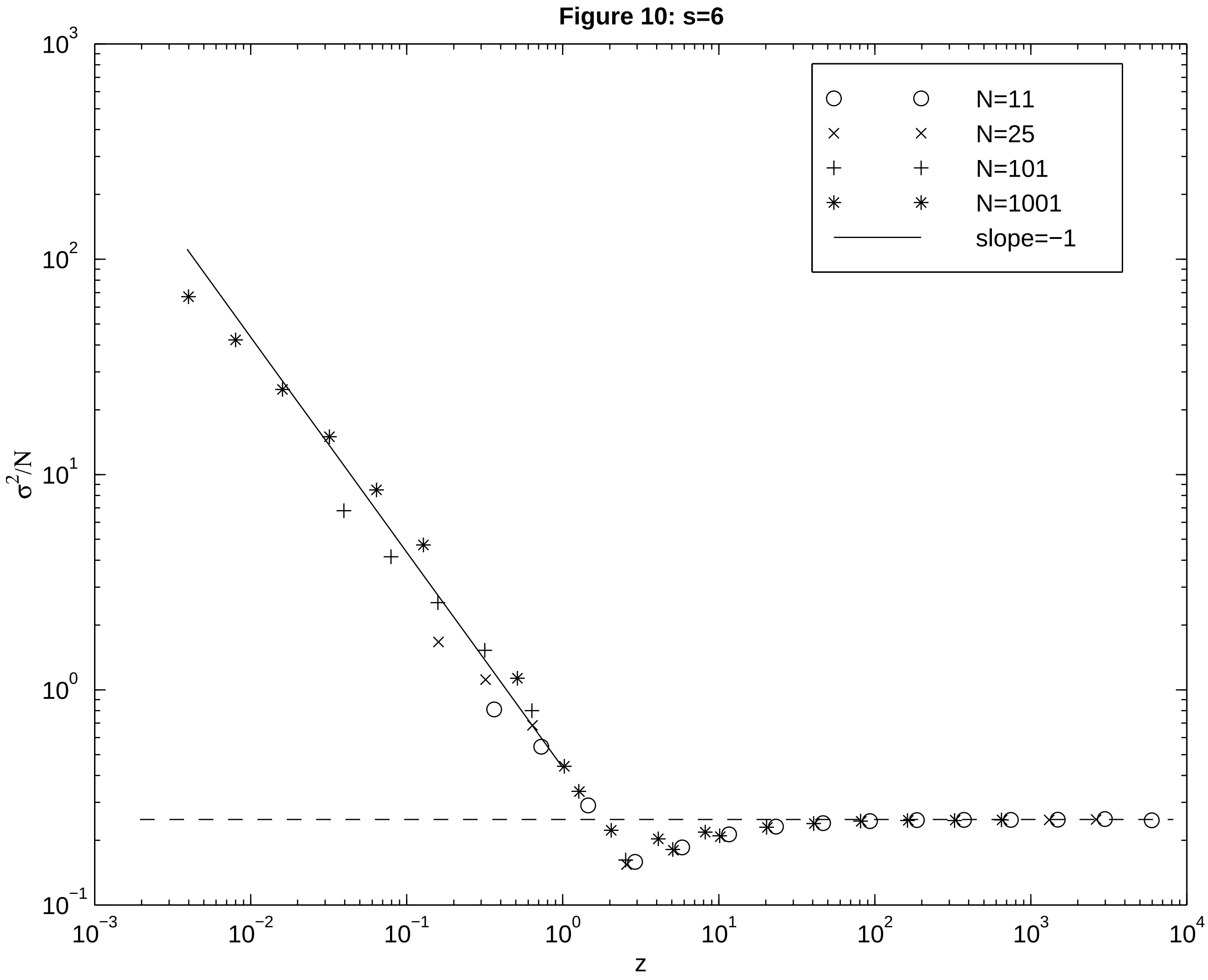

**Figure 11.**

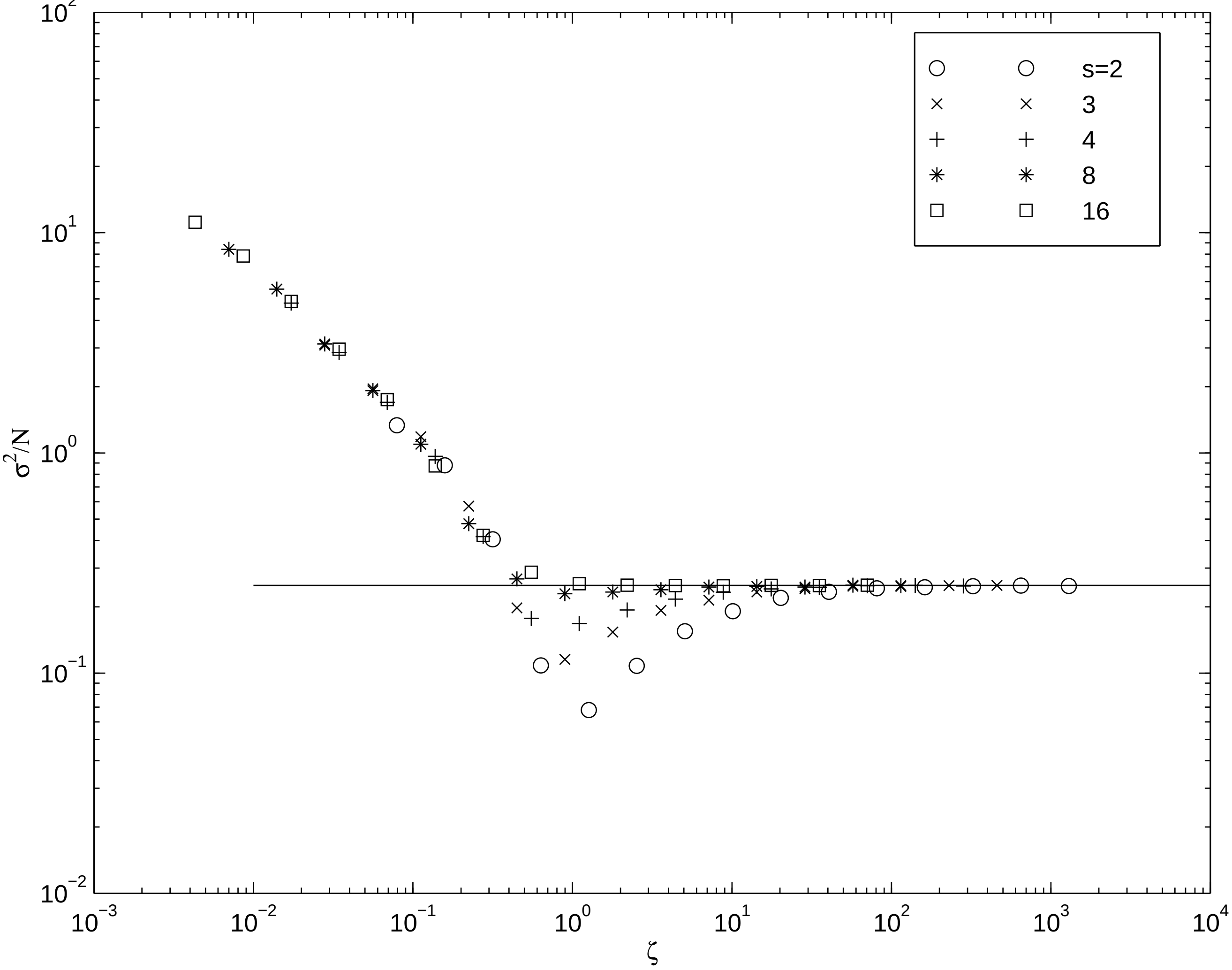

Figure 12a: m=4, s=6, N=101

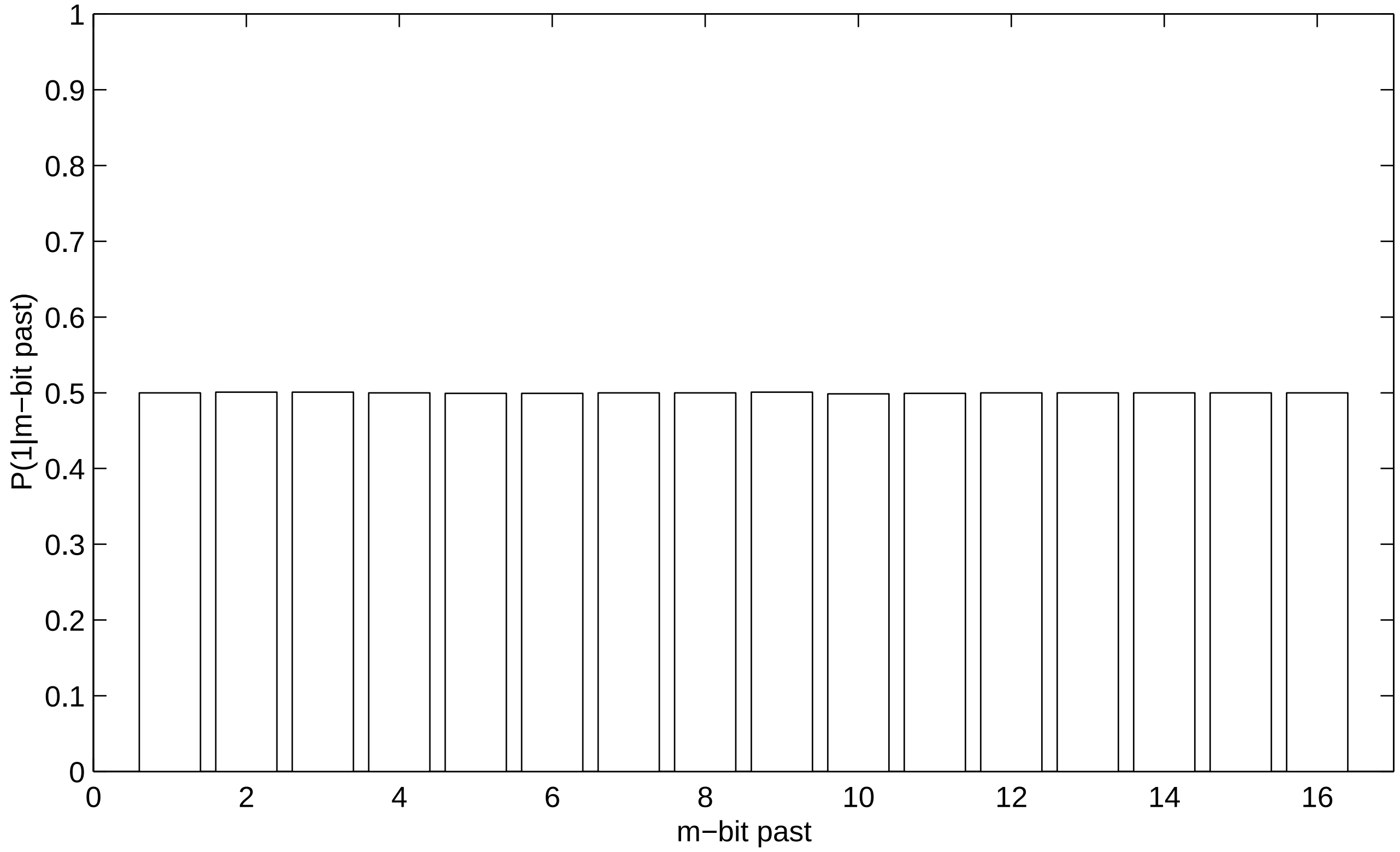

Figure 12b: m=4, s=6, N=101

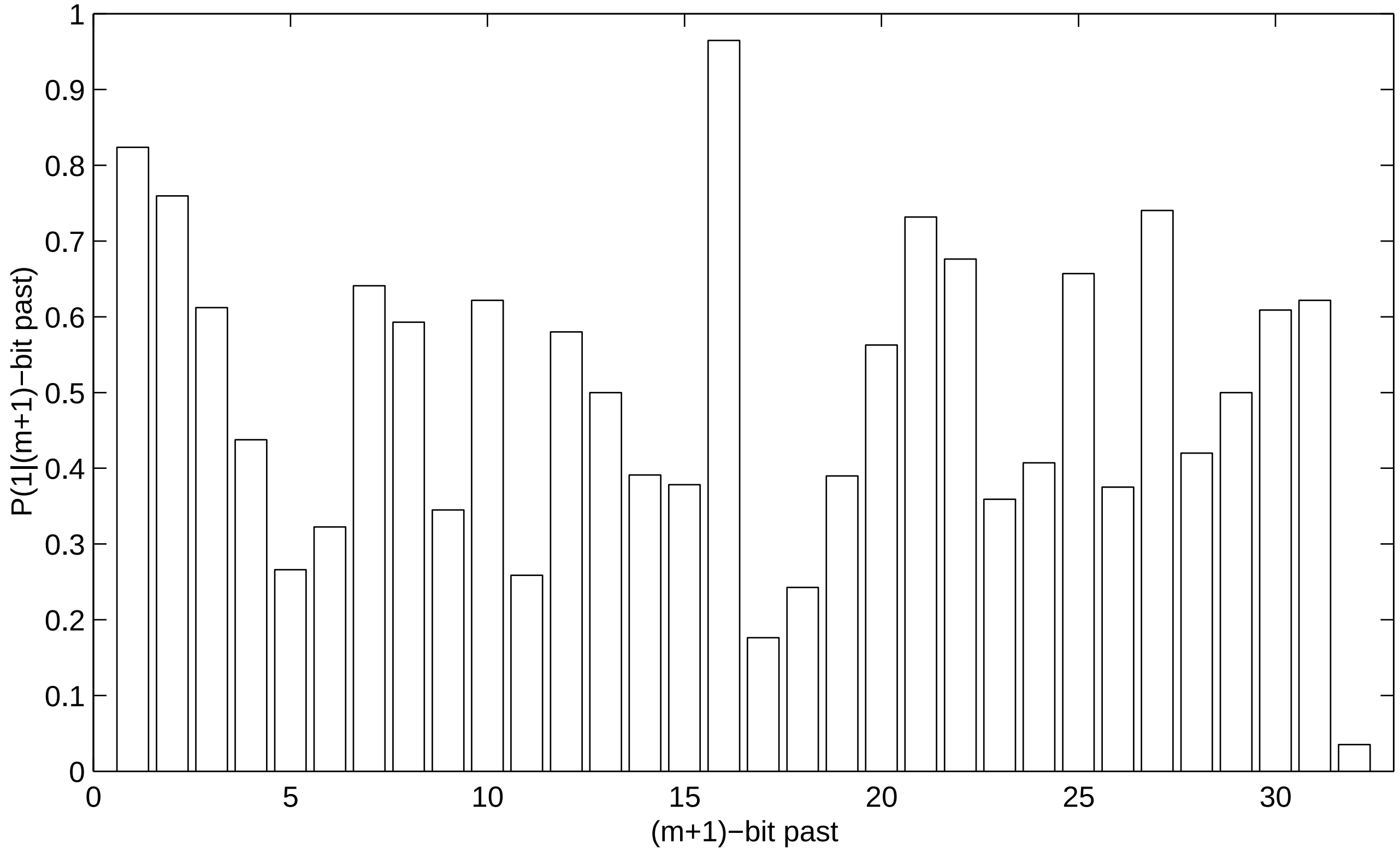

**Figure 13: m=7, s=6, N=101**

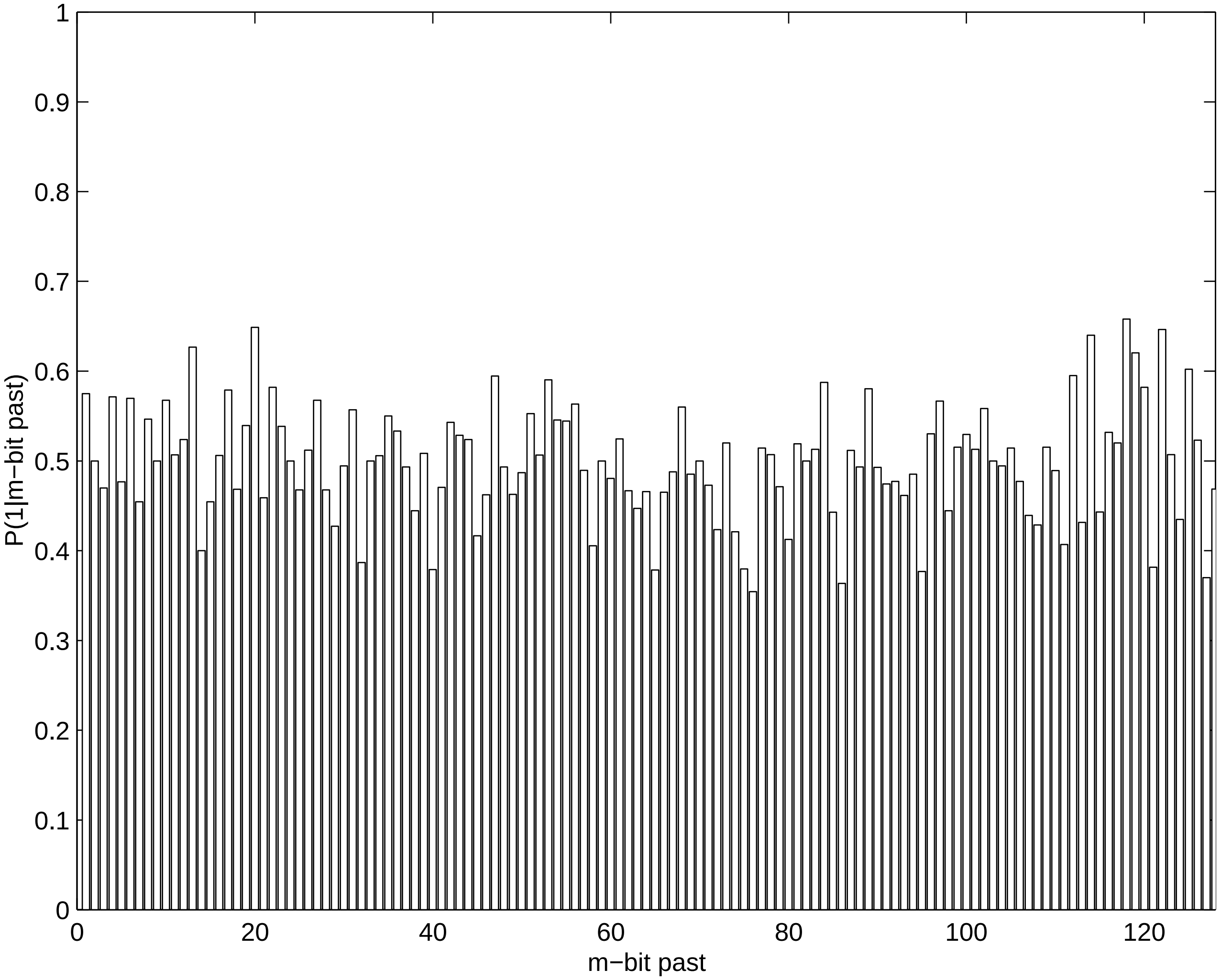

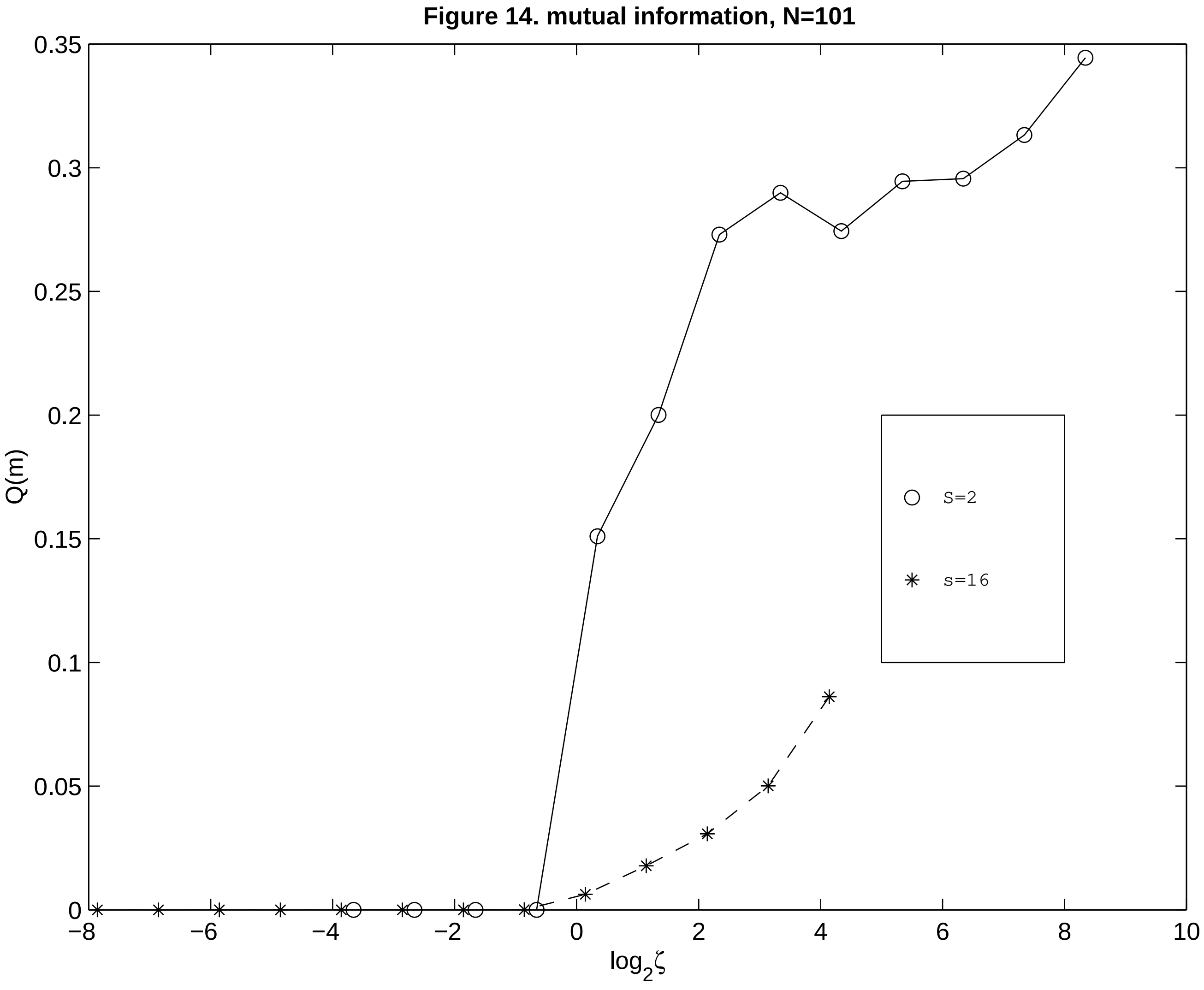

Figure 15a, Segment of time series of $L_1$, for m=2, s=2, N=101

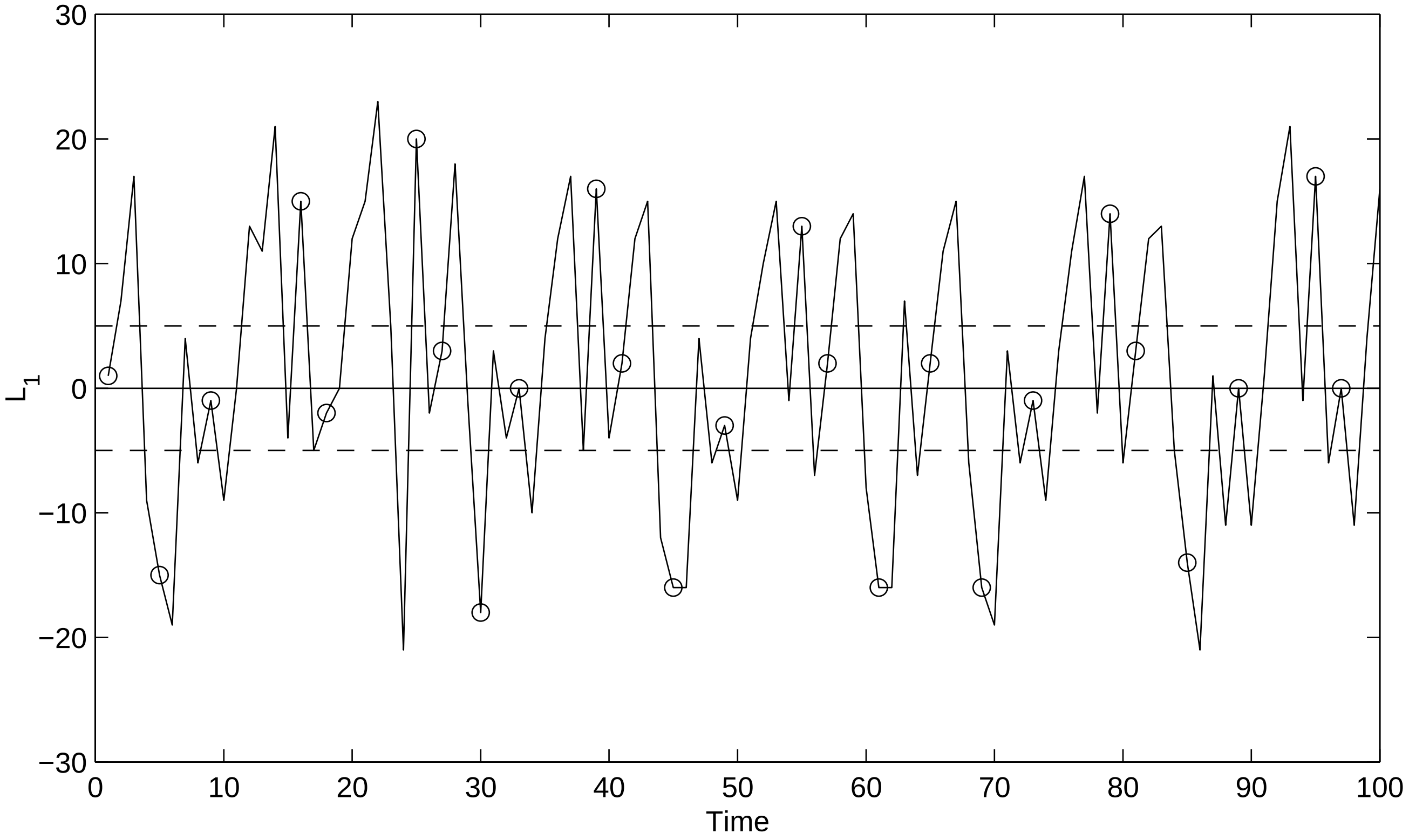

Figure 15b, Segment of time series of $L_1$, for m=8, s=2, N=101

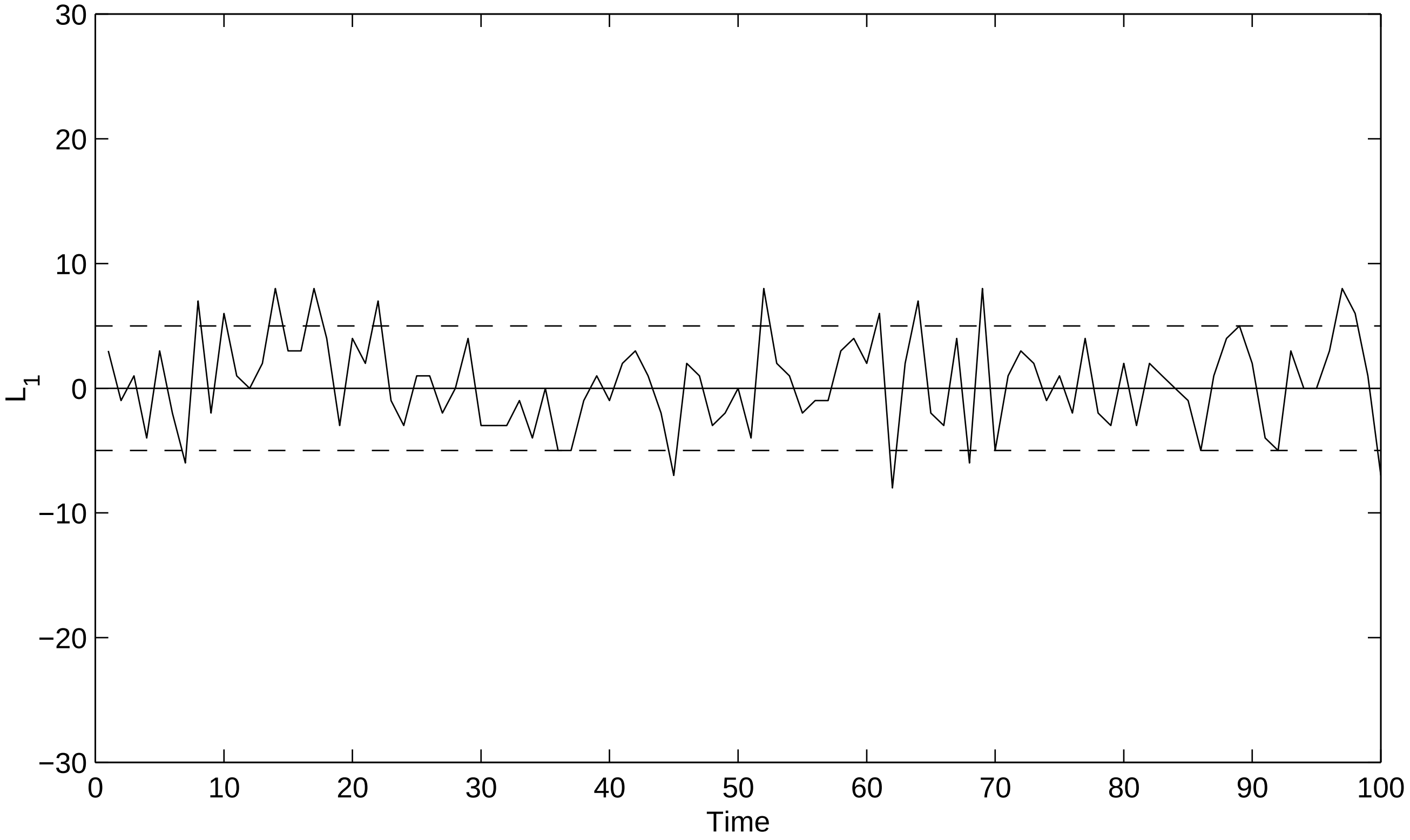

**Figure 16, Segment of time series of $L_1$, for m=2, s=6, N=101**

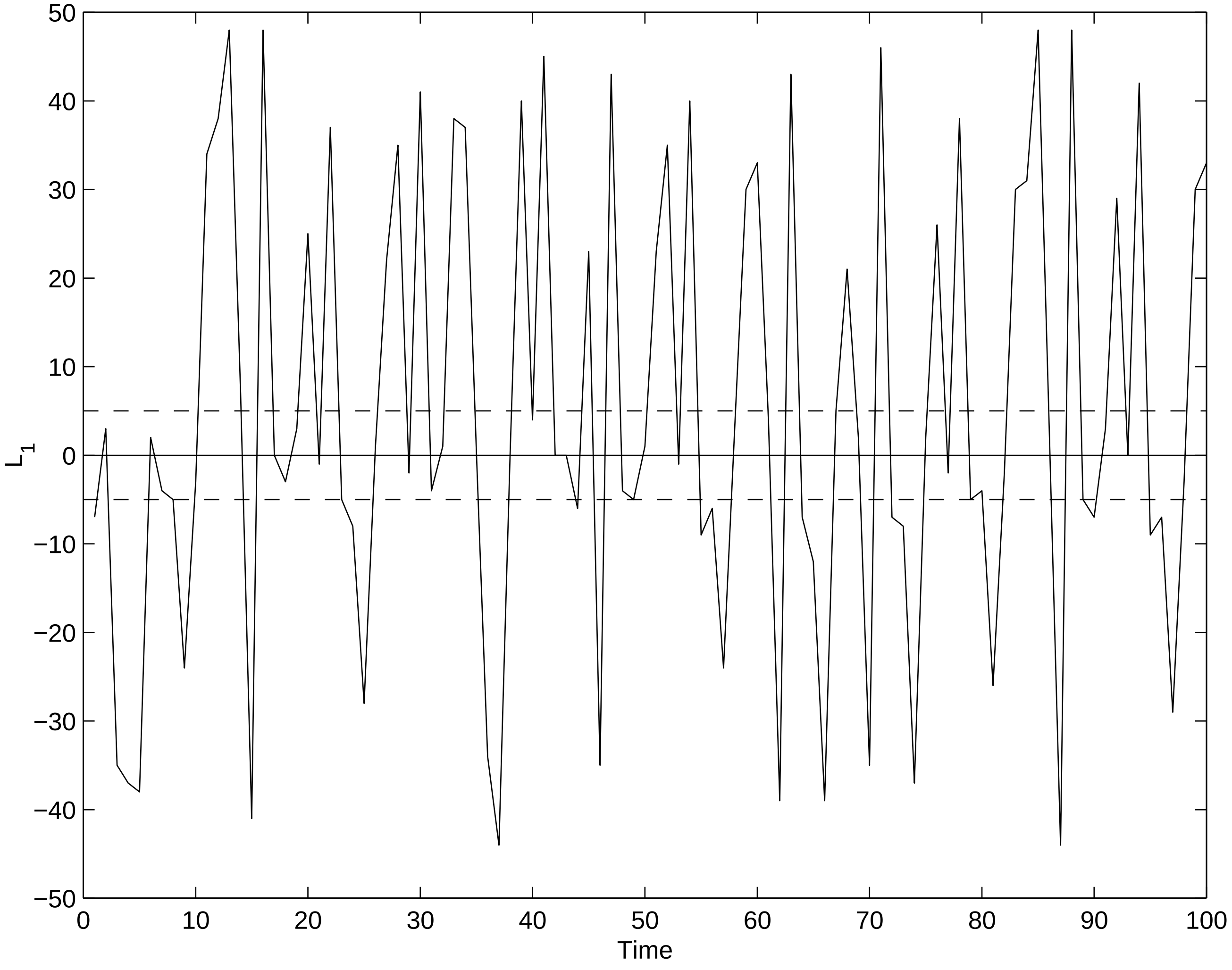

Figure 17: N=101 s=2

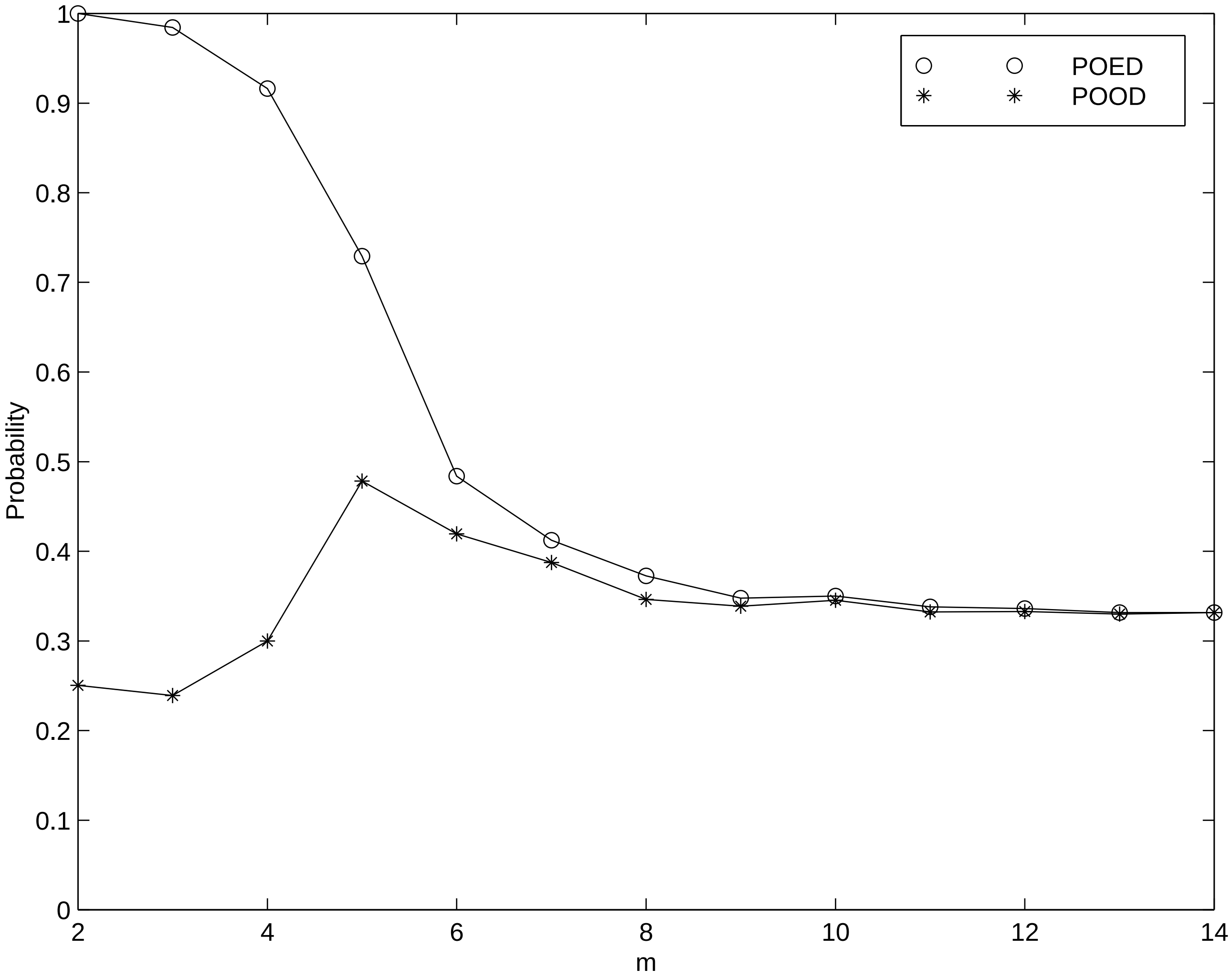

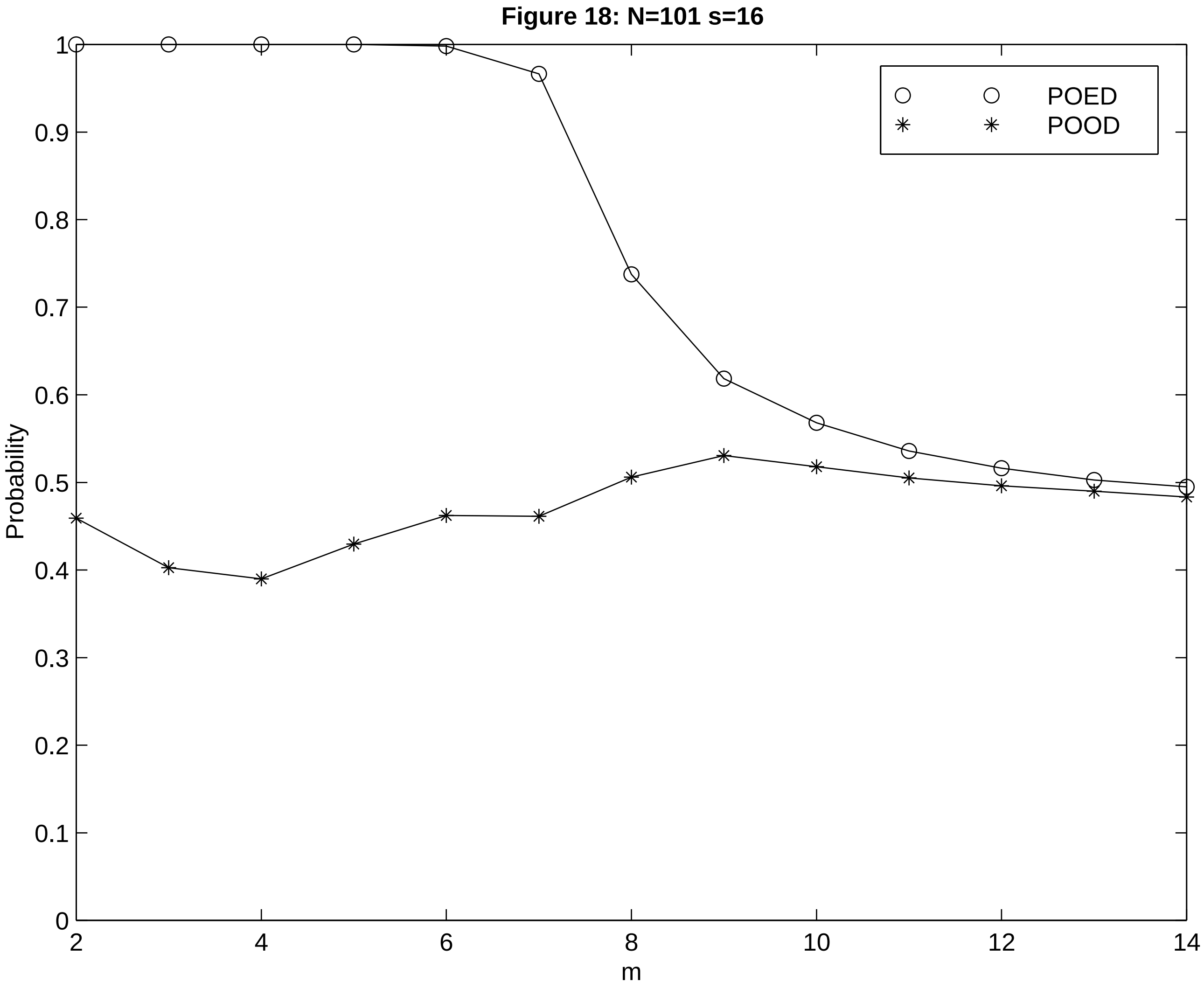

**Figure 19, Segment of time series of $L_1$, for m=5, s=2, N=101**

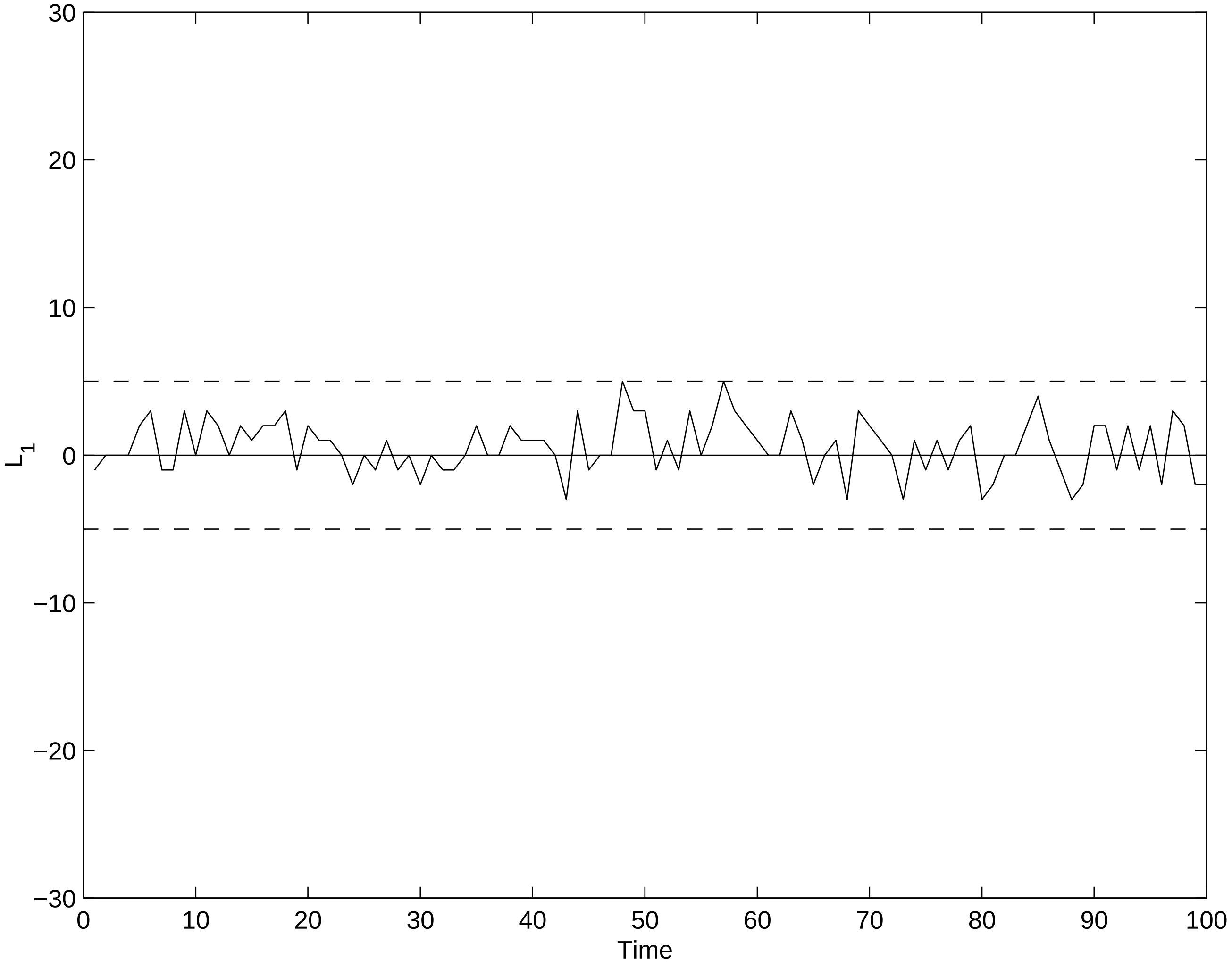

**Figure 20. An m=3.46 strategy.**

| m-string | Prediction |
|---|---|
| 0000 | 0 |
| 0010 | 1 |
| *011 | 1 |
| *110 | 0 |
| 0111 | 0 |
| 1000 | 1 |
| *001 | 0 |
| 1010 | 1 |
| *100 | 0 |
| *101 | 0 |
| 1111 | 1 |

**Figure 21a. s=2, N=101**

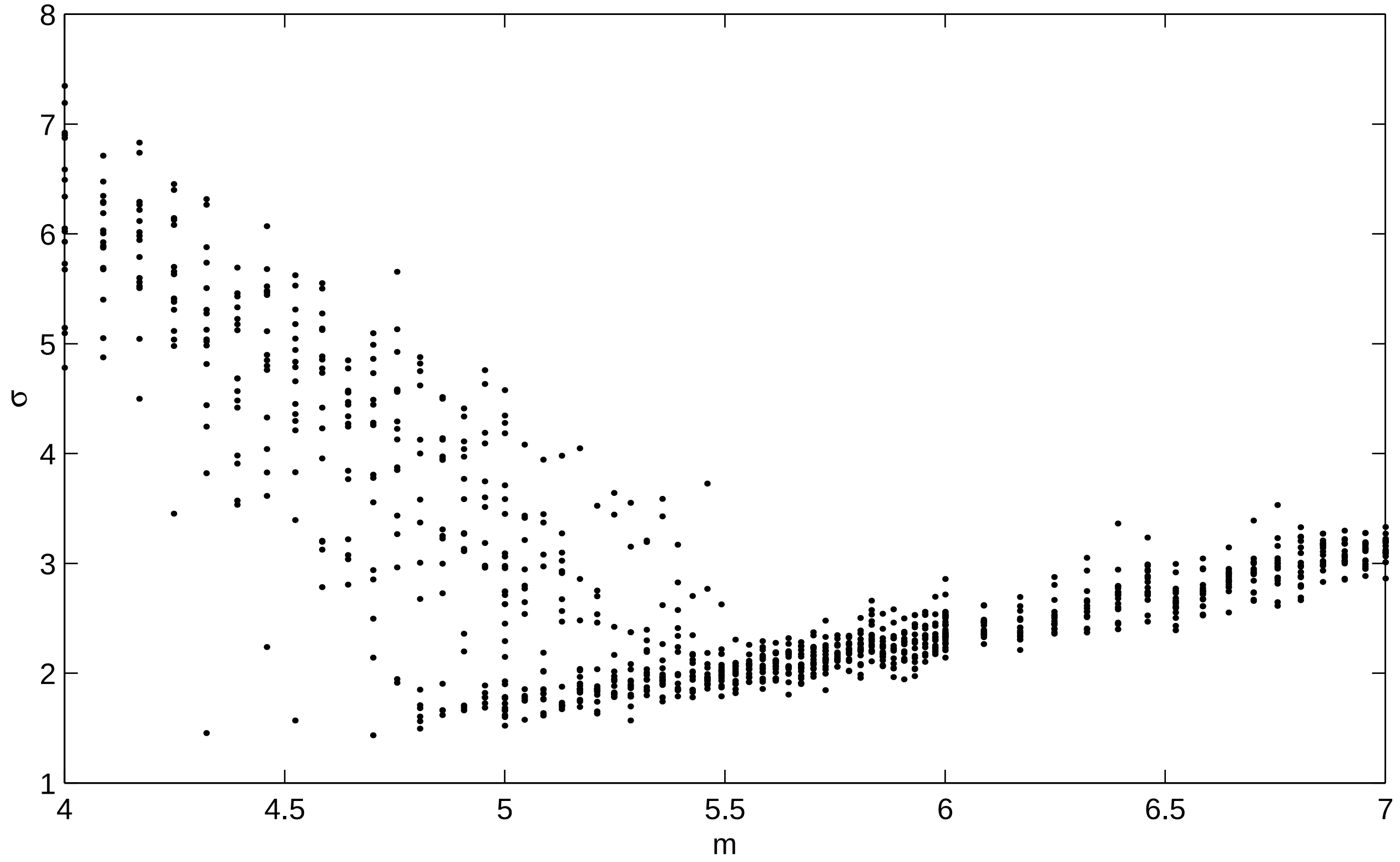

**Figure 21b. s=2, N=101**

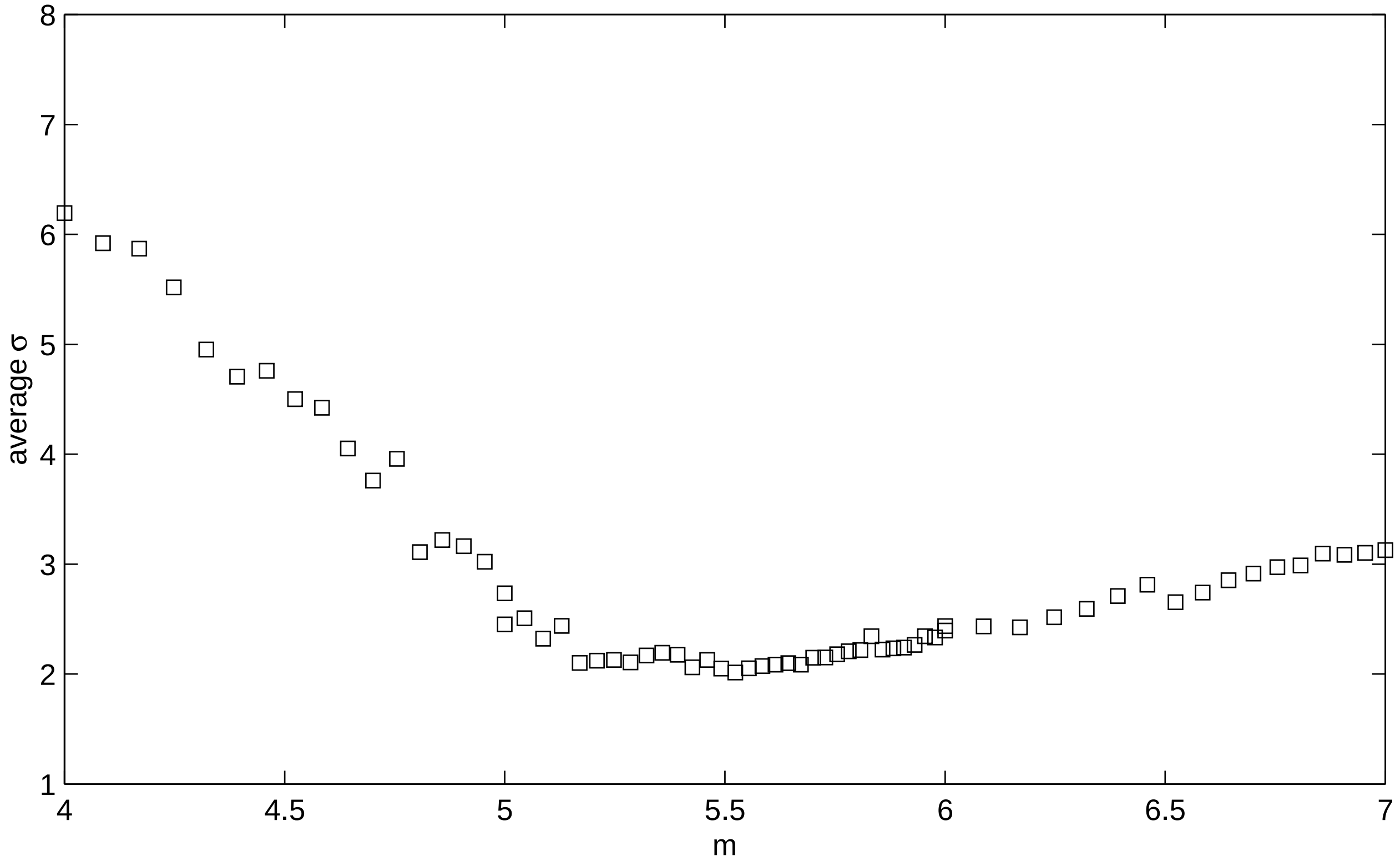

**Figure 22. s=2, N=101**

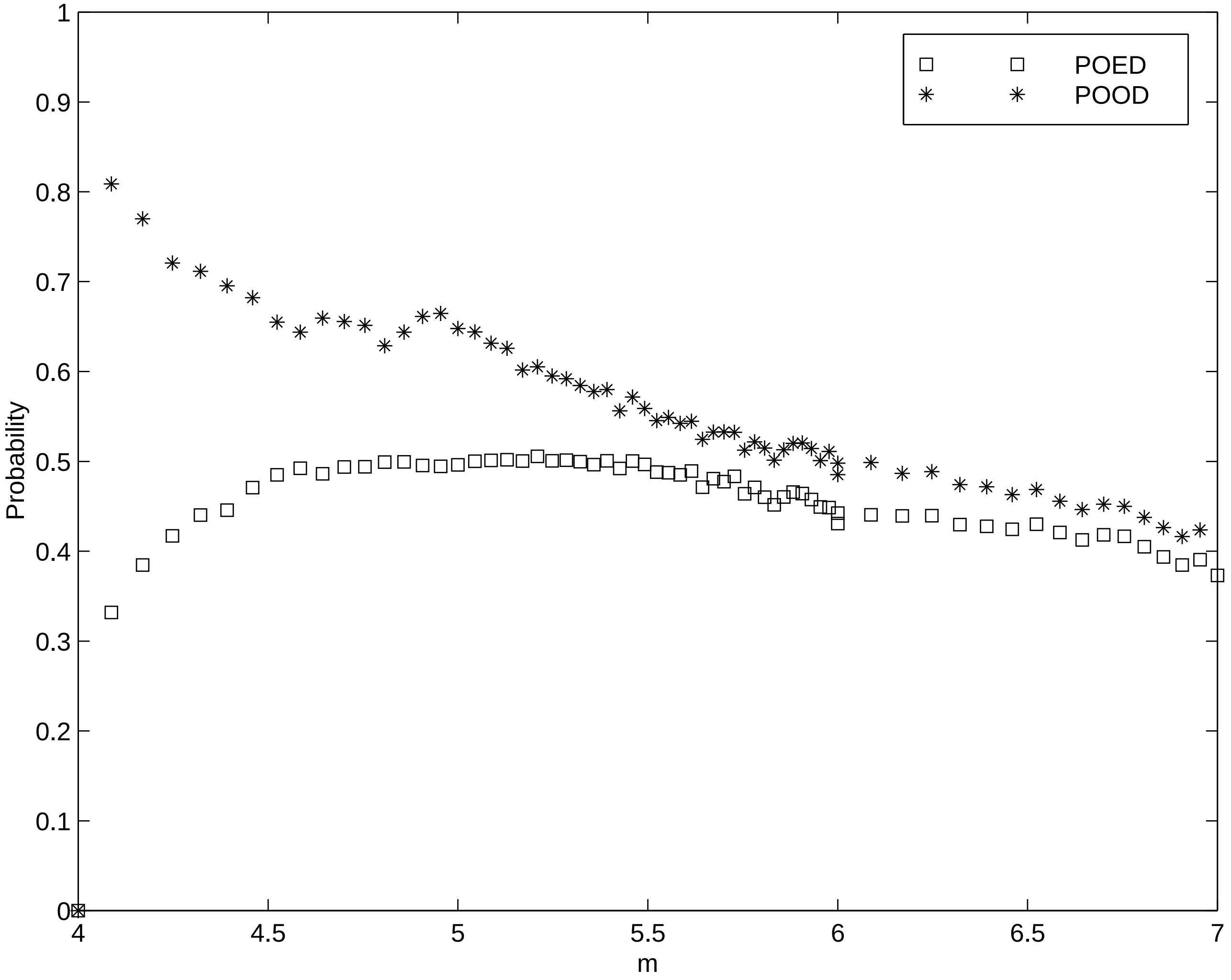

**Figure 23. $I_m^{[2]}$ vs. m, s=2 N=101.**

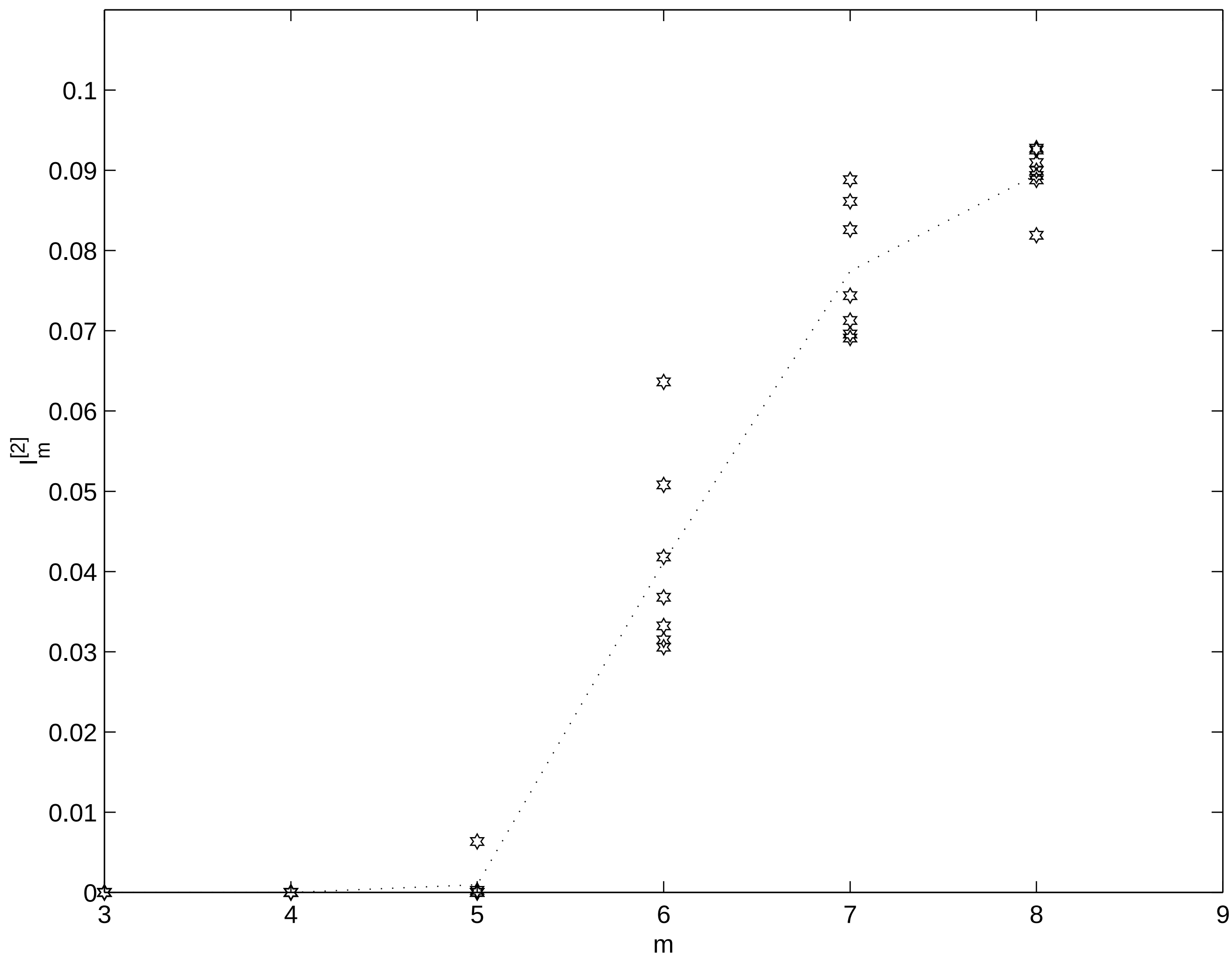

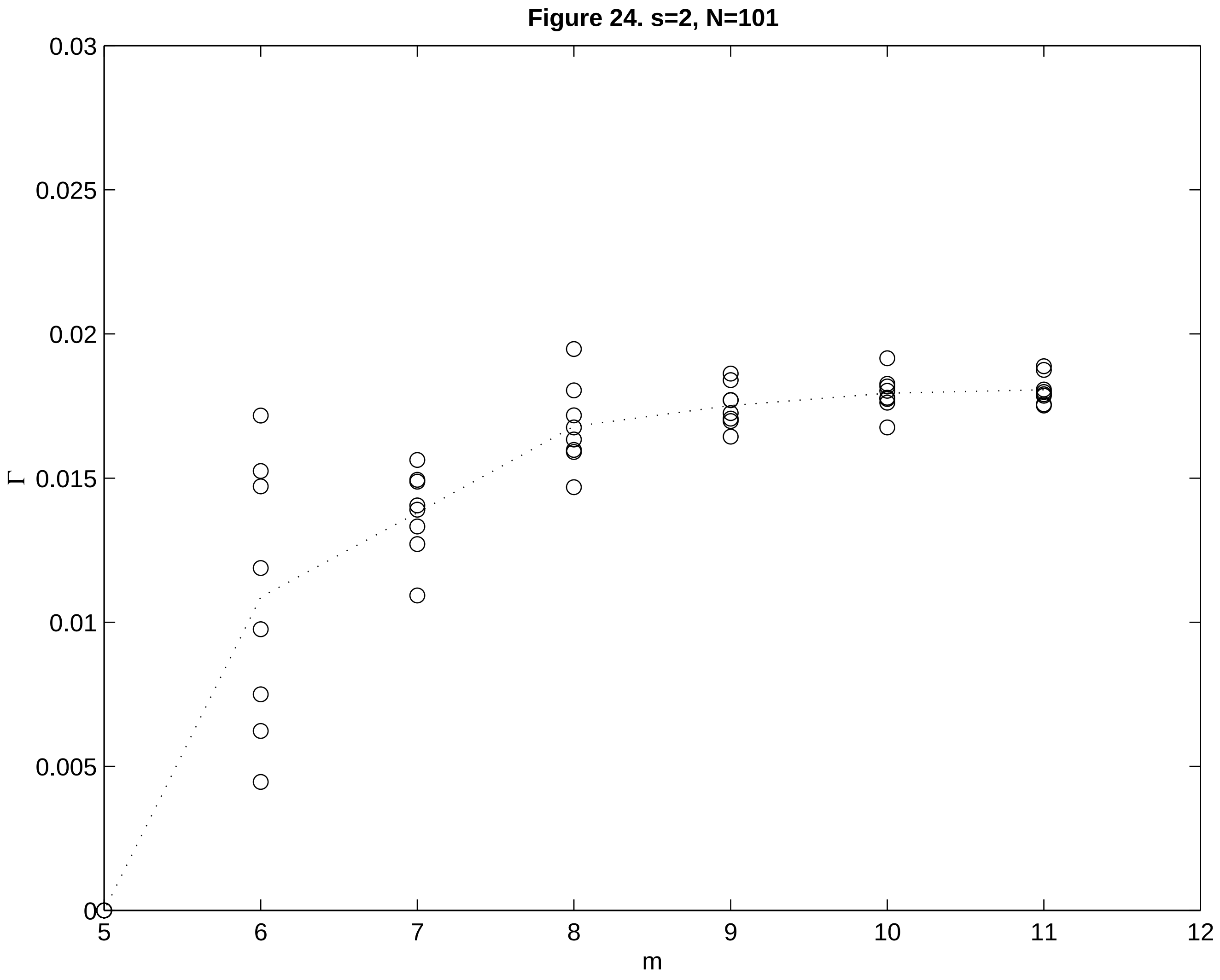

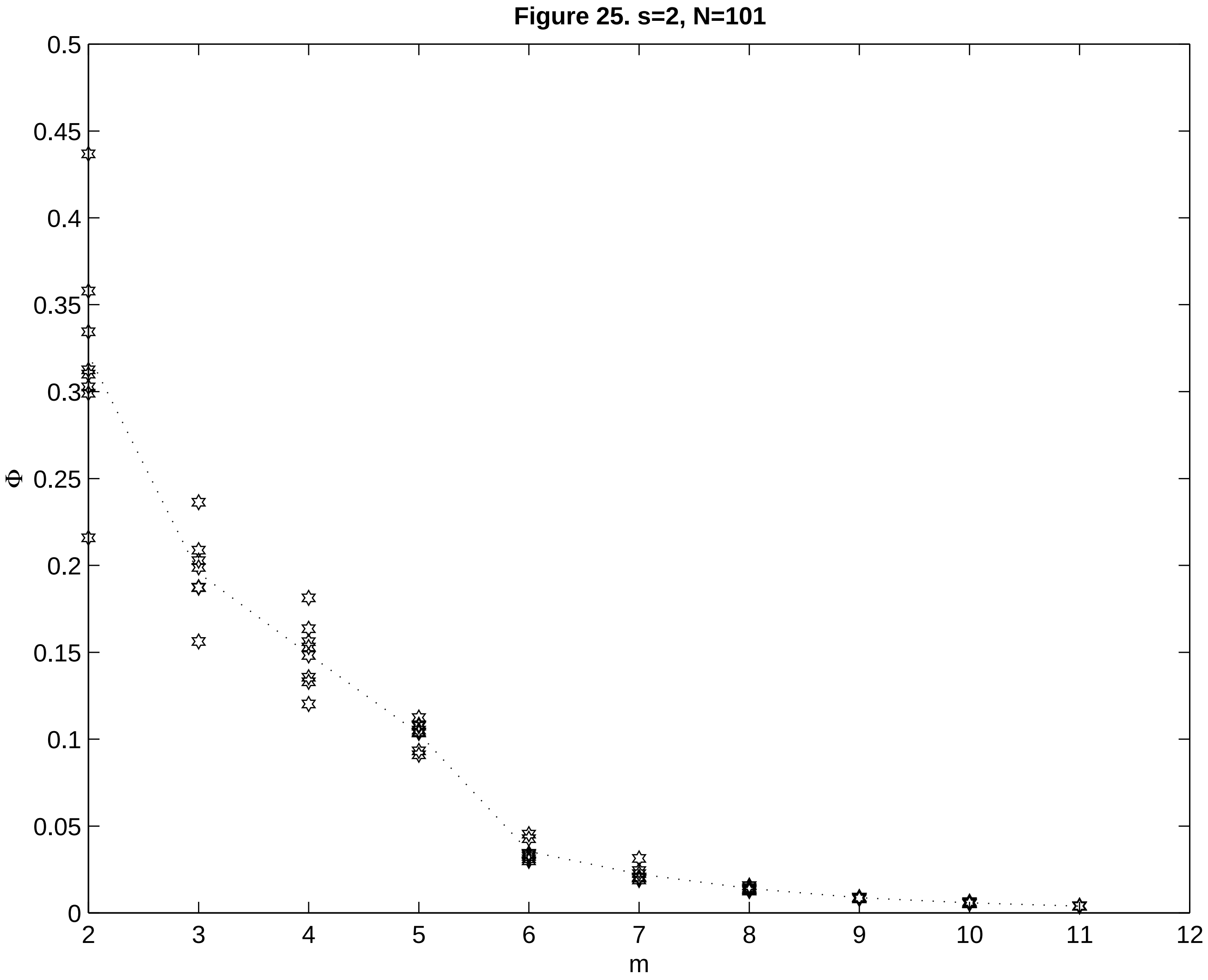

**Figure 26a. N=101, s=2, m=3**

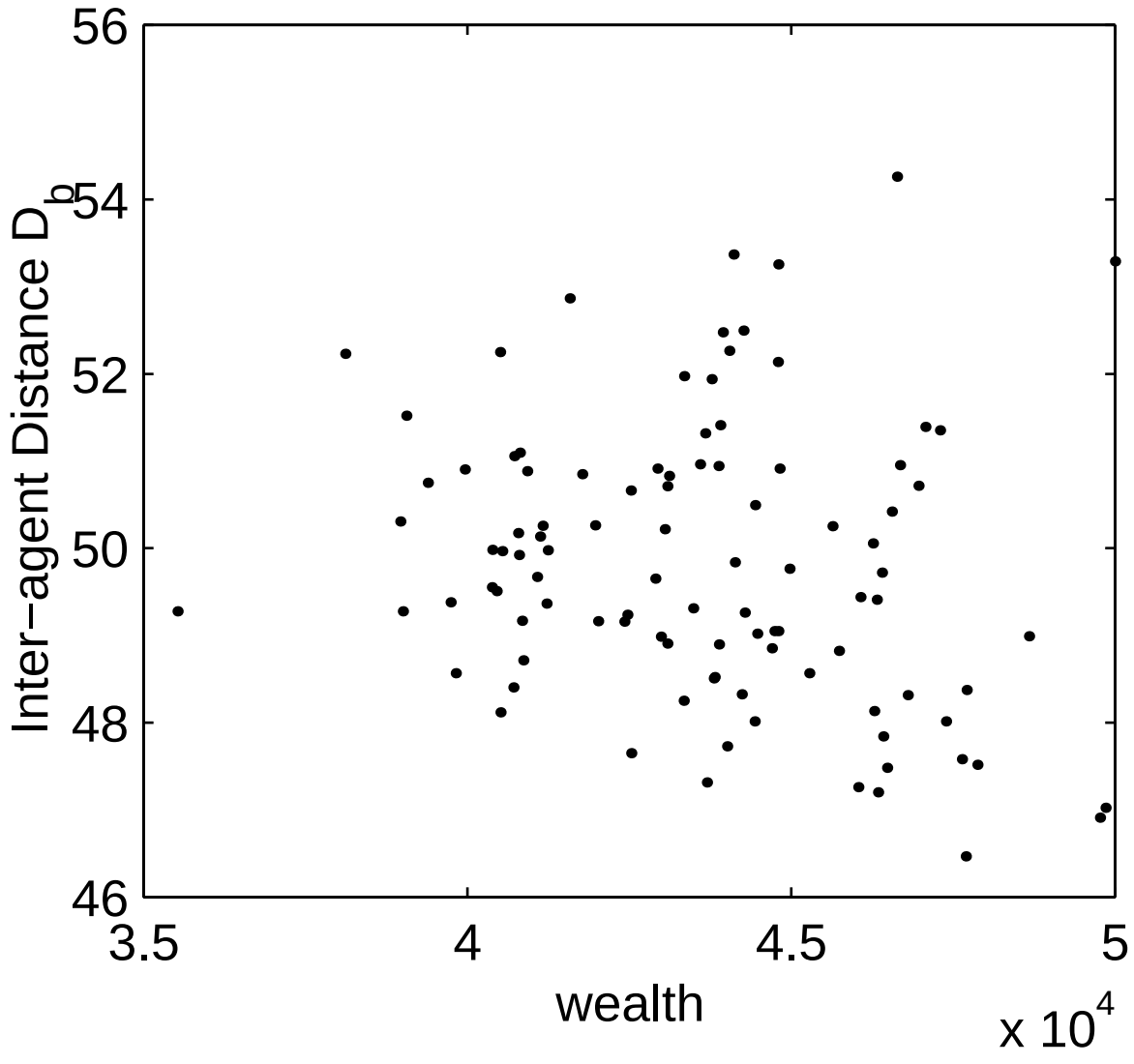

**Figure 26b. N=101, s=2, m=3**

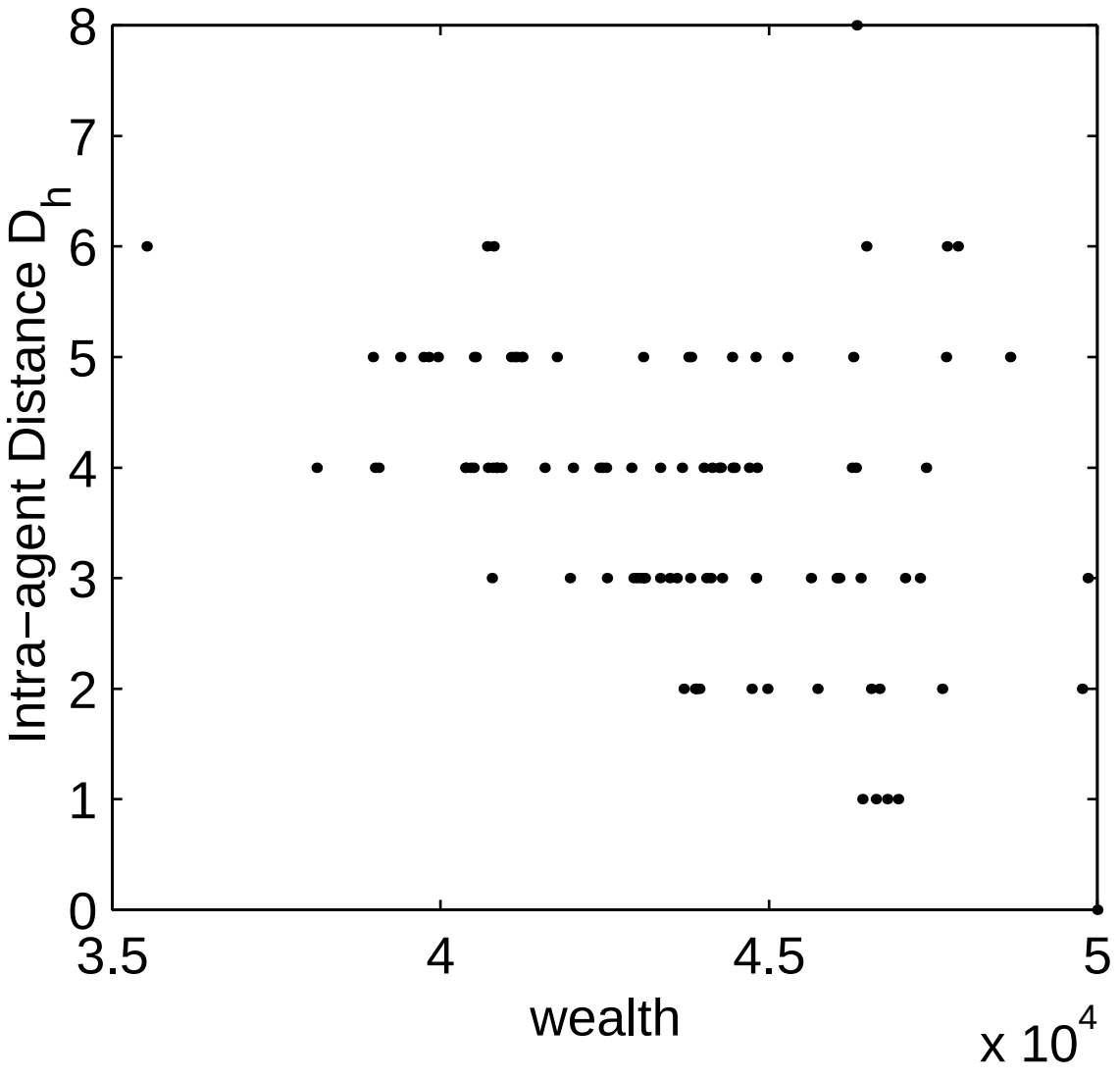

**Figure 26c. N=101, s=2, m=4**

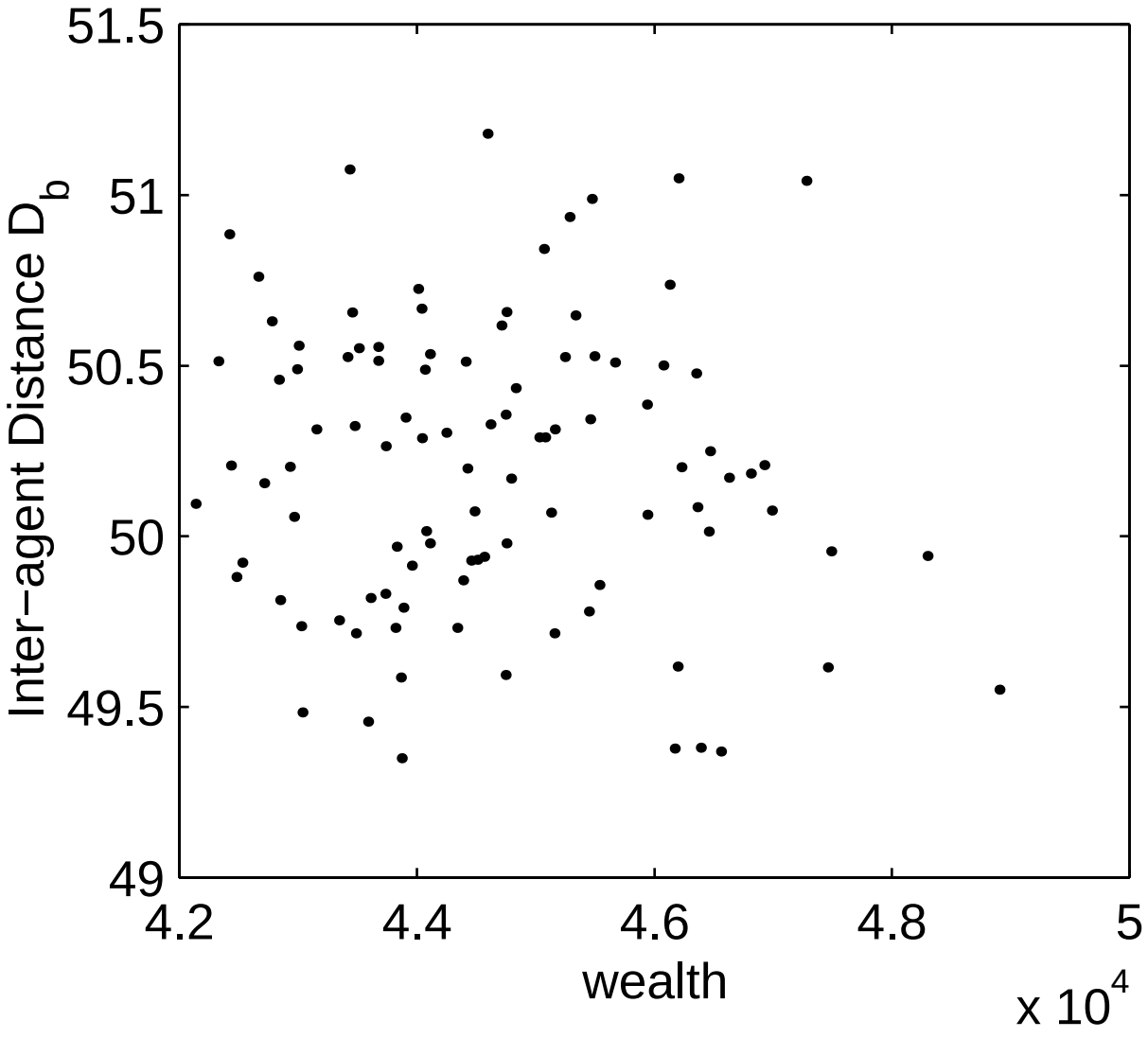

**Figure 26d. N=101, s=2, m=4**

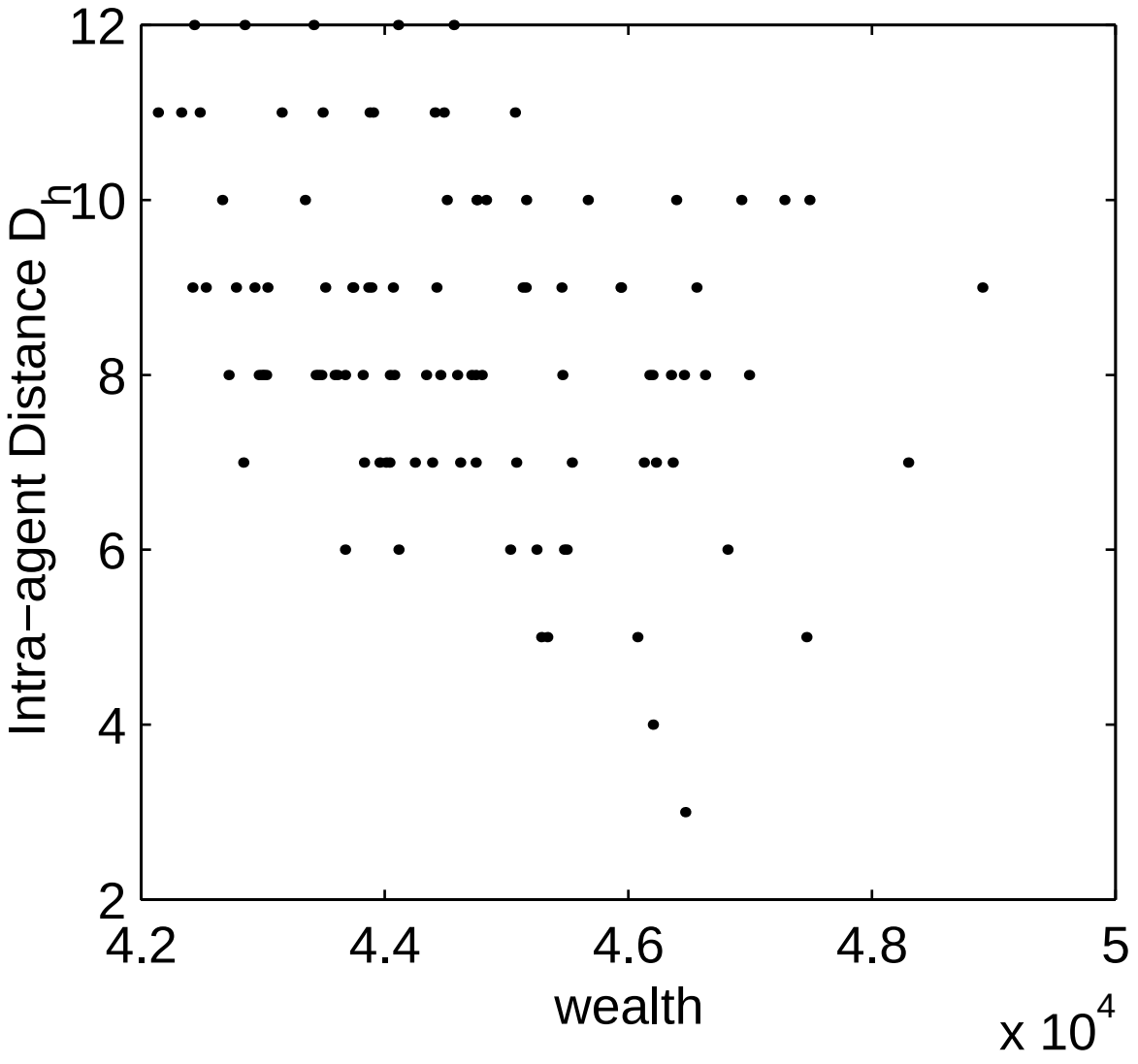

**Figure 26e. N=101, s=2, m=5**

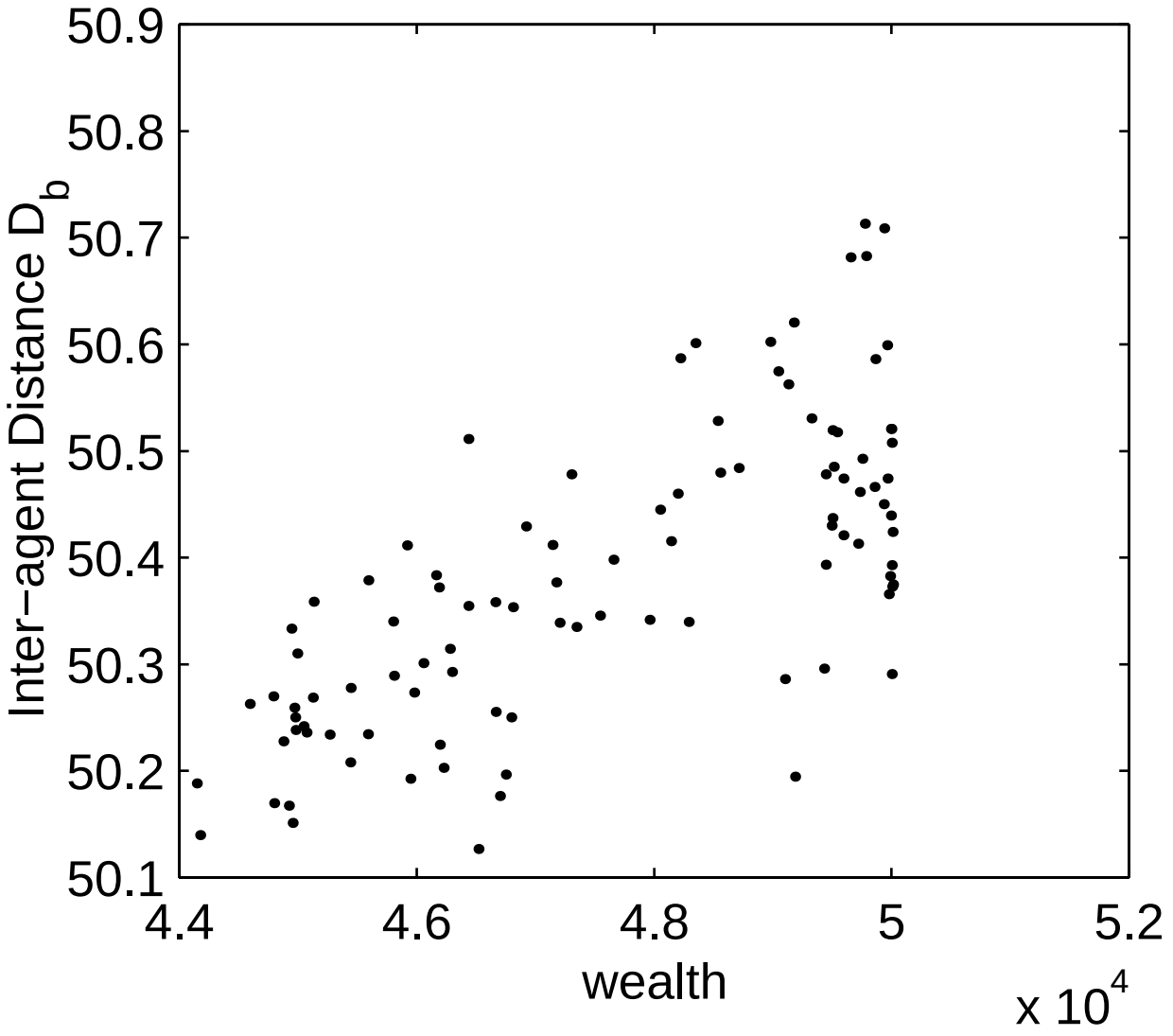

**Figure 26f. N=101, s=2, m=5**

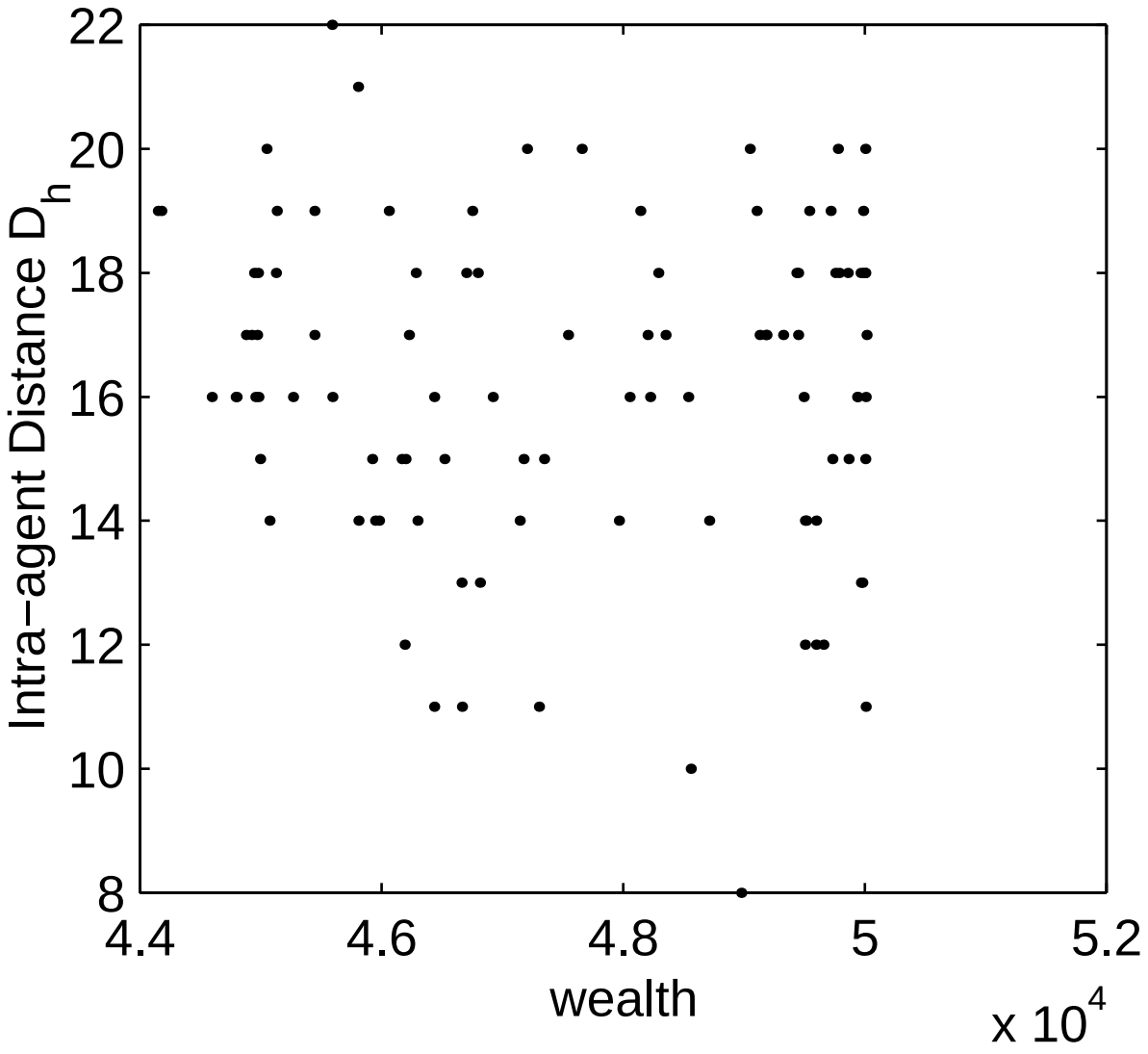

**Figure 26g. N=101, s=2, m=6**

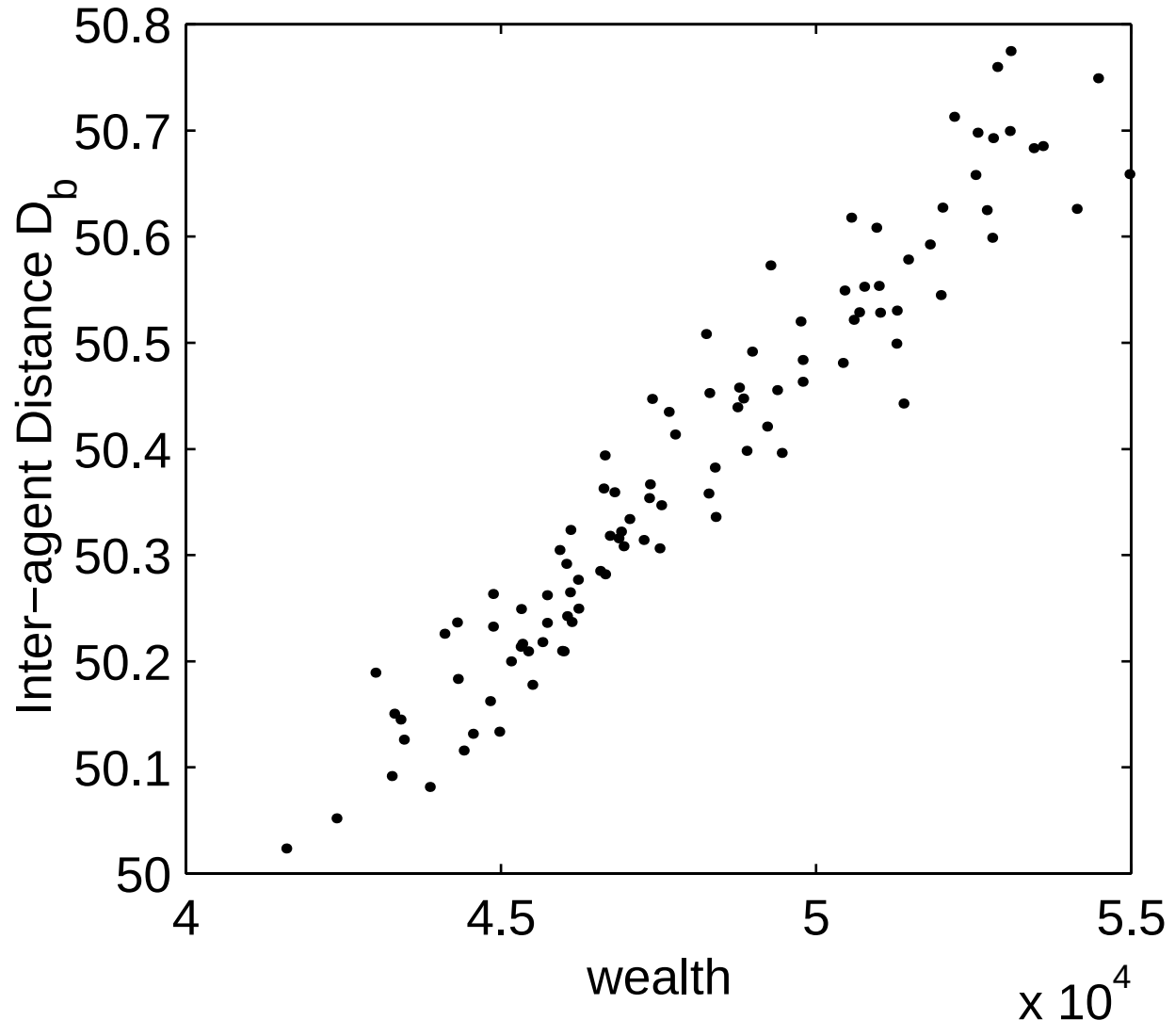

**Figure 26h. N=101, s=2, m=6**

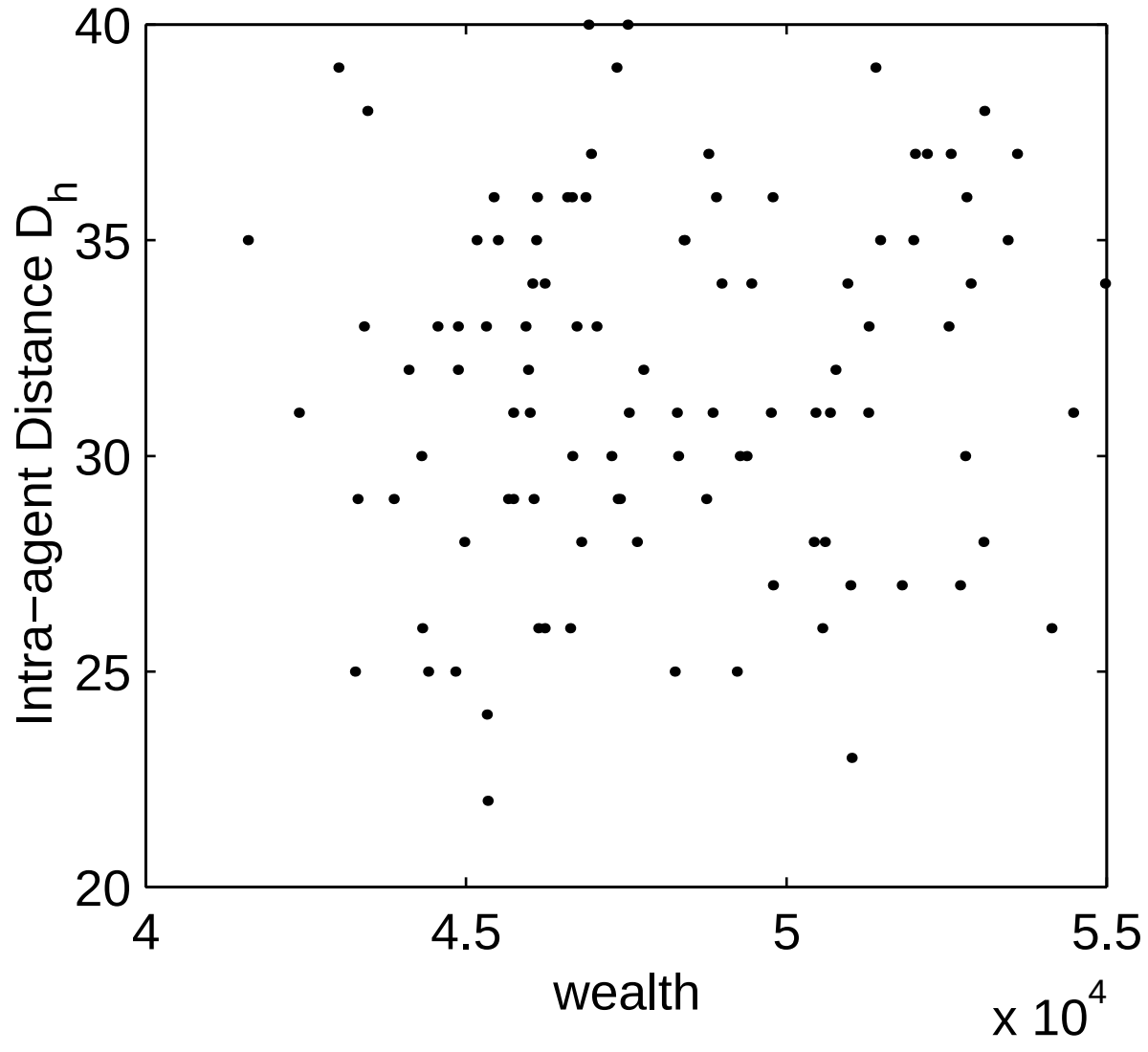

**Figure 26i. N=101, s=2, m=7**

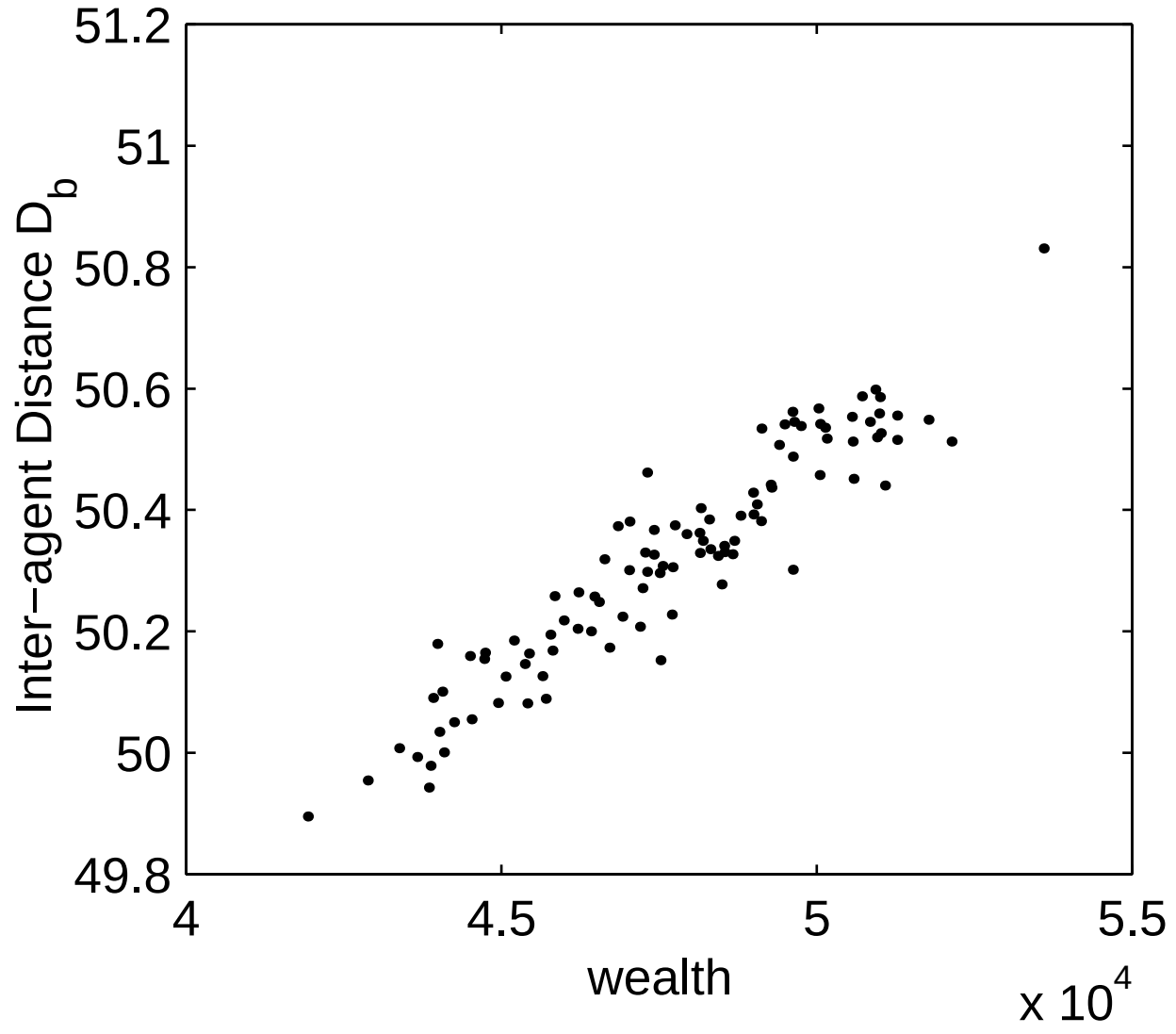

**Figure 26j. N=101, s=2, m=7**

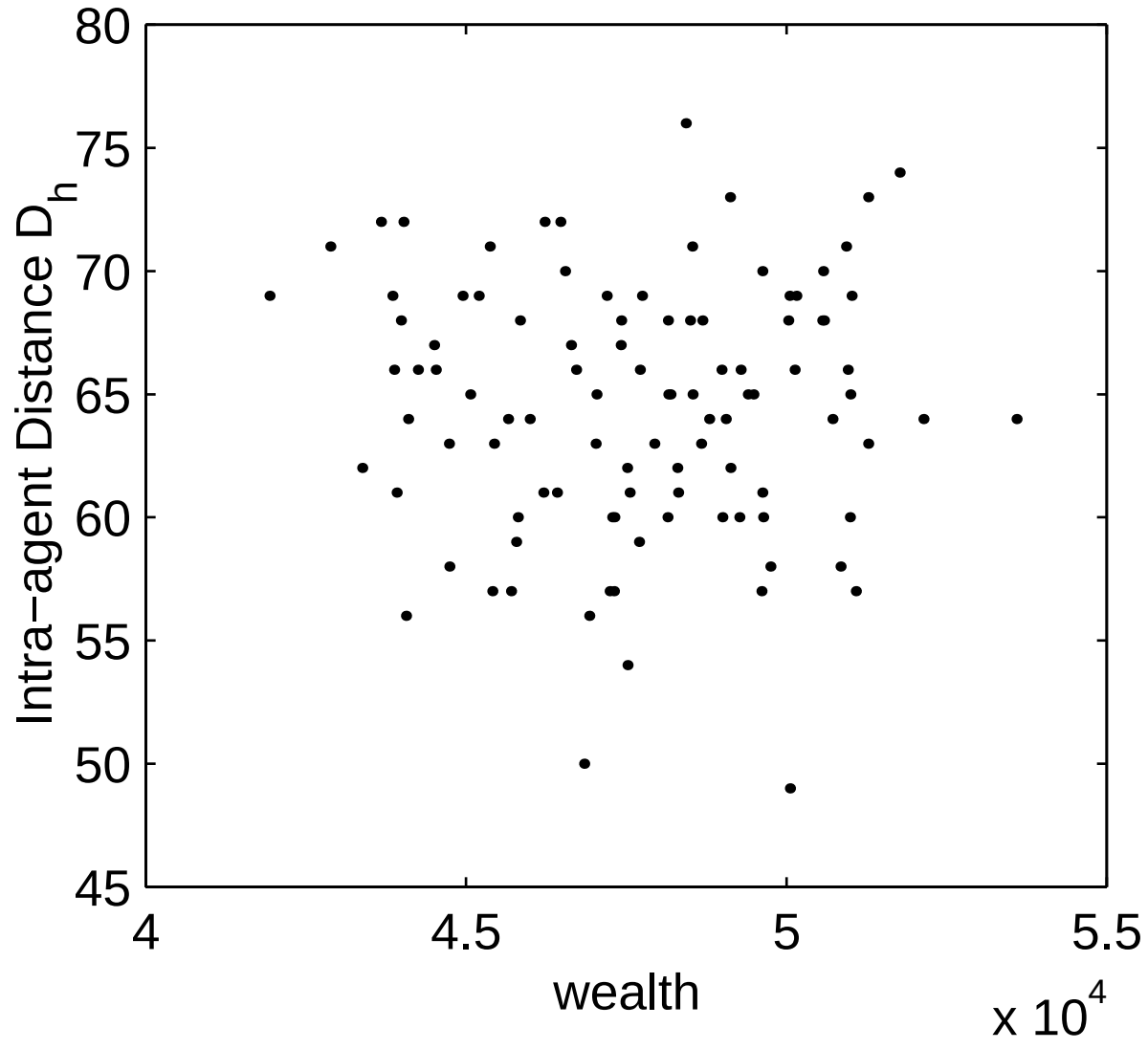

**Figure 26k. N=101, s=2, m=8**

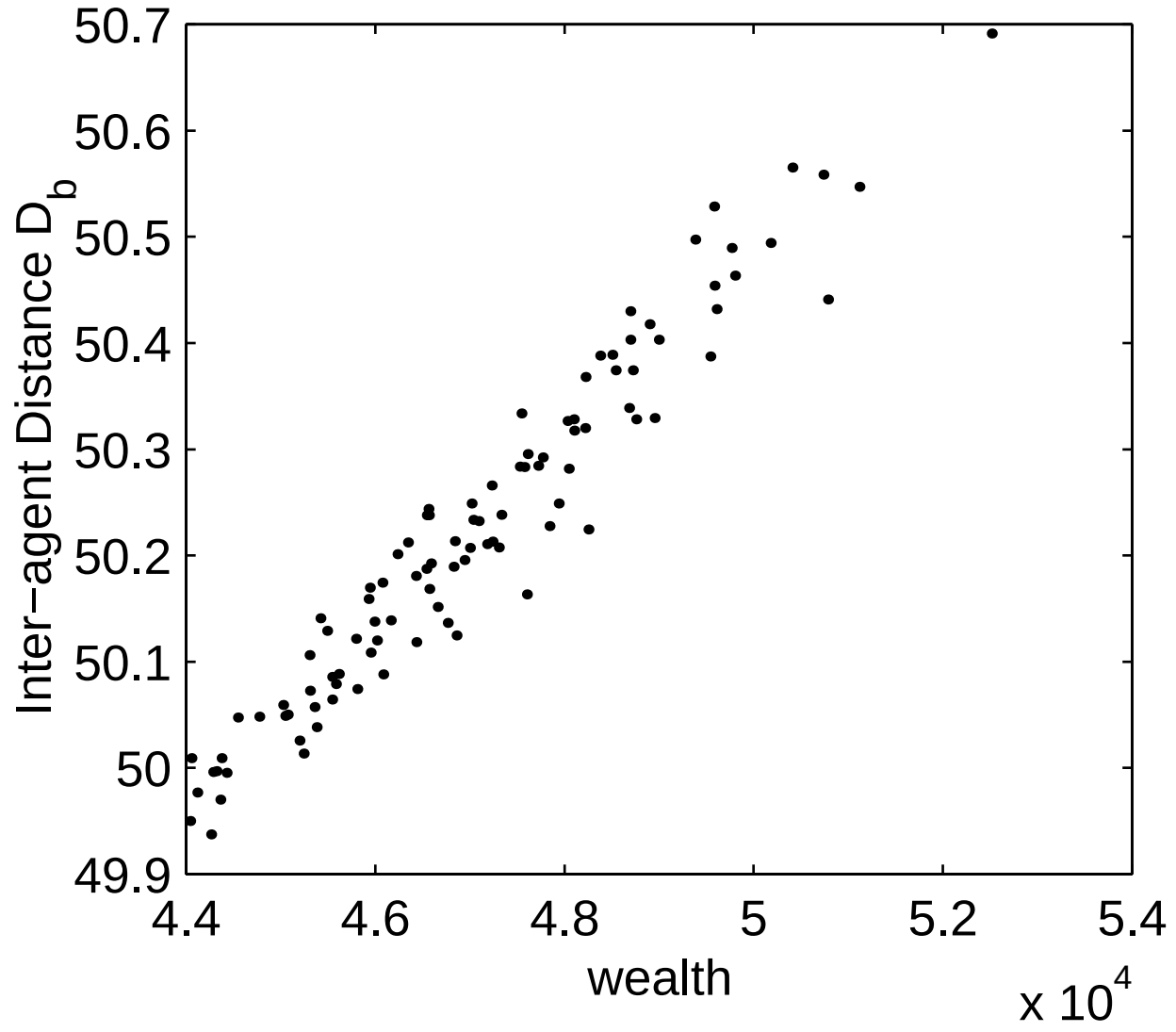

**Figure 26l. N=101, s=2, m=8**

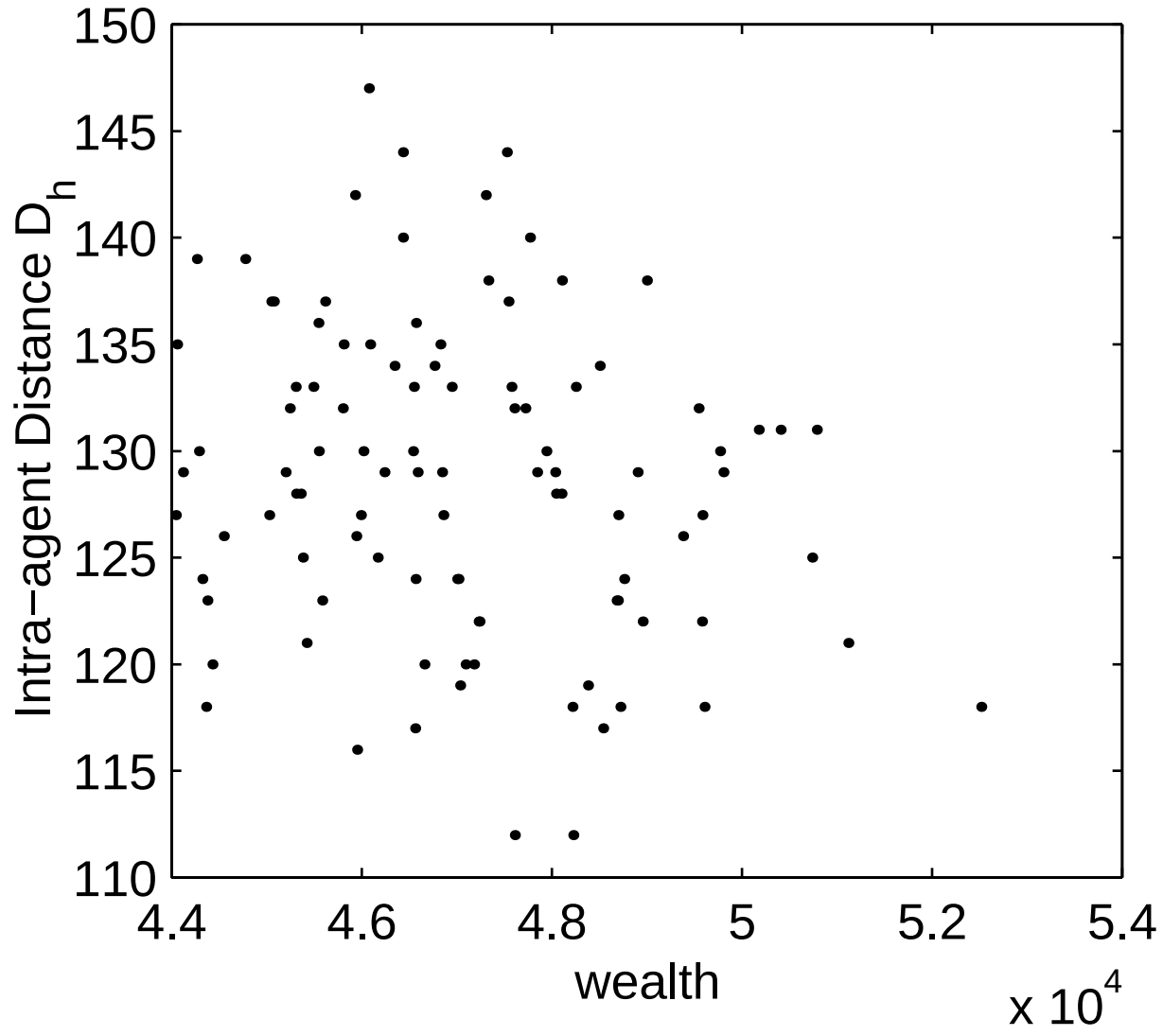

**Figure 27**

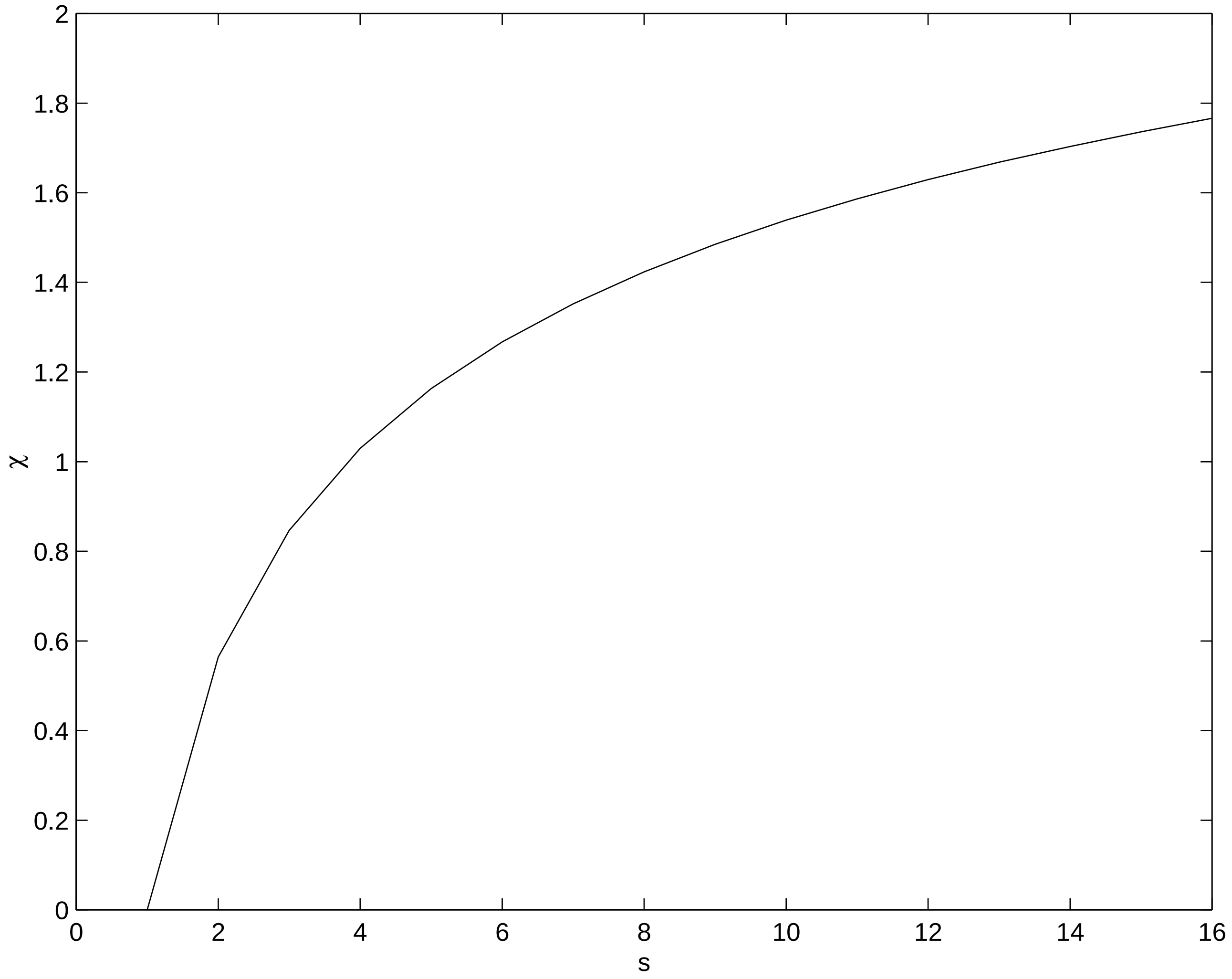

# The Structure of Adaptive Competition in Minority Games

Radu Manuca+, Yi Li, Rick Riolo and Robert Savit*
Program for the Study of Complex Systems and Physics Department
University of Michigan
Ann Arbor MI 48109

Abstract

In this paper we present results and analyses of a class of games in which heterogeneous agents are rewarded for being in a minority group. Each agent possesses a number of fixed strategies each of which are predictors of the next minority group. The strategies use a set of aggregate, publicly available information (reflecting the agents' collective previous decisions) to make their predictions. An agent chooses which group to join at a given moment by using one of his strategies. These games are adaptive in that agents can choose, at different points of the game, to exercise different strategies in making their choice of which group to join. The games are not evolutionary in that the agents' strategies are fixed at the beginning of the game. We find, rather generally, that such systems evidence a phase change from a maladaptive, informationally efficient phase in which the system performs poorly at generating resources, to an inefficient phase in which there is an emergent cooperation among the agents, and the system more effectively generates resources. The best emergent coordination is achieved in a transition region between these two phases. This transition occurs when the dimension of the strategy space is of the order of the number of agents playing the game. We present explanations for this general behavior, based in part on an information theoretic analysis of the system and its publicly available information. We also propose a mean-field like model of the game which is most accurate in the maladaptive, efficient phase. In addition, we show that the best individual agent performance in the two different phases is achieved by sets of strategies with markedly different characteristics. We discuss implications of our results for various aspects of the study of complex adaptive systems.

* To whom correspondence should be addressed. e-mail: savit@umich.edu.
+ Current address: College of William and Mary, Physics Department, Williamsburg, VA 23187-8795

## I. Introduction

Complex adaptive systems of agents with co-adaptive or co-evolving strategies are systems of agents which compete for some resource, or set of resources, and by their competition alter their environment. In such systems agents are typically endowed with heterogeneous strategies, beliefs and behaviors. The collective action of these agents in general alters their environment, and so, as time goes by, agents must reevaluate and possibly change their strategies to more effectively compete in the altered environment. Markets, ecologies, political structures and competition among businesses are all examples of such systems. Indeed, it is difficult to imagine an interesting system in the biological or social sciences which does not encompass elements of co-evolving competitive systems.

Despite their central role, not much is understood about the nature or behavior of such systems in general. Although many such systems have been studied by a variety of methods, and much has been learned about those specific systems, there is currently no general theory of N agent competitive systems with co-evolving strategies. In fact, there is not even a well-established epistemology. We do not know, in general, what range of behaviors such systems can exhibit, nor under what general circumstances various behaviors will be manifest. In fact, we do not even know what the sensible questions are (by which we mean questions that can have sensible, well-defined answers) which we can ask about such systems.

This latter comment may require some explanation. Depending on the nature of the dynamics, there are well-known limitations to questions that can be reasonably asked of various physical and mathematical systems. We know, for example, that if a system is chaotic, it is not sensible to ask for a precise prediction of its behavior over a long time. As another example, we know that in quantum mechanics one cannot sensibly ask for a precise, simultaneous determination of the position and momentum of a particle. Given the general dynamical architecture of N-agent complex adaptive systems with co-evolving strategies, we do not know if there are analogous limitations (and if so, what they are), to the questions we can sensibly ask of these systems.

One common approach to trying to understand the behavior of N-agent complex adaptive systems with co-evolving strategies is driven by the specific disciplinary interests of researchers. In this approach, one tries to model specific systems with some degree of realism. While the study of such models may indeed illuminate the object of their study, there are problems with this approach: For example, with many mechanisms incorporated *ab initio* into a model, it is difficult to understand precisely which aspects of the model are responsible for various results. Here the epistemological problem also becomes important. It may be, for example, that some of the results gleaned from very specific models may not be robust or meaningful, casting doubt on inferences drawn from these results.

A different approach, that we find more useful, is to study in detail the simplest possible models that capture at least some of the most basic underlying dynamics that may appear in a wide range of N-agent co-evolving systems. Such very simple models, while not necessarily accurately reflecting the ingredients of any specific system, have the virtue of being controllable and potentially understandable. These models can help us address the most basic issues of complex adaptive systems giving us great insight, not only into what emergent behaviors we can expect from various fundamental underlying dynamics, but also what the appropriate questions are that we can ask of such systems.

In this paper, we analyze one such simple (and we believe, paradigmatic) system. This system is a repeated game of N players, choosing (independently) according to some set of strategies to join one of two groups. At each time step only members of the minority group are rewarded. As explained in more detail below, this system has a most remarkable behavior.[1] The system manifests a phase change from a phase which is in a certain sense, efficient, but maladaptive, to a phase which is, in a certain sense, inefficient, but in which the system as a whole utilizes it's resources effectively. As one increases the size of the strategy space available to the agents, one moves from the former to the latter phase. The dynamics whereby this transition takes place are subtle and fascinating, and are the result of an emergent coordination in the decisions taken by the agents. This coordination is mediated only by aggregate, publicly available information, and does not result from any detailed knowledge of the behavior of one agent by any other. The system also has some intriguing properties reminiscent of a spin-glass in statistical mechanics.

---

[1] R. Savit, R. Manuca and R. Riolo, University of Michigan Program for the Study of Complex Systems preprint no. 97-12-001. (1997). Available at http://xxx.lanl.gov/abs/adap-org/9712006. Submitted for publication.

In the next section we define the model, and offer some simple interpretations to motivate its relevance. In section III we present the results of computer experiments performed with this model. In particular, we will demonstrate that the model, despite its incredible simplicity, has a remarkable and surprising richness of behaviors. In section IV we will study the results of section III with an eye toward understanding the general system-level behavior of these systems. In section V, we briefly touch on some aspects of individual agent wealth accumulation in these games. In particular, we study what properties of the agent's strategies are important for accumulating wealth. In section VI, we will present a mean-field like approximation to the dynamics of the model which will allow us to explain the behavior of the model in some of the parameter space. The mean-field approximation, will also help elucidate the nature of the dynamics in parameter regimes where the approximation breaks down. The paper ends with section VII, in which we present a summary and discussion of our work.

## II. The Model and its Motivations

The simple model of competition we discuss here consists of N agents playing a game[2]. The rules of the game are as follows: At each time step of the game, each of the N agents playing the game joins one of two groups, labeled 0 or 1. Each agent that is in the minority group at that time step is awarded a point, while each agent belonging to the majority group gets nothing. An agent chooses which group to join at a given time step based on the prediction of a strategy. (In general, different agents use different strategies, as we shall explain below.) A strategy of memory m is a table of $2^m$ rows and 2 columns. The left column contains all $2^m$ combinations of m 0's and 1's, and each entry in the right column is a 0 or a 1. To use a strategy of memory m, an agent observes which were the minority groups during the last m time steps of the game and finds that entry in the left column of the strategy. The corresponding entry in the right column contains that strategy's prediction of which group (0 or 1) will be the minority group during the current time step. An example of an m=3 strategy is shown in Fig. 1. Thus, a strategy uses information from the historical record of which group was the minority group as a function of time, which is the information publicly available to all the agents.

In each of the games discussed here, all strategies used by all the agents have the same value of m. At the beginning of the game each agent is randomly assigned s (generally greater than one) of the $2^{2^m}$ possible strategies of memory m, with replacement. For his current play the agent chooses from among his strategies the one that would have had the best performance over the history of the game up to that time. Ties among an agent's strategies can be decided either by a coin toss, or by choosing to play the strategy that was most recently played.[3] Following each round of decisions, the cumulative performance of each of the agent's strategies is updated by comparing each strategy's latest prediction with the current minority group. Because the agents each have more than one strategy, the game is adaptive in that agents can choose to play different strategies at different moments of the game in response to changes in their environment; that is, in response to new entries in the time series of minority groups as the game proceeds. Because the environment (i.e. the time series of minority groups) is created by the collective action of the agents themselves, this system has very strong feedback.

---

[2] D. Challet and Y.-C. Zhang, *Physica A*, **246**, 407 (1997). A related model has also been considered by N. Johnson, S. Jarvis, R. Jonson, P. Cheung, Y. Kwong and P. Hui, *Physica A* **256**, 230 (1998).

Although this system is adaptive, the versions we analyze here are not, strictly speaking, evolutionary. The strategies do not evolve during the game, and the agents play with the same s strategies they were assigned at the beginning of the game. Evolutionary mechanisms can be simply incorporated into this model, and will be discussed elsewhere.

While this model is very simple, and far from a realistic model of any particular biological or social system, it is not difficult to see the relevance of the dynamics incorporated in the model. First, the model was inspired, in part, by the El Farol problem[4]. El Farol is a bar in Santa Fe, New Mexico that plays Irish folk music on Thursday nights. Many people want to go to the bar to hear the music, but only if there are fewer than a certain number of people at the bar. Otherwise, there will be too much noise, and the patrons will be unable to enjoy the music. In our model, we can let 0 represent attendance at the bar, while 1 means staying home. Then, one benefits (and is awarded a point) if one attends the bar on a night when fewer than half the agents attend. One also benefits (and is awarded a point) if one stays home on a night when more than half the agents attend the bar.[5]

---

[3] The latter is the default for our games, but, as we shall discuss, the system-wide behavior does not generally depend on whether tie-breaking is achieved stochastically or deterministically.

[4] W. Brian Arthur, *Amer. Econ. Assoc. Papers and Proc.* **84**, 406 (1994).

[5] We are saddened to report that a fire has taken a serious toll on this fabled establishment. However, reliable first hand accounts indicate that it is again operational. (E. Bonabeau, private communication).

This game can also be interpreted as a toy model of a market. In this market, each agent has an indefinite supply of money and widgets. At each time step, each agent must either buy or sell one widget. Call group 0 the buyers and group 1 the sellers. The price of the widgets can take on only two values, and is determined by a simple supply-demand rule: If there are more buyers than sellers, the price is high, and if there are more sellers than buyers, the price is low. If group 0 is in the minority, then there are more sellers than buyers, the widget price is low, and the buyers (the minority group, 0) do well, in that they buy at a low price. Therefore, they each get a point. If group 1 is in the minority, then there are more buyers than sellers, the widget price is high, and the sellers (the minority group, 1) do well, in that they sell at a high price. Therefore, they are awarded a point.

From these two interpretations it is clear that the dynamics associated with minority membership plays an important role in adaptive competitive systems. Other examples of systems that are, at least in part, driven by minority-membership dynamics include vehicular traffic on roads, in which each agent would prefer to be on an uncongested road,[6] packet traffic in networks, in which each packet will travel faster through lesser used routers,[7] and ecologies of animals looking for food, in which individuals do best when they can find areas with few competitors[8]. Of course, for any particular system, care must be taken to understand how minority dynamics is to be imposed, and what other ingredients may be necessary. Nevertheless, studying minority dynamics is clearly of great interest for understanding the fundamental general behavior of adaptive competitive systems.

## III. Basic Results

In this section, we will present a set of system-level results which demonstrate a remarkable and rich phase structure for minority games. We will first describe the general methods used for our experiments. We will then describe basic results for (a) one strategy per agent, (b) two strategies per agent, and (c) more than two strategies per agent.

---

[6]K. Nagel, S. Rasmussen, and C. Barrett, Network Traffic as a Self-Organized Critical Phenomena *in Self-organization of Complex Structures: From Individual to Collective Dynamics*, p. 579, F. Schweitzer (ed.) (Gordon and Breach, London. 1997).

[7] *Coordination of the Internet*, B. Kahin and J. Keller (eds.) (MIT Pres, Cambridge MA, 1977); B. Huberman and R. Lukose, *Science*, **277**, 535 (1997).

[8] See, for example, Ecology and Evolution of Communities, M. L. Cody and J. M.Diamond (eds) (Harvard University Press), Cambridge MA, 1975); Individual-Based Models and Approaches in Ecology : Populations, Communities, and Ecosystems, D. DeAngelis and L. Gross (eds), (Chapman & Hall, New York, 1992).

Note that in this section and in most of the rest of this paper, we will primarily present and discuss experiments done with integer m. However, it is easy to generalize to non-integer m, and, in Section IVB, we shall present and discuss some results for non-integer m. The general case of non-integer m, which also has other implications, will be discussed more extensively in a forthcoming publication.[9]

Unless stated otherwise, the experiments we describe were performed with the following default settings: Generally, for each value of m, N and s, 32 different runs were performed. A different random distribution of strategies to the agents was assigned for each run. In addition, a random initial history of m+1 minority groups was necessary for the initial evaluation of the agents' strategies.[10] In most figures in which a single value of a quantity is presented for a given m, N and s, that quantity is an average over 32 runs. For most of our runs, in which N ranged between 11 and 1001, we found that 10,000 time steps were generally enough to sufficiently eliminate transients for the purpose of estimating quantities. The largest values of N, however, required runs of 100,000 time steps. Typically results were based on the final 20%-30% of data from each run. The reason for these fairly long runs is discussed in subsection B, below.

A. One strategy per agent, s=1

To begin, we consider the case of only one strategy per agent. In this exceptional case, there is no adaptivity. The response of each agent to a given string is always the same. Because of this, it is not difficult to see that the time series of minority groups will eventually become periodic with a period that depends on the particular (random) distribution of strategies to the agents.

An important quantity for us will be the standard deviation, $\sigma$, of the time series of the number of agents belonging to group 1 ($\equiv L_1$). (This information is not available to the agents but it is available to the researchers.) The mean of this series is generally close to (but slightly less than) 50% for all values of N, m and s (we shall return to this point below), and so the standard deviation, $\sigma$, of this time series is a measure of how well the commons do: The smaller $\sigma$, the more total points are awarded to all agents combined. That is, if there are typically many fewer than 50% of the agents in the minority, then $\sigma$ will be large and there will be few total points awarded. On the other hand, if $\sigma$ is small, then most of the time the minority group will consist of only slightly fewer than half of the agents, and more total points will be awarded.

---

[9] Y. Li, R. Manuca, R. Riolo and R. Savit, in preparation.

[10] The overall statistical properties of each run do not depend on the initial history of minority groups. However, more detailed aspects of the results may. For example, under some circumstances the relative wealth of a specific agent may depend on this initial history. This will be discussed in more detail in Ref. 9, above.

In Fig. 2, we plot $\sigma$ for a number of experiments with s=1, N=101 and various values of m. The horizontal line on this graph is the value that $\sigma$ would have if each agent made his choice randomly and independently (with equal probabilities for 0 and 1) at each time step. We call this the random choice game (RCG). In this figure we note that there is a wide dispersion in $\sigma$ for a given m, but that the mean of the $\sigma$'s is near the RCG result.[11] This is not surprising. Each run of the game effectively represents a small temporal sample, (the duration of which is equal to the period of that particular game, determined by the initial distribution of strategies) of a random game. Thus, we expect, in general, some dispersion about the RCG result, which is what we see. This s=1 case is trivial, but provides a good comparison for the considerably more interesting games with s≥2.

B. Two strategies per agent, s=2

Consider next the case s=2 which is the simplest case of adaptivity. The reason that the game must typically be run for thousands of time steps is that the agents must have sufficient time to "learn" about each other through the collective, publicly available information of the time series of minority groups. Only with this learning is it possible for the agents to coordinate their strategy choices. This was not an issue in the s=1 experiments in which there was no learning, and no possibility of agents coordinating their choices.

To begin to understand the behavior of this system, consider, as before, $\sigma$, the standard deviation of the time series of the number of agents $L_1$. The behavior of $\sigma$ is quite remarkable. In Fig. 3, we plot $\sigma$ for these time series as a function of m for N=101 and s=2. For each value of m, 32 independent runs with different initial distributions of strategies were performed. The horizontal dashed line in this graph is at the value of $\sigma$ for the random choice game (RCG), i.e. for the game in which all agents randomly choose 0 or 1, independently and with equal probability at each time step. The comparison with Fig. 2 is startling.

---

[11] Although the results shown in Fig. 2 are spread near the result of the RCG, it is clear that there is a bias toward smaller $\sigma$. This is just a trivial consequence of the fact that these games have a finite periodic structure in which $L_1$ is also a periodic function.

Note the following features:

1. For small m, the average value of σ is very large (much larger than in the random case). In addition, for m<6 there is a large spread in the σ's for different runs with different (random) initial distributions of strategies to the agents, but with the same m.

2. There is a minimum in σ at m=6 at which σ is less than the standard deviation of the random game. We shall refer to the value of m at which the σ vs. m curve (for fixed N) has its minimum as $m_c$.[12] Thus, Fig. 3, indicates 6 as an approximation to $m_c$. Also, for $m \geq m_c$, the spread in the σ's appears to be small relative to the spread for $m < m_c$.

3. As m increases beyond 6, σ slowly increases, and for large m approaches the value for the random choice game.

The system clearly behaves in a qualitatively different way for small and large m. To further study the dynamics in these two regions, we consider the time series of minority groups, (≡G), the data publicly available to the agents. We want to study the information content of strings of consecutive elements of this time series of various lengths (including strings of length m) for different values of m and N. Since the strategies use only the information contained in the most recent m-time steps of G, it is natural to ask how much information is accessible to the strategies. To do this, we consider the conditional probability $P(1|u_k)$. This is the conditional probability to have a 1 immediately following some specific string, $u_k$, of k elements of G. For example, P(1|0100) is the probability that 1 will be the minority group at some time, given that minority groups for the four previous times were 0,1,0 and 0, in that order. Recall that in a game played with memory m, the strategies use only the information encoded in strings of length m to make their choices. In Fig. 4, we plot $P(1|u_k)$ for G, the time series of minority groups generated by a game with m=4, N=101 and s=2. Fig. 4a shows the histogram for k=m=4 and Fig. 4b shows the histogram for k=5. Note that the histogram is flat at 0.5 in Fig. 4a, but, remarkably, is not flat in Fig. 4b. Thus, any agent using strategies with memory (less than or) equal to 4, will find that those strings of minority groups contain no predictive information about which group will be the minority at the next time step. But recall that this time-series was itself generated by agents playing strategies with m=4. Therefore, in this sense, the market is efficient (at least with respect to the strategies)[13] and no strategy playing with memory (less than or) equal to 4 can, over the long run, accumulate more points than would be accumulated randomly[14]. But note also that the time series of minority groups is not a random (IID) string. There is information in this string, as indicated by the fact that the histogram in Fig. 4b is not flat. However, that information is not available to the strategies used by the agents in playing the m=4 game which collectively generated that string in the first place. The histogram for k=m is flat for $m< m_c$ (although it does begin to show some small $O(T^{1/2})$ fluctuations for m just below $m_c$).

[12] In general, $m_c$ is not an integer. Since this experiment was performed only for integer m, 6 must be considered an estimate of $m_c$. In section IVB we will show how to interpolate to non-integer m, and we will see that for N=101 $m_c$ is approximately in the range 5.2~5.7. In the remainder of this paper, values of $m_c$ deduced from experiments using only integer m, should be understood to be rough approximations to $m_c$.

We can repeat this analysis for $m\geq 6$ (N=101, s=2). For this range of m, the corresponding histogram for k=m is not flat, as we see in Fig. 5 for the m=6 game. In this case, there is significant information available to the strategies of memory m and the "market" is not informationally efficient with respect to the strategies.[15] Indeed, some individual agents using their strategies do accumulate significantly more points than they would by simply making random guesses. However, for m<5, no agent ever achieves more than 50% of the possible available points. The wealth distribution of the agents in the two phases is very interesting and will be discussed in section V, below, and in detail in forthcoming publication[9].

How does the system behavior change as we change the number of agents? One can repeat the calculation of $\sigma$ for different N, again with s=2. One finds, plotting $\sigma$ vs. m for fixed N, that in all cases one obtains a graph similar to that in Fig. 3, but in which the position of the minimum, $m_c$, is proportional to lnN. In addition, σ and the spread in σ behave in very simple ways with changes in N which differ depending on whether m is greater than or less than $m_c$. In Fig. 6 we study the behavior of σ as a function of N for m=3 and m=16. For the range of values of N used in these figures, m=3 is to the left of the minimum in the curve of σ vs. m (i.e. $3< m_c$ for all these values of N) and m=16 is to the right of the minimum ($16> m_c$). In Fig. 6a we plot σ vs. N on a log-log scale. We see that for m=3 σ is proportional to N, while for m=16, σ is proportional to $N^{1/2}$. This is typical: For fixed m, and integer $m< m_c$, σ is proportional to N, while for fixed m and $m> m_c$ σ is proportional to $N^{1/2}$. In Fig. 6b, we plot, again for m=3 and 16, $\Delta\sigma/\sigma$ as a function of N, on a log-log scale . Here Δσ is the standard deviation of the σ 's for runs with the same value of m, N and s, but different initial (random) distributions of strategies. The horizontal lines indicate that the dependence of Δσ on N for fixed m is the same as that of σ, namely for m=3 Δσ is proportional to N, while for m=16, Δσ is proportional to $N^{1/2}$. As before, this behavior is representative of the two behaviors seen for values of $m< m_c$ and values of $m\geq m_c$, respectively.

---

[13] Actually, the structure in this phase is a little more subtle. The system is informationally efficient with respect to the strategies, but is not efficient with respect to the agents. In fact, there are strong correlations between the minority groups over long times of order $2^m$ time steps, which results in the very poor performance of the system in this regime. This will be discussed in more detail in section IVA, below. Note also that these long time correlations are not to be confused with the non-IID conditional probabilities (that indicate information over time scales of order m+1) discussed in the text and illustrated, for example, in Fig. 4b. Despite these subtleties this phase does partake of the usual idea that efficiency implies an absence of usable information. As we shall see below when we study the inefficient phase, the use of the idea of efficiency is quite appropriate in this context. For a discussion of the idea of efficient markets as traditionally understood in economics, see E. F. Fama, *J. Finance,* **25**(2), 383 (1970) and E. F. Fama, *J. Finance*, **46**(5), 1575 (1991).

[14] In fact, asymptotically, for $m<m_c$, no strategy can be right more than 50% of the time. If the behavior for these values of m were truly random, there would be fluctuations of order $T^{1/2}$ in the number of times a strategy won over a time T. But in our case, these Gaussian fluctuations are absent, and strategies are limited to being correct no more than 50% of the time. A corollary to this is that the histogram in Fig. 4a is extremely flat, and does not contain even the random fluctuations that would be present for an IID series. The reason for this will be explained below.

[15] Note that the particular shape of the histogram in Fig. 5 as well as in Fig. 4b depends on the initial distribution of strategies to the agents. But in all cases with $m>m_c$ or with $m<m_c$ and k>m, the histograms do indicate the presence of significant predictive information.

The transition between these very different behaviors is at $m_c$~lnN. To see this explicitly, we plot, in Fig. 7, $\sigma^2/N$ as a function of $z\equiv 2^m/N$ on a log-log scale for various N and m (with s=2). We see first that all the data fall on a universal curve. (As we shall discuss in section VI, we have also found, using mean-field-like arguments that are most accurate in the low m phase, that for fixed s, $\sigma^2/N$ is a function only of z.) The minimum of this curve is near $2^{m_c}/N\equiv z_c\cong 0.5$, and separates the two different phases.[16] The slope for $z<z_c$ approaches -1 for small z, while the slope for $z>z_c$ approaches zero for large z, consistent with the results of Fig. 6. Because $\sigma^2/N$ depends only on z, it is clear that for fixed z, σ is proportional to $N^{1/2}$ for any fixed z, both above and below $z_c$. In addition, it can be shown that, for fixed z, Δσ is approximately independent of N, approaching a z-dependent constant as $N\to\infty$. The $N\to\infty$ limit of Δσ is large for small values of z and appears to decrease monotonically with increasing z. It is unclear whether or not Δσ is non analytic at $z_c$.

C. More than two strategies per agent, s>2

[16] As before, since these experiments involve only integer m, this value of $z_c$ should be considered only an estimate. Experiments with non-integer m indicate a value of $z_c$ closer to 1/3. See section IVB, below.

We now study the case in which each agent has more than two strategies among which to choose. In Fig. 8, we present a plot like that of Fig. 3, but for the case s=6. We see here that the qualitative features seen in Fig. 3 persist: $\sigma$ is large for small values of m ($m<m_c$), first decreases as m increases, reaches a minimum at some value of m, then increases, and approaches, asymptotically, the result of the RCG. There is also a spread in $\sigma$, $\Delta\sigma$, which has the same qualitative behavior as in the s=2 case, namely, for $m<m_c$ $\Delta\sigma$ and $\sigma$ are proportional to N, while for $m> m_c$ $\Delta\sigma$ and $\sigma$, are proportional to $N^{1/2}$, as demonstrated in Fig. 9. For s=6, there is also a scaling result similar to that shown in Fig. 7 for the s=2 case. In Fig. 10, we plot, for s=6, the mean value of $\sigma^2/N$ as a function of z for various values of m and N. As in Fig. 7, we see that all the points lie on a universal curve. Notice, however, two important differences with the s=2 case: First, the dip in Fig.10 is not as deep as the corresponding dip in Fig. 7 for the s=2 case. Second, the position of the minimum is slightly offset from the value for the s=2 case. In Fig. 11 we plot, these universal scaling curves for various values of s on a log-log plot. Here the horizontal axis is $\zeta=[z/z_C(s)]$, so that the minima of all the curves are at $\zeta=1$. For large and small values of $\zeta$, the curves coincide, approaching the RCG result for large $\zeta$, and falling on a curve with a slope of -1 for small $\zeta$. However, for intermediate values of $\zeta$, there are differences. As s increases, the value of $z_c$ (i.e., the value of z at which the curve has its minimum) appears to drifts slightly to somewhat larger values of z as can be seen in Fig. 11 and also by comparing Figs. 7 and 10. Even more importantly, the dip at $z_c$ becomes increasingly shallow, and for very large s, the universal curve approaches two straight lines, one with a slope of -1 (small z), and the other with a slope of 0 at the value of the RCG. These two lines intersect at $\zeta =1\cdot$

The information content above and below $z_c$ for s>2 is similar to that for the s=2 case. In Fig. 12, we present $P(1|u_k)$ for G, the time series of minority groups generated by a game with m=4, N=101 and s=6. Fig. 12a shows the histogram for k=m=4 and Fig. 12b shows the histogram for k=5. As in Fig. 4, the histogram is flat in Fig. 12a, but not flat in Fig. 12b, showing that there is no information available to the strategies playing the m=4 game with N=101 and s=6, but that the sequence, G, is not IID. Similarly, in Fig. 13, we plot $P(1|u_k)$ for k=6 generated in the m=7 game with s=6. As in Fig. 5, this histogram is not flat, demonstrating that there is information available to the strategies in the $z>z_c(s)$ phase, even when s>2.

Another very convenient way to display the information content that is accessible to the strategies in the time series of minority groups, is to compute the mutual information[17] between the current minority group and the sequence of the last m minority groups. The mutual information between two data sets can be thought of as a measure of the extent to which there is common information between the two sets. For the game played with memory m, we consider the mutual information defined as

$$Q(m)=H(m) + H(1) - H(m+1) \qquad (3.1)$$

where H(j) is the entropy of the distribution of strings of length j. In Fig. 14 we plot Q(m) as a function of $\zeta$ on a semi-log plot for N=101 and s=2 and 16. In both cases, Q=0 for $\zeta<1$, and Q>0 for $\zeta>1$. Note also that for positive $\zeta$, Q is much larger for s=2 than for s=16. This is an indication that, for a given m and N, the strategies have more information available to them for smaller s. It is also related to the observation that the dip in Fig. 11 is deeper for smaller s.

## IV. Explication of the Phase Structure

In this section we will examine in more detail the behavior of the system in three regions: (A) the low m, strategy-efficient phase, (B) the transition region, and (C) the informationally inefficient region. We will present explanations for the behavior of the system in these regions, relying on the results of section III, as well as on some more detailed results to be presented in this section. Topics that will be covered include an explanation of the poor system-wide performance and strategy-efficiency in the low-m phase, qualitative arguments for the position of the phase transition, a more detailed look at the transition region, and a discussion of the degradation of system performance as m increases beyond $m_c$. In addition, some of these topics will be further elucidated when we examine some aspects of individual agent wealth, in section V. In subsection B, we will introduce a method of interpolating to non-integer values of m, for the purpose of studying the transition region in greater detail.

### A. Dynamics in the efficient phase.

To begin to understand our system, let us first look more closely at the dynamics in the efficient (low m) phase. In Fig. 15 we show a short segment of the time series. Figure

[17] *Time Series Prediction: Forecasting the future and Understanding the Past*, A Weigend and N. Gershenfeld (eds.) (Addison-Wesley, Reading, MA, 1994), and references therein.

15a shows a segment for N=101, s=2 and m=2, which is in the efficient phase. Fig. 15b shows a segment for N=101, s=2 and m=8, which is in the inefficient phase. For comparison, the horizontal dashed lines indicate the range of one standard deviation for N=101 in the RCG. The first thing we notice is that Fig. 15a shows a bursting structure with segments of large variation separated by segments of smaller variation, a feature which is absent in Fig. 15b. In Fig. 15a we have further marked, by open circles, those times at which the last two minority groups were 01. We note that the odd occurrences of this string elicit responses from the agents that result in relatively large minority groups (i.e. small deviations in the population of group 1 from 50%), while for even occurrences there are large deviations in $L_1$.

We can understand this as follows: When m is relatively small, then there is a reasonable probability that an agent will have two strategies that are fairly similar. Suppose, for example, that two m=2 strategies differ in only one (say the 01) entry, and are otherwise the same. The first time the sequence 01 appears in G, either strategy *a priori* may be chosen, with equal probability to be played by its agent, and so we expect that the population of group 1 will be close to 50% of the total number of agents. Suppose that the minority group happens to be group 1. If an agent has two strategies which differ only in the 01 entry, then the strategy that predicts 1 following an occurrence of 01 in G will have one extra point. Now suppose the game continues. Up until the next occurrence of 01 in G, both strategies will accumulate the same number of points. At the next occurrence of 01 in G, the strategy that predicted 1 for the previous occurrence of 01 will have an extra point, and so the agent will play that strategy. If there are a significant number of agents which have strategies that are similar (aside from their response to the string 01), then there will be a large population in group 1 at this occurrence of 01. Consequently, the minority group will be 0, and the two strategies that differ only in their responses to 01 will again have an equal number of points. Thus, at the next occurrence of 01, the population of the minority group will once again be relatively close to 50% and the cycle will continue. That is, odd occurrences of a given string produce a minority population which is close to 50% (and consequently a small deviation in $L_1$ from the mean), while even occurrences produce a large deviation in $L_1$ from the mean, and a minority group opposite that of the preceding odd occurrence. It is the responses of the system to the even occurrences of strings that give rise to the large deviations in $L_1$, and are responsible for the fact that in the efficient, small m phase, $\sigma$ (as well as $\Delta\sigma$) is proportional to N for fixed m. Thus, the typical choices of the agents may be said to be maladaptive, in that their choices lead, half of the time (at the even occurrences of a given string) to very small minority groups, and consequently to a poor utilization of the resources. This is a clear manifestation of herd following or faddish behavior often seen in social systems.

This dynamics also explains the flat conditional probability distribution of Fig. 4a. Alternating occurrences of a given string produce opposite responses in the sequence of minority groups. Thus, at any given moment in a game, the number of times a 1 follows a given m-string of minority groups will differ by at most one from the number of times that same m-string is followed by a 0. Consequently the conditional probabilities will be very close to 0.5 for all m-strings.

Notice that because of the way in which the agents evaluate their strategies, the dynamics in this phase induces a new time scale. The time scale over which the strategies operate is m--i.e. each strategy looks back only m time steps to produce its prediction. But the agents evaluate their strategies from the beginning of the game, and so the dynamics produce a time scale of order $2^m$ (i.e., the typical time between occurrences of the same string) which is the time scale relevant for the adaptive dynamics.

Qualitatively, this period-two dynamics persists as long as there is a reasonable probability that the relative ranking of an agent's two strategies will not be altered between successive occurrences of a given string. Clearly, as m increases it becomes increasingly likely that the relatively rankings of an agent's strategies will be altered, since the time between successive occurrences of the same string ($\sim 2^m$) increases. Consequently, we expect that the bursty structure in $L_1$ and the maladaptivity will become less pronounced with increasing m.

Up to what values of m should we expect the period two, maladaptive dynamics to dominate? In section VI we shall present a quantitative description of the low m-phase that addresses this point, but here we offer a useful semi-quantitative discussion. The back-of-the-envelope argument presented here is not quantitatively correct, but it does capture much of the nature of the dynamics that leads to the cross-over from the efficient phase as m approaches $m_c$. First, consider two randomly chosen strategies of memory m. Suppose that following the occurrence of some string, $\xi$, their relative rankings differ by one. We ask, what is the probability that, after a time of order $2^m$ (which is the mean time between successive occurrences of $\xi$), they continue to differ by one in their rankings?

Roughly speaking, each of the other strings will be sampled once during a time of order $2^m$. Because the strategies are randomly chosen, each occurrence of one of the intervening strings will cause a change in their relative ranking of +1, -1, or 0, with probabilities 1/4, 1/4 and 1/2 respectively. Consequently, the change in their relative rankings over a time of order $2^m$ will be given by the excursion of a one-dimensional random walk of $2^m$ time steps, whose probability of moving to the left is 1/4, of moving to the right is 1/4, and of standing still is 1/2. The standard deviation of such a walk grows like $(2^m)^{-1/2}$, and the probability of ending up at the origin after $2^m$ time steps decreases like $(2^m)^{-1/2}$. If there are N agents, then there will be about $N/(2^m)^{-1/2}$agents the relative rankings of whose strategies will not change between successive occurrences of a given string. Now, absent any other dynamics, the number of agents voting for a given group at a given time will be of the order of N/2, with fluctuations of the order of $N^{1/2}$. The period two structure will dominate (so that the minority group on an even occurrence of $\xi$ is the opposite of the minority group at the preceding odd occurrence of $\xi$) if there is a sufficiently large number of agents, the relative rankings of whose strategies is unchanged. If so, then the deterministic period two dynamics will dominate over the order $N^{1/2}$ random statistical fluctuations in the number of agents voting for a group at an even occurrence of $\xi$. I.e., for the period-two dynamics to dominate, we must have $N/(2^m)^{-1/2} > N^{1/2}$, or $2^m/N < 1$. This gives us a correct order of magnitude estimate for the phase transition between the efficient and inefficient phases, and also gives us one argument that leads, naturally, to the scaling variable, $z=2^m/N$.

The period two dynamics at small m is even more pronounced as s increases. In Fig. 16 we plot a portion of the time series for m=2, but now with s=6. The bursting phenomenon is much more pronounced in this case, than in Fig. 15a. To understand this, consider the case of infinite s (or, for practical purposes, $s >> 2^{2^m}$). In this case, all agents possess all possible strategies. (Note that even if $s=2^{2^m}$ not all agents will possess all strategies, since the strategies are drawn by the agents randomly with replacement, so that agents may have more than one copy of the same strategy.) At the first occurrence of some string, $\xi$, each agent will have a subset of its strategies (at least two strategies), each of which has the largest number of points at that time step. For each strategy, t', in this subset which predicts a 1 following the string, $\xi$, there will be another strategy in the subset, identical with t', but which predicts a 0 following the string $\xi$. Suppose each agent's decision about which group to join is determined by a coin flip, when its strategies rankings are tied. In that case, the population of group 1 will be near 50%, within random statistical fluctuations. Suppose the minority group at this time step turned out to be group 0. Now consider the next occurrence of the string $\xi$. Because each agent has all strategies available, it is easy to see that the top ranked strategy must be the one whose prediction following $\xi$ is 0. To see this, suppose that the top ranked strategy was one whose response to $\xi$ was 1. Since each agent has all strategies available, there is another strategy, identical to that one, but whose response to $\xi$ is 0. Since the minority group following the last occurrence of $\xi$ was 0, that strategy will have one extra point. Thus, on even occurrences of each string, all agents must join the group which was the minority group on the last odd occurrence of that string. But since all agents join that group (in our example, group 0), then the other group will turn out to be the minority group. This will even up the points among strategies that differ only in their response to string $\xi$, so that the next time string $\xi$ occurs, strategies which differ only in their responses to $\xi$ will once again be tied in their rankings, and agents will choose which group to join randomly. It is not difficult to show that the variance of $L_1$ in this infinite s limit is $\sigma^2 = N(N+1)/8$, independent of m. Note that this result is obtained by first taking s to infinity. If s is large, but much less than $2^{2^m}$, then $\sigma$ does depend on z, as shown in Fig. 11.

B. The transition region

We will define the transition region in this system to be $z_o \leq z \leq z_c$, where $z_o$ is the value of z at which $\sigma$ first becomes less than $\sigma$ in the RCG. It is over this range of z that there are dramatic qualitative changes in the behavior of the system It is, therefore, clearly advantageous to be able to study this region in detail. To that end we will introduce a new set of strategies that define strategy spaces with non-integer m. This will allow us to sample the behavior of the system at much finer intervals in z than would be possible were we restricted to only integer values of m. Aside from its use in studying the transition region, this generalization is very important for studying the robustness of the general phase structure of this system, to changes in the nature of the strategy space. This issue will be discussed in detail in a forthcoming publication[9]. In what follows, we will first study some useful quantities over a broad range of integer m. Later in this section, we will introduce the generalized strategy space for the purpose of studying the transition region in greater detail.

We begin our discussion of the transition region by looking at the behavior of two useful quantities over a wide range of m. From the discussion in the preceding subsection, we know that the period two dynamics becomes less pronounced as z approaches $z_c$ from below. To see this explicitly, we have plotted in Figs. 17 and 18 the probability, averaged over all m-strings, that the minority group following an even occurrence of a given string is opposite of the minority group following the preceding odd occurrence of the same string (POED). These figures plot this quantity as a function of integer m for N=101 and two different values of s, s=2 and s=16. On the same graphs, we also plot the probability that the minority groups following two successive odd occurrences of a given string are different from each other (POOD). Refer first to Fig. 17. We note that the upper curve (POED) is close to one for small m, which is a reflection of the dominance of the period-two dynamics at small m. As m increases, POED decreases, reaching about 0.7 for m=5. This is what we expect: For m close to $m_c$, not all the strategies maintain their relative rankings from one odd occurrence of a string to the subsequent even occurrence. Consequently, the populations of the minority groups associated with even occurrences of strings are generally closer to half the agents than is the case for smaller values of m. This means that the probability that the minority group following an even occurrence of a string is opposite that of the previous occurrence of the same string will no longer be close to one. As further confirmation of this picture, we plot in Fig. 19 a short segment of $L_1$ for N=101, m=5 and s=2. We see no large bursts in this time series as we saw, for example, in Fig. 15a. This, of course, is consistent with the fact that $\sigma$ is below that for the RCG for m=5, N=101.

Next, look at the lower curve in Fig. 17, POOD. This curve is always less than 0.5, rising to near 0.5 for integer m=5 and then decreasing. The reason that POOD is generally less than 0.5 is a consequence of the fact that the agents' strategies are fixed, and do not change over the course of the game. This can be understood by recalling that the strategies represent a fixed random sample of responses to different strings. Although, on average there will be as many strategies that respond 0 to a given string as respond 1, for a given distribution of strategies to the agents, there will be statistical fluctuations in this number. Thus, absent any other dynamics, if the minority group is, say, 1 following a given string, then *a priori*, the probability will be greater than 0.5 that a subsequent occurrence of that same string will result in 1 again being the minority group. Of course, the period two dynamics in the low m-phase swamps this *a priori* expectation, and results in POED being close to 1. But since the period two dynamics does not affect POOD, that probability is still less than 0.5 in the low m phase. Note also that both POED and POOD decrease for m> $m_c$, and approach 0.33 as m gets large. We shall explain this result in the next subsection.

We now want to examine the region near the phase transition in more detail. To do that we need to generalize strategy space to a dimension of $2^m$ with non-integer m. To begin, in a strategy of memory m, consider two strings of length m, say A and B, which differ only in their last bits (i.e. in the bit referring to the minority group m time steps earlier). In one of these strings (say A), replace the last bit, (0 or 1) with a “don’t care”, denoted by an asterisk (*). To avoid possible inconsistencies in the strategy table, it is then necessary to remove from the strategy table string B. Thus, for every asterisk inserted in this way into the strategy table, the dimension of the strategy space is reduced by 1, allowing m to take on non-integer values. An example of a strategy from such an interpolating space is shown in Fig. 20. In this way we can interpolate between a strategy space with a dimension $2^M$, (M an integer) and one with dimension $2^{M-1}$. To construct a strategy space with non-integer m, therefore, one randomly chooses a number of strings of length m, replaces the last bit with an asterisk, and eliminates the partner string. The remaining strings (some with asterisks, some without) define the dimensions of the strategy space. The value of m for such a strategy space is just $\log_2$[the number of different strings in the strategy table]. So, for example, the strategy shown in Fig. 20 is drawn from an 11-dimensional space yielding a value of $m=\log_2[11]\cong 3.46$. The dimension of the strategy space decreases as the number of asterisks used increases, and if each string contains as asterisk, then the dimension of the strategy space will have been reduced from $2^M$ to $2^{M-1}$. A given set of strings is used for the left column of all the strategies of a given run. That is, in these games, all strategies look at the same set of information to make their decisions[18] We have studied strategy spaces constructed in this way over a broad range of m and N, and have found that this procedure does produce results which smoothly interpolate between those presented above for integer m.

Let us now take a closer look at the transition region using the method described above to interpolate between integer values of m. In the results of Figs. 21 and 22 discussed below, 16 runs were performed for each non-integer value of m. In each run, an appropriate sized strategy space was constructed by distributing asterisks randomly as described above (a different random distribution of asterisks for each run). For each run, random strategies in the strategy space so defined were assigned to the agents in the usual way.

In Fig. 21 we show $\sigma$ as a function of m for $4<m<7$, $N=101$ and $s=2$. In Fig. 22 we plot POOD and POED for the same parameters. In both these figures, we have divided both the ranges $4<m<5$ and $6<m<7$ into 16 bins each. The range $5<m<6$ was divided into 32 bins to show more detail. In Fig. 21a we show $\sigma$ for 16 different runs for each value of m. In Fig. 21b, we plot the average $\sigma$ averaged over all 16 runs for each value of m shown in Fig. 21a. First, in Fig. 21b, we see that the minimum in $\sigma$ does not occur at an integer value of m, rather, we see that $\sigma$ has a broad minimum extending from $m\approx5.2$ to $m\approx5.7$. This suggests that a reasonable estimate for $m_c$ would lie in this range. In Fig. 22 we see that that POOD also has a broad maximum reaching 0.5 over a similar range, from $m\approx4.8$ to $m\approx5.4$.

[18] The very interesting situation in which not all strategies use the same set of information to make their predictions, will be discussed in Ref.9, above.

Look now at Fig. 21a. With the detail provided by this graph, we see an effect that was not evident when we studied only integer m. In this figure it appears that there are two distinct effects driving the phase change in the region of $m_c$. There is a one dynamic which produces a broad distribution of $\sigma$'s for a given m, and which dominates for low m, and a second dynamic which produces a $\sigma$'s that lie in a narrow range, and dominates for larger m. The coexistence of the two dynamics in the region $4.8<m<5.5$ is evident in this figure. This suggests that the phase change (which we have loosely referred to as a phase transition) may actually be accomplished by a smooth cross-over from one dynamic to another. In analogy with physical systems, one may think of two local "energy" minima representing different states of the system, one associated with the strategy-efficient dynamics of the low m phase, the other with the inefficient, cooperative dynamics of the high m phase. For any given parameter setting, the system prefers to be in a minimal energy state. For small m, the efficient minimum is also the global minimum, and so the system prefers to be in that minimum. For large m, the inefficient minimum is the global minimum, and so the system is preferentially in that minimum. As m passes through the transition region, these two minima smoothly change their relative values. When their depths are nearly the same, the system will sometimes occupy one minimum and sometimes another, as suggested in Fig. 21a for $4.8<m<5.5$. That there is a phase change of the system through the transition region is clear. What its analytic structure is is not so clear.[19]

[19] Two comments are useful here. First, it seems likely, at least as $s\to\infty$, that there is a singularity in at least some quantities, for example, Q(m) in Fig. 14. It may be, therefore, that for finite s the phase change is accomplished by a smooth crossover which becomes singular as $s\to\infty$. Second, it is interesting to remark that the transition between the two phases is accompanied by a change in the nature of the system from essentially periodic to apparently non-periodic. In the low-m phase, for example, $L_1$ shows a distinct periodic structure. As m approaches the region near $m_c$, the periods in $L_1$ become longer and more complex. In the inefficient, high m phase, there is no apparent periodicity in $L_1$.

Now let us turn again to Fig. 22. First, we remark that, although not plotted on this graph, PEED, the probability that the minority group following two successive even occurrences of the same string are different from each other, forms the same curve as POOD. In particular, both PEED and POOD are 0.5 at $m_c$. Second, it is interesting that POED is greater than 0.5, even at $m_c$. Furthermore, PEOD, the probability that the minority group following an odd occurrence of a given string is different than that following the preceding even occurrence, is also about 0.6 at $m_c$[20]. These results paint a picture of cooperative dynamics in which responses to a given string are anti-correlated over times of order $2^m$, but are uncorrelated over times of order $2^{m+1}$.

Given the argument that *a priori*, POOD and PEED should be less than 0.5, it is most remarkable that they reach the value of 0.5 at $m_c$. The cooperative dynamics is evidently significant enough to overcome the *a priori* tendency for minority groups to be the same following any given string. On the other hand, it is not clear what the dynamical origin is of the fact that POED and PEOD are both greater than 0.5 at $m_c$. One possibility is that this just indicates a remnant of the period two dynamics that dominated the efficient phase. In this case, the observed transition at $m_c$ may be more accurately thought of as a cross-over effect, as might be suggested by Fig. 21a. On the other hand, it is possible that these correlations are intrinsically bound up with the cooperative dynamics. We are currently agnostic about this issue, but its resolution is of significance. It is certainly important to understand the nature of the transition in as much detail as possible. But in addition, one may be interested in attempting to control the system to improve the degree of cooperation (lower $\sigma$). Then, it may make substantial difference whether it is possible to identify a reasonably separable dynamics (e.g. the remnant of the maladaptive period two dynamics) which can be tuned down near $m_c$ to further enhance the agents' emergent cooperation.

We also note in Fig. 22 fluctuations in POED, particularly near $m \approx 5$. This is apparently a real detail, not a statistical effect, whose origin we do not fully understand. The smaller fluctuations in POED may also have dynamical significance.

Our studies with non-integer m indicates that $z_c$ is closer to 0.35 than to the value of 0.5 that was deduced from the studies performed with integer m. We have examined a number of cases with different values of N. We find in all cases that the results for non-integer m smoothly interpolate between the results for integer m, that $\sigma^2/N$ still scales with z, and that the results still lie on a universal curve, but that the minimum of the curve is near $z_c \approx 0.35$.

[20] However, PEOD does differ from POED for lower m, as we would expect from our discussion of the period two dynamics in section IVA.

To close our discussion of the transition region, we refer briefly to Fig. 18, in which we plot (for integer m) POED and POOD for the case N=101 and s=16. Here we see the same qualitative behavior as in Fig. 17, except that POED stays close to 1 throughout the efficient phase, and both POED and POOD approach a value close to 0.5 as m gets large. The behavior of POOD for small m is easy to understand on the basis of the discussion of the last subsection: If the agents have more strategies from which to choose, they will be more likely to find pairs of similar strategies for higher m, and so the period two dynamics will be more pronounced in the efficient phase. The asymptotic values of POOD and POED for large m will be explained in the next subsection.

C. Dynamics in the inefficient phase

For $z>z_c$, the system is in a quite different phase. The most obvious hallmark of this phase, as shown in Fig. 7, is that the system is more effective at generating resources than would be the case in the RCG, but increasingly less so as z increases. For z near $z_c$, the agents are able to achieve maximal coordination and thus generate resources most effectively, but coordination evidently becomes less effective as z increases still further. To understand what happens, we will consider the information in G for various values of z. To illustrate, we will again use our canonical example of games with N=101 and s=2. In this section it will also be sufficient to consider only integer m.

There are a number of different measures that we could use to characterize the information in G, including the entropy, but for our purposes a simpler measure is more appropriate. We are interested, in particular, in characterizing the information in G that is available to the strategies. Thus, we consider

$$I_m = \frac{1}{2^m}\sum_{u_m} I(u_m) = \frac{1}{2^m}\sum_{u_m}(P(1\,|\,u_m) - \frac{1}{2}) \qquad (4.1a)$$

and the sum over the squares of $I(u_m)$,

$$I_m^{[2]} = \frac{1}{2^m}\sum_{u_m} I^2(u_m) = \frac{1}{2^m}\sum_{u_m}(P(1\,|\,u_m) - \frac{1}{2})^2 \qquad (4.1b)$$

The sum here is over all strings of length m. Clearly, the more $P(1|u_m)$ differs from ½, the more information the string $u_m$ contains for predicting the next minority group. In Fig. 23 we plot $I_m^{[2]}$ (4.1b) as a function of m for the case N=101 and s=2. As we expect from our discussion of the dynamics of the efficient phase, $I_m$ is zero for $m<m_c$. For $m>m_c$ $I_m$ is non-zero and increases for increasing m. As m increases, the system becomes increasingly *inefficient* with respect to the strategies, in that increasing amounts of information are left in G.

It is not difficult to understand what information is contained in G. First, recall that the distribution of strategies to the agents represents a small random sample of all the $2^{2^m}$ possible strategies of memory m. Suppose we ignore any dynamics or cooperation among the agents, and just consider the probability that a 1 will follow any given m-string in G. In the absence of any coordination, each agent will choose randomly between its two strategies. Given a random choice between each of the agent's two strategies, the probability to have a 1 following a given m-string will not be exactly 0.5. The reason is that for any particular distribution of strategies, some agents will have the same response to a given string in both their strategies, thus skewing $P(1|u_m)$ away from 0.5, even in the absence of non-trivial dynamics. We can measure the skewness of the distribution of strategies by considering

$$\Psi_m = \frac{1}{2^m}\sum_{u_m}\Psi(u_m) = \frac{1}{2^m}\sum_{u_m}\frac{\nu(0,0\,|\,u_m) - \nu(1,1\,|\,u_m)}{N} \qquad (4.2)$$

where $\nu(0,0|u_m)$ and $\nu(1,1|u_m)$ are the number of agents both of whose strategies predict, respectively, 0 and 1 following $u_m$. The more positive (negative) $\Psi(u_m)$ is, the more likely it is that the minority group following $u_m$ will be 1 (0).

Now consider the correlation between $\Psi(u_m)$ and $I(u_m)$. In Fig. 24 we plot $\Gamma_m$ vs. m, where

$$\Gamma_m = \langle I(u_m)\Psi(u_m)\rangle - \langle I(u_m)\rangle\langle\Psi(u_m)\rangle$$
$$= \frac{1}{2^m}\sum_{u_m} I(u_m)\Psi(u_m) - I_m\Psi_m \qquad (4.3)$$

We see very clearly that $\Gamma_m$ increases monotonically with m, indicating that the information left in G is increasingly correlated with the skewness associated with the agents' strategies. That is, as m increases G contains increasingly accurate information about certain features of the agents' strategies.

Since the agents' choices are apparently less coordinated as m increases beyond $m_c$, we expect that for very large m, the agents will choose between their two strategies randomly at each time step. If that is so, then we should be able to compute a good estimate of $P(1|u_m)$ directly from $\nu(0,0|u_m)$ and $\nu(1,1|u_m)$. If the agents choose between their two strategies randomly and with equal probability, then it is not difficult to see that the conditional probabilities become

$$P_R(1\,|\,u_m) = 2^{-\omega}\sum_{k=\alpha}^{\omega}\frac{\omega!}{k!(k-\omega)!} \qquad (4.4)$$

where $\omega=N-\nu(0,0|u_m)-\nu(1,1|u_m)$,
$\alpha=N/2-\nu(0,0|u_m)$,

and $P_R(1|u_m)$ are the conditional probabilities assuming that the agents choose between their two strategies randomly, independently, and with equal probability.

In Fig. 25, we plot the average squared deviation between the observed conditional probabilities, $P(1|u_m)$, and $P_R(1|u_m)$ as a function of (integer) m,

$$\phi_m = \left[\frac{1}{2^m}\sum_{u_m}\frac{[P(1\,|\,u_m)-P_R(1\,|\,u_m)]^2}{P^2(1\,|\,u_m)}\right]^{1/2}. \qquad (4.5)$$

Here we see that as m increases, P and $P_R$ are increasingly similar, differing only by a few percent by the time m=8, and being nearly identical for m=11. Thus, the information contained in G at large m is a simple reflection of the particular preferences embodied in the agent's strategies.

It may seem counter-intuitive that the system as a whole behaves more like the random choice game while the information in G increases. The point here is that the time series of minority groups is endogenous. It is created by the choices of the agents themselves. The fact that more information is left in G, just reflects that fact that the agent's choices are less coordinated, so that those choices use less of the information inherent in the strategies of the agents themselves. For this system, uncoordinated choices means that each agent chooses randomly, independently and with equal probability between his two strategies at each time step. Note, however, that this is not the same as the RCG in which each agent chooses 0 or 1 randomly, independently and with equal probability at each time step. In the latter case G would be an IID sequence with no information, (and the system would be in a mixed-strategy Nash equilibrium) but in the former case G contains significant information about the structure of the agents' strategic preferences (and the system is not in a Nash equilibrium). The existence of this fixed strategic preference structure is the reason that the system becomes increasingly inefficient (so that there is more information in G) as m increases, while at the same time the agents' choices become as uncoordinated and as random as possible.

The fact that the agents' choices among their strategies becomes increasingly random as m increases also explains the behavior of POOD and POED in the large m limit seen in Figs. 17 and 18. It is not difficult to show that a random choice between two fixed strategies will yield a value of 0.33 for the probability that any two responses to the same string result in opposite minority groups. As s increases, the large m value of these quantities approaches 0.5. The reason is, that for larger s, an agent can choose from among more strategies in response to a given string. Thus, an agent will respond to a given string with either minority group with equal probability and with relative fluctuations of order $s^{-1/2}$. Consequently, POOD and POED should approach 0.5 as m increases for fixed large s, as we observe in Fig. 18.

In general, the inefficient phase for larger s should have much the same qualitative structure as the s=2 example discussed above. We do expect, however, that the information in G for fixed m and N in the inefficient phase should decrease for larger s. The reason is that the information left in G reflects the conditional probabilities $P_R(1|u_m)$, and as s increases, these conditional probabilities should approach 0.5. At the same time, the agents look increasingly similar as s increases, for fixed m and N. But a homogeneous population cannot coordinate their strategy choices to produce large minority groups. Therefore, we also expect to see less effective emergent coordination in the inefficient phase, as s increases, which is indeed what we observe in Fig. 11.

## V. Agent Wealth

Given the interesting dynamics in the efficient and inefficient phases, it is natural to ask about the distribution of wealth (i.e. the number of points) to the agents. In a forthcoming publication,[9] we will present a detailed analysis of agent wealth distribution, but here we wish to emphasize one important point. We ask, with what features of the agents' strategies is agent wealth associated. In particular, is there something intrinsic about strategies (or sets of strategies used by an agent) that are very successful, or do they gain their success only in the context of the other strategies being played in the game.

To address this question, we define two distance measures. One is in an intra-agent distance, $D_h(i)$ defined as the Hamming distance between the $i^{th}$ agent's two strategies. The second is a distance in "behavior space", $D_b(i)$ and may be understood to be the average behavioral distance of the $i^{th}$ agent from all other agents playing the game. In particular, if $T_i^{[j]}(u_m)$ denotes the response (0 or 1) to the string, $u_m$, of the $j^{th}$ strategy of agent I, then

$$D_h(i) = \sum_{u_m} \left| T_i^{[1]}(u_m) - T_i^{[2]}(u_m) \right| \tag{5.1}$$

and

$$D_b(i) = \sum_{j} \sum_{k,l} \sum_{u_m} P(u_m) \left| T_i^{[k]}(u_m) - T_j^{[l]}(u_m) \right| . \tag{5.2}$$

In Fig. 26 we plot accumulated agent wealth as a function of $D_h$ and $D_b$ for games played with s=2 N=101 and various values of m. One first notes that for $m<m_c$, agent wealth is quantized. This follows from the fact that the dynamics in the low m-phase is periodic, an effect that will be discussed in more detail elsewhere[9]. But the point we wish to emphasize is the fact that for $m<m_c$, agent wealth is strongly correlated with $D_h$. In this phase, the more similar an agent's two strategies, the wealthier the agent tends to be. Note also that for $m<m_c$ there is apparently no correlation between an agent's wealth and $D_b$. That is, in this phase it does not matter what all the other agents are doing. It only matters how similar an agent's two strategies are.

For $m>m_c$, the situation is quite the opposite. Now there is no correlation between agent wealth and $D_h$, but there is a strong correlation between agent wealth and $D_b$. The further an agent is, on average, from all other agents in behavior space, the wealthier he tends to be. This is quite reasonable in a game in which points are rewarded for being in the minority. Why, then, is there no correlation between $D_b$ and agent wealth for $m<m_c$?

The reason is that for $m<m_c$, there is no predictive information available to the agents' strategies and consequently no emergent coordination among agents' choices. For $m>m_c$ the agents use the real information in G to coordinate their choices. Those agents possessing strategies that allow them to behave maximally differently from other agents will more often find themselves in the minority group, and will accumulate more points. On the other hand, as we have demonstrated, for $m<m_c$ there is no real information available to the agents' strategies. Therefore, any attempt to use apparent information, and to make choices on the part of the agents, invariably leads to maladaptive herding behavior, which is not rewarded in the minority game. For $m<m_c$, agents do best when then ignore, as much as possible, the misleading signals being sent by the collective behavior of the other players. In our game, a surrogate for ignoring those misleading signals is for an agent not to be able to make choices, and that is accomplished when the agent's strategies are as similar as possible. (If the agent's strategies were identical, the agent would effectively have only one strategy, and could not respond to the misleading signals of the collective by choosing to join different minority groups at different times, in response to a given string.) Although for $m<m_c$ the information in G available to the strategies is zero, in a very real sense, the information in G available to agents (i.e. information that extends over times greater than $2^m$) is negative, in that it leads to maladaptive decisions.

**VI. Mean-field description of the minority game**

In this section, we present a mean-field approximation for the behavior of the minority game. This description is most accurate for the behavior of the system for s extremely large ($s>2^{2^m}$). In addition, for smaller s, the description is applicable to the low m, efficient phase ($z \lesssim z_c$). In both the large s and low m regions, the key feature which allows us to write a mean-field approximation, is that the system response to an occurrence of a given m-string of minority groups is nearly independent of its responses to the other m-strings. (Recall that it is this independence that is necessary for the period-two dynamics that dominates the low-m phase.)

In formulating this mean-field description, we shall use the language of statistical mechanics, in which we will identify a set of independent variables and associate "energies" with various states of objects in our system. In particular, we shall identify m-strings of minority groups as variables in the system, and shall associate energies with strategies. The lower the energy of a strategy, the more likely it is to be chosen by its agent to be played. This is analogous to states of a statistical mechanical system, in which a system is more likely to occupy states with lower energy. In our case, there is simple relationship between the energy of a strategy and the number of points it has accumulated. One reason for introducing the language of statistical mechanics and ranking strategies by energy, is that this allows us to introduce a temperature into the system. This in turn lets us perform an interesting interpolation between the game as it was originally defined (the zero temperature limit), and a game in which the agents choose among their strategies randomly (the infinite temperature limit). This will be discussed elsewhere.

Now, the primary objects of interest to us are the strategies. We will associate an energy with each strategy at each time step, T. Specifically, if a strategy predicts the next minority group correctly, its energy will be lowered by 1, and if it predicts incorrectly, its energy will be raised by 1. It is easy to see that a strategy's energy, E, is related to the number of points accumulated by that strategy, $\lambda$, by $2\lambda=T-E$, where T is the time step of the game.

The energy, $E_t$, of a strategy, t, can be written in terms of the contributions to that energy associated with the response of the strategy to different m-strings of minority groups. Thus, consider a string, $\xi$. Define the "string energy", $e_\xi$, of $\xi$ as follows:. Every time the string $\xi$ is followed by a 1, $e_\xi$ decreases by one and every time $\xi$ is followed by a 0, $e_\xi$ increases by one. Then, the cumulative contribution to the energy, $E_t$, of any strategy, t, due to its responses to the string $\xi$, is equal to $e_\xi$, or $-e_\xi$, depending on that strategy's response to the string, $\xi$. I.e.,

$$E_t = \Sigma_\xi \; p_\xi(t)(-e_\xi) \qquad (6.1)$$

where $p_\xi(t)$ is -1 if the strategy t predicts 0 following the string ξ, and is 1 if the strategy t predicts 1 following the string ξ.

We can now define our mean-field approximation of the minority game. The key to the simple treatment of the game in the infinite s limit (and in the low m-phase for finite s) is the independence of the response of the system to different strings, resulting in the simple period-two dynamics for each string. Thus in the mean-field theory the m-strings of minority groups are considered the independent degrees of freedom.

The observation of the period two dynamics suggests that in this phase it is sufficient to allow all the string energies to take on only values of 0, 1, or -1, independently, and with probabilities 1/2, 1/4, and 1/4, respectively. Under these assumptions, we can easily compute P(E;n), the probability for a strategy of memory m to have a total energy, E, where $n=2^m$. This is just given by the trinomial distribution,

$$P(E;n) = \sum_{k=|E|}^{(n+|E|)/2} \frac{n!}{k!(k-|E|)!(n+|E|-2k)!4^{2k-|E|}2^{n+|E|-2k}} \quad ; \quad n=2^m. \qquad (6.2)$$

Now, it is clear that the leading term in the variance of $L_1$ in the low-m phase comes from the large bursts in this time series. These are the times at which the system responds to a string, ξ, where the string energy, $e_\xi$ is 1 or -1. I.e., it is those times at which there is an even occurrence of ξ. Therefore, we will be interested in computing an estimate of the number of agents joining group 1 at an even occurrence of ξ. For specificity, lets assume that $e_\xi=-1$. We now calculate $p_1$, the probability that an agent joins group 1 in response to the string ξ, if $e_\xi=-1$.

First, consider the case s=2. The probability that the lower energy of two randomly chosen strategies of memory m is E, is given by

$$W_2(E) = P(E;n)\left[P(E;n) + 2\sum_{E'>E} P(E';n)\right] \qquad (6.3)$$

Now, if $2^m >>1$, we can approximate P(E;n) by a normal distribution:

$$P(E;n) \cong \sqrt{\frac{4}{3\pi n}} \exp\left(\frac{-4E^2}{3n}\right) \qquad (6.4)$$

and the probability that the lower of the two energies is E becomes

$W_2(E;n)=2P(E;n)Q(E;n)$, (6.5)

where

$$Q(E;n)=\int_E^\infty P(E';n)dE' . \tag{6.6}$$

This result can be generalized to the case of s strategies, so that the probability that E is the lowest energy among s randomly chosen strategies is given by

$$W_S(E;n)\cong -\frac{\partial}{\partial_E}Q^S(E;n) . \tag{6.7}$$

Now, given that $e_\xi$ = -1, the probability that a strategy with energy E will choose group 1 in response to string $\xi$, is just given by

$$\frac{P(E+1;n-1)}{P(E+1;n-1) + P(E-1;n-1)} . \tag{6.8}$$

The reasoning is as follows: A strategy with n elements can be decomposed into two parts: The first, it's response to the string $\xi$, and the second, the remaining n-1 elements. For a randomly chosen strategy, this latter part may be considered a random strategy with n-1 elements. There are two ways in which a strategy with n elements can achieve an energy E, given that $e_\xi$ =-1. If the response of the strategy to $\xi$ is 1, then the remaining n-1 elements of the strategy must contribute an energy of E+1 to the total energy of the strategy. Conversely, if the strategy's response to $\xi$ is 0, then the remaining n-1 elements of the strategy must contribute an energy of E-1. Thus, the probability that a strategy with n elements and energy E chooses group 1, is just P(E+1;n-1), normalized by the total probability to find a strategy of energy E.

Consistent with our restriction that the $e_\xi$ independently takes on only values 1,-1 and 0, (i.e., because of the period two dynamics) $E\sim O(n^{1/2})$, if n>>1. In that case, expression (6.8) can be approximated as

$$\frac{P(E+1;n-1)}{P(E+1;n-1)+P(E-1;n-1)}\cong\frac{1}{2}-\frac{4E}{3n} \tag{6.9}$$

We can now estimate $p_1$, the probability that an agent will choose group 1, given that $e_\xi$=-1. It is

$$p_1 = \frac{1}{2}\int_{-\infty}^{\infty}\left(1-\frac{8E}{3n}\right)W_s(E;n)dE = \frac{1}{2}\left[1+\sqrt{\frac{8}{3n}}\chi(s)\right] \tag{6.10}$$

where

$$\chi(s) = \int_{-\infty}^{\infty} xdx\frac{d}{dx}\left(\frac{1}{\sqrt{2\pi}}\int_x^{\infty} e^{-\frac{y^2}{2}}dy\right)^S . \tag{6.11}$$

The function $\chi(s)$ is plotted in Fig. 27. Note that $\chi(s)$ is zero for s=1, and varies relatively slowly for s>5.

Now, the average number of agents joining group 1, given that $e_\xi$=-1 is just $p_1N$. So long as the fluctuations in this number are small enough, the number of agents choosing 1 will always be greater than N/2, and the period-two dynamics will obtain. This will certainly be the case if $N\chi(s)^2/2^m=\chi(s)^2/z>>1$. Thus, under these circumstances, a majority of agents will choose group 1 when $e_\xi$=-1, and the variance in the attendance from N/2 given that $e_\xi$= -1 will be

$$\sigma^2 = \frac{N}{4} + \frac{N(N-1)}{2}\left(p_1 - \frac{1}{2}\right)^2 \cong \frac{N}{4} + \frac{N^2}{3\bullet 2^m}\chi^2(S) . \tag{6.12}$$

This result is a good approximation as long as the second term in (6.12), $N\chi(s)^2/2^m=\chi(s)^2/z>>1$, or z<<1. For z $\gtrsim$ 1, this picture breaks down. The $e_\xi$ take on values other than 0, 1 and -1, and the response of the system to different m-strings lose their independence. Since this picture, and the approximation in equation (6.9) breaks down when z $\gtrsim$ 1, we expect this to be the point at which the phase transition occurs, consistent with Fig. 7. Note also that the form of the second term in equation (6.12) is consistent with our earlier argument for the position of the phase transition.

## VII. Summary and Discussion

### A. Summary

In this paper we have presented an overview of the rich and complex behavior that very simple minority games display. Our main results are:

1. As a function of the size of the strategy space from which agents draw their strategies, minority games display two phases separated by a clear transitional region. The first (small strategy space) is an informationally strategy-efficient phase in which no predictive information about the next minority group is available to the agents' strategies. The second (large strategy space) is a phase in which information is available to the agents' strategies.

A. For small strategy spaces the system does poorly at generating resources. The dynamics in this phase is dominated by a simple period two dynamics for which it is possible to write a mean-field theory. This dynamics leads to agents' choices that are maladaptive resulting in worse than random performance. The system is also strategy-efficient, in that *there is no information available to the agent's strategies* that can be used to determine which group, at any given time step will be the minority group.

B. For larger strategy spaces the agents can coordinate their decisions to produce better than random generation of resources. As the strategy space increases in size, agent coordination becomes less effective and, for very large strategy spaces, over-all system performance approaches that of a random choice game. At the same time, the information in the data set that the agents use to determine their strategy choices becomes greater as the size of the strategy space increases. For large strategy spaces, this information directly reflects the probability structure of choices embodied in the agents' strategies.

C. The best resource generation is obtained in the transition between the two phases, when the dimension of the strategy space is on the order of the number of agents playing the game.

2. For different numbers of agents, and different sizes of the strategy space, this general phase structure is a function only of the ratio of the dimension of the strategy space divided by the number of agents.

3. Agent wealth is correlated with different strategy characteristics in the two phases.

A. In the small strategy space, maladaptive phase in which the agents exhibit herding behavior, agents whose strategy sets are composed of very similar strategies do best, without regard to the nature of those strategies, or to the nature of the strategies of the other agents.

B. In the large strategy space phase, agents whose strategies are maximally distant in behavior space from all other strategies do best. Thus, the value of a given strategy is dependent on details of that strategy in the context of all other strategies in the game.

4. All of this structure depends on the agents having a small to moderate degree of adaptivity. In the case in which there is only one strategy per agent, this structure disappears, and the game becomes trivial and periodic. On the other hand, as the number of strategies available to each agent increases beyond two, coordination becomes more difficult. The general phase structure still obtains, but the degree of coordination diminishes and the system generally performs more poorly at the transition and in the inefficient phase than it does when each agent has fewer strategies (but still more than one).

B. Discussion

The very rich behavior seen in these simple minority games is quite remarkable, as is its apparent robustness. Although not reported in detail in this paper, we have also studied the robustness of the general phase structure of these games to changes in the information set used by the agents' strategies. Of course, the interpolated strategy spaces introduced to study the detailed behavior in the transition region do represent one significant alteration in the information set available to the agents. The fact that the results smoothly interpolate between the strategies spaces with integer m speaks to the generality of our results. But in addition, we have found that when other kinds of information is given to the agents—for example, more specific information about the number of agents in a group as a function of time—the general phase structure of the system remains the same. There is a single phase transition between a poorly performing strategy-efficient phase, and a better performing strategy inefficient phase. Moreover, the transition still occurs when the ratio of the dimension of the strategy space divided by the number of agents in the game is a number is of order one[9].

Nevertheless, there are a number of effects that could alter the system's dynamical structure significantly. One feature of the games described in this paper is that all of the agents are looking at the same signals. That is, the strategy space is the same for all strategies. It may be that the existence of a set of common signals on which the strategies base their predictions allows the agents to coordinate their strategy choices in a relatively simple way. One straightforward way to change this condition, is to allow different strategies to use different memories. We are currently studying such games to assess the importance of a universally shared strategy space.

Another very important issue to address is the role of evolution in such games. Although the systems discussed in this paper were adaptive, they were not evolutionary, in the sense that the agents' strategies were fixed from the beginning of the game. It is unclear how the system will behave when evolution is introduced and the agents are allowed to change their strategies. It does seem likely, however, that the results will depend on some of the characteristics of the way evolution is introduced into the system. In particular, it is clear that if evolution takes place on a very fast time scale (significantly shorter than $2^m$) then the system may devolve into random behavior, since the agents will not have time to learn and coordinate their strategy choices with the other agents. For evolution on a slower time scale, though, the outcome is unclear. One intriguing possibility, however, is that the system would evolve to a set of strategies with memory near $m_c$.

A natural question to ask about our results is their relation to Nash equilibria. The minority games have many Nash equilibria. For example, one Nash equilibrium is a rule in which the first (N-1)/2 agents always join group 0, the last (N-1)/2 agents always join group 1, and the remaining agent alternates between the two groups. There are clearly many variations on this rule that are also Nash equilibria. The mixed strategy rule in which each agent randomly, independently and with equal probability chooses between the two groups (i.e. the RCG) is also a Nash equilibrium. However, it does not appear that the states achieved by the dynamics of the games described here are Nash equilibria. Even for very large m, where the system-wide behavior is that of the RCG, the system is not in a Nash equilibrium. Although the agents do choose between their strategies randomly, they do not choose the minority groups randomly. It would be possible, therefore, for some agent to choose some other deterministic ordering of choices for his strategies and increase his wealth. That this is possible, in principle, follows from the fact that G is not an IID sequence. Thus, there is information that can be exploited to improve an agent's wealth. However, the dynamics of this game do not allow the agents to exploit that information. These observations, of course, are just special cases of the general idea that dynamics in an adaptive system can prevent the system from achieving equilibria, and in particular, Nash equilibria. In our opinion, they are important indicators of the need to seriously consider dynamics in the analysis of economic and other complex adaptive systems, and not rely solely on equilibrium analyses.

It is also interesting to consider again the nature of the strategies and their success in these games. In particular, consider the inefficient phase in which the strategies in play represent only a small sample of the total number of strategies in the strategy space. Here some strategies perform better at predicting minority groups than others. But *a priori* there is nothing special about any particular strategy. They are all just randomly generated strings of 0 and 1. Nevertheless, in the context of the other strategies represented in that particular game, certain strategies perform better than others. A necessary condition for the emergence of this ordering is the fact that the strategies in play represent only a small sample of the total strategies possible. Indeed, increasing the number of strategies in play, either by increasing N or s sufficiently, puts the system into a region in which no strategy performs better than any other. Thus, there is a sense in which there is an emergent meaning to specific *a priori* random strategies, which meaning emerges contextually in an environment representing a small random sample. We believe that a sparsely sampled configuration space may be a general condition for the emergence of meaning. The simplest example of this is the observation that it is always possible to find structure and patterns in a finite string of random numbers. As the length of the string grows, however, finite length patterns lose their significance. The kinds of meaning that emerge in different contexts, and the criteria for what constitutes a "small" random sample will certainly depend on the dynamics associated with the system. Nevertheless, we believe that a sparse sampling of the configuration space of the system may well be a necessary condition for the emergence of meaning[21].

This system also has some characteristics that are reminiscent of a spin-glass in statistical mechanics. In a spin-glass[22], the interactions among particles are mediated by a set of fixed, randomly distributed pair-wise interactions, which may differ in sign, some being ferromagnetic (causing spins to align in the same direction) and others being anti-ferromagnetic (causing spins to align in opposite directions). In general the system may not be able to satisfy all these tendencies for a given distribution of interactions, while at the same time achieving a low energy. This is the phenomenon of *frustration*, and leads to a very rich behavior of the system. In our game, the strategies that are distributed to the agents typically represent a small random sample of all possible strategies. They are fixed for the duration of the game, and they endow agents with strategic preferences which, in general, cannot all be satisfied, while at the same time achieving an optimal utilization of resources. Much of the rich behavior of our game is intimately bound up with the frustration associated with highly constrained agent strategic preferences. It will be interesting to study the ways in which frustration effects change if evolutionary dynamics are introduced into the system, so that agents are allowed to alter their strategies in response to selective pressure. The general dynamic of frustration is likely to be a very important one in the study of adaptive competition and evolution in biological and social systems[23], and deserves more intensive study in that context.

[21] Of course, this discussion can be put into the larger context of symmetry breaking. In that case, one can suggest a number of mechanisms that might break a pre-existing symmetry and cause associations and meaning to emerge. Which of these mechanisms can be subsumed under the rubric of a sparsely sampled space is a semantic question. But in the context of many social and biological systems, a sparsely sampled configuration space (interpreted more narrowly) may turn out to have been a precondition for the emergence of meaning.

[22] See, for example, *Spin Glasses*, K. Fischer and J. Hertz (Cambridge University Press, Cambridge, 1991); *Spin Glass Theory and Beyond*, M. Mezard, G. Parisi, and M. Virasoro (World Scientific, Singapore, 1987).

Finally, it is clear that our work has implications for the study of a wide range of complex adaptive systems. The apparent generality of the phase structure suggests that this may be a persistent feature of a large number of specific systems, even when more realistic details are included. At the very least, this underlying phase structure is likely to be a base upon which the behavior of more complicated specific systems are built.[24] In addition, we believe our work raises questions about the underlying epistemology of complex systems. There are currently few, if any, established general principles for the emergent behavior of complex systems. In fact, we do not even know what the proper questions are to ask about such systems. For example, we do not know, in general, what features of systems are specifically dependent on the details of the systems, and what are more generic. In our simple model, for instance, one might not have expected that $\sigma$ in the efficient phase would have been so strongly dependent on the particular distribution of strategies to the agents. This absence of guiding principles, of course, has profound implications for the veracity of conclusions that can be drawn from more realistic models of specific systems. We believe that an epistemology for complex systems will emerge from the aggregation of insights garnered over time from a range of research in this area. The investigation of very simple prototypical systems, such as the one presented in this paper, has much to contribute to the evolution of an epistemology of complex systems.

---

[23] See, for example, *The Economy as an Evolving Complex System*, P. Anderson, K. Arrow and D. Pines (eds.), (Addison-Wesley, Reading MA, 1988); *The Complexity of Cooperation*, R. Axelrod, (Princeton University Press, Princeton NJ, 1997).

[24] In this regard, it is interesting to note that the SFI artificial stock market described in W.B. Arthur, J. Holland, B. LeBaron, R. Palmer, and P. Tayler, Santa Fe Institute Working Paper 96-12-093 (1996) describes two phases as a function of the rate at which agents are allowed to evolve their strategies. The SFI artificial market model is much more complex than the games described in this paper. However, it may be that the evolution rate in the SFI market model can be thought of as a proxy for the size of the strategy space available to the agents, and that the two phases seen there are reflections of the underlying dynamics we have discussed in this paper.

Acknowledgements: One of us, (RS) would like to thank Stephen Fahy for stimulating discussions and for his hospitality at University College, Cork, where part of this work was done.

*Note added*: After this work was completed we learned of the recent work by M. de Cara, O. Pla, and F. Giunea (Los Alamos preprint archive, cond-mat/9811162) in which some studies of the efficient phase of the original minority game are reported.

## Figure Captions

**Fig. 1**. An example of an m=3 strategy.

**Fig. 2**. $\sigma$, the standard deviation of the time series of the number of agents joining group 1 ($L_1$) as a function of m, for N=101 agents, each with s=1 strategies. Each dot represent an independent run; 32 runs of 10,000 steps each were performed for each m. The horizontal line (at $\sigma$=5) is the value of $\sigma$ for the random choice game (RCG) described in the text.

**Fig. 3**. $\sigma$ as a function of m for N=101 agents, each with s=2 strategies. Note the broad spread in values of $\sigma$ from the runs done with m<6.

**Fig. 4** **a)** A histogram of the conditional probabilities $P(1|u_k)$ with k=m=4. There are 16 bins corresponding to the 16 possible combinations of four 0's and 1's. The bin numbers, when written in binary form, yield the strings $u_k$. **b)** A histogram of the conditional probability $P(1|u_k)$ with k=5 for the game played with m=4. There are 32 bins corresponding to the 32 possible combinations of five 0's and 1's.

**Fig. 5**. A histogram of the conditional probability $P(1|u_k)$ with k=m=6. There are 64 bins corresponding to the 64 combinations of six 0's and 1's.

**Fig. 6 a)** $\sigma$ as a function of N, for s=2, and for m=3 and m=16, on a log-log scale. Note that for m=3, when the system is in an informationally efficient phase, $\sigma \propto N$, while for m=16 (in the information inefficient phase), $\sigma \propto N^{1/2}$. **b)** The relative spread in $\sigma$, $\Delta\sigma/\sigma$, as a function of N, for s=2, and for m=3 and m=16, on a log-log scale. Note that in both cases $\Delta\sigma/\sigma$ is independent of N, indicating that in the information efficient phase (m=3), $\Delta\sigma \propto N$, while in the information inefficient phase (m=16), $\Delta\sigma \propto N^{1/2}$.

**Fig. 7**. $\sigma^2/N$ as a function of $z \equiv 2^m/N$ for s=2 and for various values of N and m, on a log-log plot.

**Fig. 8**. $\sigma$ as a function of m for s=6 and N=101. Compare to Fig. 3 (s=2).

**Fig. 9a**) $\sigma$ as a function of N, at fixed s=6, for m=3 and m=16, on a log-log scale. As in Fig. 6a (s=2) for m = 3 (in the informationally efficient phase) $\sigma \propto N$, while for m=16 (in the information inefficient phase), $\sigma \propto N^{1/2}$. **b)** The relative spread in $\sigma$, $\Delta\sigma/\sigma$, as a function of N, for s=6, and for m=3 and m=16, on a log-log scale. As in Fig. 6b (s=2), $\Delta\sigma/\sigma$ is independent of N for m=3 and m=16, indicating that in the information efficient phase (m=3) $\Delta\sigma \propto N$, while in the information inefficient phase (m=16), $\Delta\sigma \propto N^{1/2}$.

**Fig. 10**. $\sigma^2/N$ as a function of $z \equiv 2^m/N$ for s=6 and for various values of N and m, on a log-log scale. Compare to Fig. 7.

**Fig. 11**. $\sigma^2/N$ as a function of $\zeta \equiv z/z_c(s)$, for a range of values of s, where $z \equiv 2^m/N$ (as in Figures 7 and 10), and $z_c(s)$ is the value of z at which $\sigma$ is a minimum for a given s. Thus the minima of $\sigma^2/N$ in this plot are at $\zeta = 1$.

**Fig. 12 a)** A histogram of the conditional probabilities $P(1|u_k)$ with k=m=4 and with s=6. Compare to Figure 4a, which shows the histogram for the s=2 case. **b)** A histogram of the conditional probability $P(1|u_k)$ with k=5, m=4 and s=6. Compare to Figure 4b, which shows the results for the s=2 case.

**Fig. 13**. A histogram of the conditional probability $P(1|u_k)$ with k=m=7 and with s=6. Compare to Figure 5, which shows a histogram of a game in the inefficient phase for the s=2 case.

**Fig. 14**. Q(m), the mutual information in the time series of minority groups, as a function of $\zeta \equiv z/z_c(s)$, on a semi-log plot, for values of s=2 and s=16. See equation (3.1) for the definition of Q(m).

**Fig. 15 a)** A representative short segment of a typical time series of $L_1$, the number of agents joining group 1, for the case of N=101, m=2 and s=2 (in the informationally efficient phase). The zero of the vertical scale means a population in group 1 of 50 agents. The horizontal dashed lines show the range of one standard deviation for the N=101 random choice game (RCG). The circles represent those times at which the previous two minority groups were 01, as explained in the text. **b)** A short segment of $L_1$ for a game in the inefficient phase, m=8, N=101 and s=2. The zero of the vertical scale again represents a population in group 1 of 50 agents, and the horizontal dashed lines show the range of one standard deviation for the N=101 random choice game (RCG).

**Fig. 16**. A short segment of $L_1$ for m=2, N=101 with s=6 strategies per agent. The zero of the vertical scale represents a population in group 1 of 50 agents, and the horizontal dashed lines show the range of one standard deviation for the N=101 random choice game (RCG).

**Fig. 17**. POED and POOD as a function of m, for N=101 and s=2. POED and POOD are defined in section IVB in the text.

**Fig. 18**. POED and POOD as a function of m, for N=101 and s=6.

**Fig. 19**. A short segment of $L_1$ for m=5, N=101, and s=2, (i.e., in the transition region, just below $m_c$).

**Fig. 20**. An example strategy from an 11-dimensional strategy space. This strategy space is produced by inserting five “don’t care” symbols (*) in the left-most column of a subset of length m=4 strings. This 11 dimensional strategy space yields an interpolated value of $m = \log_2[11]=3.46$.

**Fig. 21**. **a)** $\sigma$ as a function of interpolated values of m, for $4 < m < 7$, for N=101 and s=2. The ranges $4<m<5$ and for $6<m<7$ are each divided into 16 bins, and the range $5<m<6$ is divided into 32 bins. There are 16 independent runs for each (integer and non-integer) value of m. Each independent run includes a different random placement of asterisks (*) to define the strategy space, as well as a random distribution of strategies to the agents. **b)** The average of the $\sigma$’s from fig. 21a over the 16 runs for each bin as a function of interpolated m for $4 < m < 7$ and for N=101 and s=2.

**Fig. 22**. Average values of POED and POOD for interpolated values of m, $4 < m < 7$, for N=101 and s=2. The bin structure is the same as in Fig. 21. Each point represents the average of the POED and POOD over the 16 runs for each bin.

**Fig. 23**. $I^{[2]}_m$, the square summed information available to the strategies in the time series of minority groups (G), defined in eq. (4.1b), as a function of integer m for N=101, s=2. Values for 8 independent runs are plotted for each m.

**Fig. 24**. $\Gamma_m$, (eq. (4.3)) the correlation between $I_m$ (the information in G available to the strategies, defined in eq. (4.1a)) and $\Psi_m$ (the skewness in the distribution of strategies, defined in eq. (4.2)), as a function of integer m for N=101, s=2 Note that as m increases, there is increasing correlation between the information left in G and the skewness of the agents' strategies.

**Fig. 25**. $\phi_m$ (eq. (4.5)), the average squared deviation between the observed conditional probabilities $P(1|u_m)$ and calculated estimates, $P_R(1|u_m)$ defined in (4.4), as a function of m, for N=101 and s=2. $\phi_m$=0 indicates the conditional probabilities P and $P_R$ are the same.

**Fig. 26**. Relationship between agents' inter-agent strategy distances, ($D_b$), and intra-agent strategy distances, $D_h$, and their accumulated wealth in typical runs with N=101 and s=2, for various values of m. $D_h$ and $D_b$ are defined in eqs. (5.1) and (5.2), respectively. **a)** m=3, $D_b$. **b)** m=3, $D_h$. **c)** m=4, $D_b$. **d)** m=4, $D_h$. **e)** m=5, $D_b$. **f)** m=5, $D_h$. **g)** m=6, $D_b$. **h)** m=6, $D_h$. **i)** m=7, $D_b$. **j)** m=7, $D_h$. **k)** m=8, $D_b$. **l)** m=8, $D_h$.

**Fig. 27**. $\chi$(s), defined in eq. (6.11), as a function of s.

---